\documentclass[reqno,12pt]{amsart}

\usepackage{amsmath,amssymb}

\usepackage[all]{xy}
\usepackage{color}

\usepackage{graphicx}
\usepackage{hyperref}
\usepackage[mathscr]{eucal}
\usepackage{enumerate}

\numberwithin{equation}{section}

\newtheorem{Def}{Definition}[section]
\newtheorem{Thm}[Def]{Theorem}
\newtheorem{Prp}[Def]{Proposition}
\newtheorem{Lemma}[Def]{Lemma}

\newcommand{\beq}{\begin{equation}}
\newcommand{\eeq}{\end{equation}}
\newcommand{\Proof}{\begin{proof}}
\newcommand{\QED}{\end{proof} \noindent}

\newcommand{\R}{\mathbb{R}}

\newcommand{\N}{\mathbb{N}}

\newcommand{\oneone}{C^{1,1}}
\newcommand{\oone}{C^{0,1}}

\allowdisplaybreaks[1]

\begin{document}

\title[Spacetime is Locally Inertial]{Spacetime is Locally Inertial at Points of General Relativistic Shock Wave Interaction between Shocks from Different Characteristic Families}


\author[M.\ Reintjes]{Moritz Reintjes}
\address{Departamento de Matem{\'a}tica \\ Instituto Superior T{\'e}cnico \\ 1049-001 Lisbon \\ Portugal}
\email{moritzreintjes@gmail.com}

\begin{abstract}
We prove that spacetime is locally inertial at points of shock wave collision in General Relativity. The result applies for collisions between shock waves coming from different characteristic families in spherically symmetric spacetimes. We give a constructive proof that there exists coordinate transformations which raise the regularity of the gravitational metric tensor from $C^{0,1}$ to $C^{1,1}$ in a neighborhood of such points of shock wave interaction and a $C^{1,1}$ metric regularity suffices for locally inertial frames to exist. This result was first announced in \cite{ReintjesTemple2} and the proofs are presented here. This result corrects an error in our earlier publication \cite{ReintjesTemple}, which led us to the wrong conclusion that such coordinate transformations, which smooth the metric to $C^{1,1}$, cannot exist. Our result here proves that regularity singularities, (a type of mild singularity introduced in \cite{ReintjesTemple}), do \emph{not exist} at points of two interacting shock waves from different families in spherically symmetric spacetimes, and this generalizes Israel's famous 1966 result to the case of such shock wave interactions. The strategy of proof here is an extension of the strategy outlined in \cite{ReintjesTemple}, but differs fundamentally from the method used by Israel. The question whether regularity singularities exist in more complicated shock wave solutions of the Einstein Euler equations still remains open.
\end{abstract}

\maketitle

\tableofcontents

\section{Introduction}



In Einstein's theory of General Relativity (GR), the gravitational field is accounted for by the curvature of space-time, and Special Relativity together with Newton's theory of gravity are recovered locally under the assumption that space-time is {\it locally inertial}, \cite{Einstein}.  That is, locally, General Relativity describes the physics of Special Relativity, perturbed by second order acceleration effects due to spacetime curvature (gravity).  But the assumption that spacetime is locally inertial is equivalent to assuming the metric $g$ is smooth enough at each point $p$ to admit coordinates in a neighborhood of $p$, coordinates in which $g$ is exactly the Minkowski metric at $p$, such that all first order derivatives of $g$ vanish at $p$, and all second order derivatives of $g$ are bounded almost everywhere, (so-called locally inertial coordinate frames). This requisite level of smoothness for the gravitational metric tensor is $C^{1,1}$, \cite{SmollerTemple}. However, the Einstein equations determine the smoothness of the gravitational metric tensor by the evolution they impose. Therefore, regularity theorems for the Einstein equations are required to determine whether space-time is locally inertial. 

In the presence of shock waves, it remains an open problem whether space-time is always locally inertial, or whether there exists {\it regularity singularities} where the Einstein equations are satisfied weakly, but the spacetime metric is no smoother than Lipschitz continuous, (i.e., $C^{0,1}$), in every coordinate system. This issue comes into sharp focus given that interacting shock wave solutions were proven in \cite{GroahTemple} to exist in $C^{0,1}$, but it remains an open problem as to whether coordinate transformations always exist that smooth the metric to $C^{1,1}$, the requisite level of smoothness for spacetime to be locally inertial. At points on a \emph{single} smooth shock wave, a famous result of Israel shows that space-time is always locally inertial \cite{Israel,SmollerTemple}, by proving that the metric is $C^{1,1}$ in Gaussian normal coordinates. But, for more complicated shock wave solutions, like those proven to exist in \cite{GroahTemple}, which can involve various types of shock wave interactions, it is unknown whether the locally inertial nature of space-time is preserved under the evolution of Einstein's equations.  Namely, at points of shock wave interaction, the shock surfaces no longer have the continuous normal vector fields required for constructing Gaussian normal coordinates, and Israel's method fails in its very first step. 

The problem whether spacetime is locally inertial at points of shock wave interaction was first laid out by the authors in \cite{Reintjes,ReintjesTemple}. There, we introduced the notion of a regularity singularity as a point in spacetime where the $C^{0,1}$ metric cannot be mapped to a $C^{1,1}$ metric by any $C^{1,1}$ coordinate transformation, which is natural in light of the shock wave solutions in \cite{GroahTemple}, where the metric is known to be $C^{0,1}$. We then addressed the existence question for regularity singularities at points of shock wave interaction in spherically symmetric spacetimes. However, there is an \emph{error} in the analysis in \cite{ReintjesTemple} at this point, and the announcement made regarding the proof of existence of regularity singularities is \emph{not} valid. 

In this paper, we correct the error in \cite{ReintjesTemple}. By doing so, we develop a proof that the gravitational metric {\it can} always be smoothed at points where shock waves from different characteristic families interact in spherically symmetric spacetimes. This result was first announced in \cite{ReintjesTemple2} and we here present its proofs. The strategies announced in our RSPA article, \cite{ReintjesTemple}, remain valid and provide the point of departure for the present paper.   Our new result does not resolve the open problem as to the existence of regularity singularities in general shock wave solutions of the Einstein Euler equations, but is a first and important step in extending Israel's result to solutions containing shock wave interactions in spherically symmetric spacetimes. 

One cannot ignore the issue whether regularity singularities exist for the Einstein Euler equations and thus in the most basic relativistic matter models for the evolution of stars, galaxies or the universe itself, since shock waves form in the Euler equations whenever the flow is sufficiently compressive \cite{Lax,Christodoulou}. Moreover, the question whether one can smooth the gravitational metric to $C^{1,1}$ is basic for the problem as to whether the Einstein Euler equations have a regular zero-gravity limit, at the level of shock waves.\footnote{The gravitational potential enters the GR Euler equations through first order metric derivatives, which, in a locally inertial frame, decrease linearly to zero as one approaches the coordinate center. From this, at least in principal, we conclude that, sufficiently close to the coordinate center, a given shock wave solutions of the Euler equations in a background spacetime is close to the solution of the flat Euler equation for the same initial data. However, if first order metric derivatives do not vanish at the coordinate center, this procedure could fail.} Thus, without a proof that space-time is locally inertial, it is unclear if the nature of shock wave interactions in GR is locally the same as it is in Minkowski space, c.f. \cite{SmollerTemple2}. Let us finally remark that a regularity singularity, if it exists, will be less severe than the well-known curvature singularities, but can nevertheless have implications to the physics of space-time. In \cite{ReintjesTemple3}, the authors explore the physical implications of the hypothetical structure of regularity singularities to gravitational radiation by studying the linearized Einstein equations in so-called approximate locally inertial coordinate systems.

Before we state the main result, we comment on our method of proof. In \cite{Israel}, Israel showed that conservation of the (bounded) matter source across a smooth shock wave is equivalent to the continuity of the second fundamental form across the surface, and using this, it follows that the gravitational metric is $C^{1,1}$ in Gaussian normal coordinates. But these coordinates fail to exist at points of shock wave interaction, because the shock surface no longer has the continuous normal vector field required for their construction. In our argument here we make no connection between conservation and the geometry of the shock surface through the second fundamental form, but rather use conservation of the sources, in the form of the Rankine Hugoniot (RH) conditions, to directly construct the Jacobians of the sought after coordinate systems. In more detail, instead of focusing on finding an explicit coordinate system where the metric is $C^{1,1}$, our new proof is based on constructing Jacobians that meet the so-called {\it Smoothing Condition}, and solve the {\it Integrability Condition}, both introduced in \cite{ReintjesTemple}. (In fact, we construct the class of all $C^{0,1}$ Jacobians that have these properties, properties which are necessary and sufficient to lift the metric regularity to $C^{1,1}$.) To solve the Integrability Condition, an existence theory is developed for a perculiar system of non-local hyperbolic PDE's. From this theory, we obtain a proof that there exists coordinate systems in which the metric can be smoothed to $C^{1,1}$ in a neighborhood of the points of shock wave interaction between shocks from different families, proving our main result.

To state our main result precisely, let $g_{\mu\nu}$ denote a spherically symmetric spacetime metric in Standard Schwarzschild Coordinates (SSC), that is, the metric takes the form
\beq\label{metric SSC}
ds^2=g_{\mu\nu}dx^{\mu}dx^{\nu}=-A(t,r)dt^2+B(t,r)dr^2+r^2d\Omega^2,
\eeq
where $d\Omega^2=d\vartheta^2 + \sin^2(\vartheta) d\varphi^2$ is the line element on the unit $2$-sphere, c.f. \cite{Weinberg}.   In Section \ref{sec: point regular interaction}, we make precise the definition of a point of regular shock wave interaction in SSC between shocks from different families.  Essentially, this is a point in $(t,r)$-space where two distinct shock waves enter and leave a point $p$, such that the metric is Lipschitz continuous across the shocks and smooth away from the shocks, the RH jump conditions hold across each shock curve continuously up to the point of interaction $p$, derivatives are continuous up to the shock boundaries, and the SSC Einstein equations hold weakly in a neighborhood of $p$ and strongly away from the shocks. This is the main result:

\begin{Thm}  \label{TheoremMain} 
Suppose that $p$ is a  point of regular shock wave interaction in SSC between shocks from different families, in the sense that condition (i) - (iv) of Definition \ref{shockinteract} hold, for the SSC metric $g_{\mu\nu}$.  Then the following are equivalent:
\begin{enumerate}[(i)]
\item There exists a $\oneone$ coordinate transformation $x^\alpha\circ(x^\mu)^{-1}$ in the $(t,r)$-plane, with Jacobian $J^\mu_\alpha$, defined in a neighborhood $\mathcal{N}$ of $p$, such that the metric components $g_{\alpha\beta} = J^\mu_\alpha J^\nu_\beta g_{\mu\nu}$ are $\oneone$ functions of the coordinates $x^{\alpha}$.
\item The Rankine Hugoniot conditions, \eqref{RHwithN1} - \eqref{RHwithN2}, hold across each shock curve in the sense of (v) of Definition \ref{shockinteract}.
\end{enumerate}
Furthermore, the above equivalence also holds for the full atlas of $C^{1,1}$ coordinate transformations, not restricted to the $(t,r)$-plane.
\end{Thm}

In our proof of Theorem \ref{TheoremMain}, for the construction of the Jacobians smoothing the metric not to break down, several identities must be satisfied sharply.   Our conclusion in \cite{ReintjesTemple}  was based on the (false) discovery that one of those identities is violated, but the violation was due to the error of neglected terms that should have been present in our ansatz for the Jacobians meeting the  Smoothing Condition.   In our corrected proof presented here, we can now verify that all these identities hold identically as a consequence of the RH conditions, which serves us as a second check that all terms have now been accounted for, and consequently a second check on the validity of the new argument. But, other than the proof, we have no intuitive, heuristic or physical argument to support the claim of the above Theorem or its opposite, which we take as an indication of the subtlety of the problem whether the metric can be smoothed to $C^{1,1}$, even in the simplest setting we considered here. A further indication of this subtlety is that our new method of proof indicates that smoothness of the metric, higher than $C^{1,1}$, must be given up away from the shocks in order to smooth the metric to $C^{1,1}$ in a neighborhood of a point of shock wave interaction, (c.f. footnote 2 below). 
 
Historically, the issue of the smoothness of the gravitational metric tensor across interfaces began with the matching of the Schwarzschild solution to the vacuum across an interface, followed by the celebrated work of Oppenheimer and Snyder who gave the first dynamical model of gravitational collapse by matching a pressure-less fluid sphere to the Schwarzschild vacuum spacetime across a dynamical interface \cite{OppenheimerSnyder}.     In his celebrated 1966 paper \cite{Israel}, Israel gave the definitive conditions for regular matching of gravitational metrics at smooth interfaces, by showing that if the second fundamental form is continuous across a single smooth interface, then the RH conditions also hold, and Gaussian normal coordinates provide a coordinate system in which the metric is smoothed to $C^{1,1}$.  In \cite{SmollerTemple}, Smoller and Temple extended the Oppenheimer-Snyder model to nonzero pressure by matching the Friedmann metric to a static fluid sphere across a shock wave interface that modeled a blast wave in GR.  Groah and Temple then addressed these issues rigorously in the first general existence theory for shock wave solutions of the Einstein-Euler equations in spherically symmetric spacetimes, \cite{GroahTemple}, which has later been extended to Gowdy spacetimes \cite{BarnesLeflochSchmidtStewart}. Note that weak solutions of the Einstein vacuum equation with a $C^{1,1}$ metric regularity have been constructed, without any symmetry assumptions on spacetime, c.f. \cite{LeFlochChen} and references therein. However, these results only address the special case when no matter fields are present and hence do not apply to shock wave solutions of the Einstein Euler equations.

In the remainder of the introduction we outline the content of this paper and thereby also our strategy of proof, the main ideas of which have partially been introduced in \cite{Reintjes,ReintjesTemple}. In Section \ref{sec: point regular interaction}, we motivate and make precise the structure we assume on the shock wave interaction we consider in Theorem \ref{TheoremMain}. Essentially this is the same structure as in \cite{ReintjesTemple}, except that we not only specify the structure of the incoming shocks but also of the outgoing ones, which excludes rarefaction waves from our analysis here. In the special case of a perfect fluid source, this structure corresponds to the incoming shock waves to lie within different characteristic families. (An outgoing shock and rarefaction wave would correspond to an interaction of incoming shock waves in the same families.)

In Section \ref{sec: about oone across} we discuss some properties of functions Lipschitz continuous across some hypersurface but smooth elsewhere. In particular, we introduce a canonical form all such functions can be represented in, c.f. Lemma \ref{characerization2}, which we apply in the proof of Theorem \ref{TheoremMain} to represent the Jacobians. Moreover, we study the connection to the RH conditions, which leads to an equivalent formulation of the RH conditions purely in terms of the jumps in the derivatives of the SSC metric and the shock speed, c.f. \eqref{jumpone}, \eqref{jumptwo} and \eqref{[Bt]shockspeed=[Ar]}. The results of this section were first stated in \cite{ReintjesTemple}.

In Section \ref{intro to method}, we derive a necessary condition on a Jacobian, $J^\mu_\alpha$, for smoothing the metric regularity to $C^1$ across a single shock surface. This so-called smoothing condition was first introduced in \cite{ReintjesTemple}  and is given by
$$
[J^\mu_{\alpha,\sigma}] J^\nu_\beta g_{\mu\nu} + [J^\nu_{\beta,\sigma}]J^\mu_\alpha g_{\mu\nu} =  -J^\mu_\alpha J^\nu_\beta [g_{\mu\nu,\sigma}],
$$
where $[\cdot]$ denotes the jump across the shock surface, each comma between indices denotes differentiation and $g_{\mu\nu}$ denotes the metric in SSC, c.f. \eqref{smoothingcondt1_SSC}. We then proceed by solving the smoothing condition point-wise for $[J^\mu_{\alpha,\sigma}]$, c.f. Lemma \ref{smoothingcondt2}, which is the first step in our construction of Jacobians smoothing the metric to $C^{1,1}$. In Section \ref{Israel's Thm with new method}, we give an alternative proof of Israel's Theorem, regarding the metric smoothing across a single shock curve, with our method of constructing Jacobians, which is a convenient and rigorous way to introduce our method, since many technical issues of the interacting shock case do not appear.

In the next step towards the construction, subject of Section \ref{Sec: Canonical Jacobian}, we define ~$(\varphi_i)^\mu_\alpha =-\tfrac12 [J^\mu_{\alpha,r}]_i$, where $[J^\mu_{\alpha,r}]_i$ is the value obtained from solving the smoothing condition across the $i$-th shock curve, $\gamma_i(t)=(t,x_i(t))$, and introduce the $C^{0,1}$ functions 
$$
J^\mu_\alpha(t,r) = \sum_{i=1,2} (\varphi_i)^\mu_\alpha(t)\, |x_i(t)-r| \,+ \, \Phi^\mu_\alpha(t,r),
$$
where $\Phi^\mu_\alpha$ denote some arbitrary smooth function, c.f. \eqref{Jacobian2}. This is our canonical form for Jacobians. The key observation here is that the above functions satisfy the smoothing conditions across each of the shock curves if and only if the Rankine Hugoniot conditions hold, c.f. Lemma \ref{Jacobian1}. The error in \cite{ReintjesTemple}  came from neglecting terms in the $(\varphi_i)^\mu_\alpha$.

For the $J^\mu_\alpha$ introduced in the above equation to be actual Jacobians, which are integrable to coordinates, we need to prove in the next step that the integrability condition,
$$
J^\mu_{\alpha,\beta}=J^\mu_{\beta,\alpha} \, ,
$$ 
holds. This is the subject of Section \ref{Sec: Integrability Condition}. From the definition of $J^\mu_\alpha$ above, since the $(\varphi_i)^\mu_\alpha$ are completely determined, it is clear that the integrability condition can only be satisfied through the free functions $\Phi^\mu_\alpha$. In fact, setting $U=(\Phi^t_0,\Phi^r_0)$ and assuming $(\Phi^t_1,\Phi^r_1)$ to be fixed smooth functions, the above equation can be written as a PDE for $U$ of the form
$$
U_t+c\,U_r=F(U),
$$ 
for some $C^{0,1}$ scalar function $c$ and for some $F$ depending linearly on $U$, $U\circ \gamma_i$ and $\frac{d}{dt}(U\circ \gamma_i)$. The above equation were a strictly hyperbolic PDE, if $F$ would not depend on $U\circ \gamma_i$ and $\frac{d}{dt}(U\circ \gamma_i)$. Through this dependence, the integrability conditions becomes a \emph{non-local} system of PDE's for $U$. We develop a $C^{0,1}$ existence theory for this type of PDE in Section \ref{Sec: Non-local PDE}. Once the existence result is established, we use the RH conditions in Section \ref{Sec: Bootstrapping} to proof that the solution of the integrability condition is indeed $C^{1,1}$ regular. It is crucial to establish this level of regularity for the free function $\Phi^\mu_\alpha$ in order to not interfere with the smoothing condition and to preserve the $C^{1,1}$ metric regularity away from the shock curves.\footnote{Interestingly, there is an inherent loss of smoothness of $\Phi^\mu_\alpha$ across the characteristic curve which passes through the point of shock wave interaction, even though this characteristic curve lies within the region of smoothness of the original metric. As a result, across this characteristic the metric in the constructed coordinates is $C^{1,1}$, but seems to be no smoother.}

In principal, the above strategy suffices to obtain Jacobians defining coordinates $x^\alpha$ in which the metric $g_{\alpha\beta}=J^\mu_\alpha J^\nu_\beta g_{\mu\nu}$ is in $C^{1,1}$. However, since the shock speeds change discontinuously at the point of interaction $p$ the shock curves are only Lipschitz continuous at $p$. Therefore, we are forced to first pursue the above construction before and after the interaction separately and then match the resulting Jacobians across the ($t=0$)-interface, assuming that the interaction happens at $t=0$. This final step in the construction is achieved in Section \ref{Sec: Matching Conditions}. To ensure that the matching does not lower the metric regularity across the ($t=0$)-interface, the matching must obey the so-called \emph{matching conditions}. In principal, these conditions could lead to an overdetermined list of conditions on the $\Phi^\mu_\alpha$, but, surprisingly, the RH conditions prevent this from happening. In Section \ref{Sec: Proof_MainThm_Wrap-up}, we finally prove that the constructed Jacobian indeed maps the SSC metric $g_{\mu\nu}$ to a metric $g_{\alpha\beta}$, which is $C^{1,1}$ in a neighborhood of the shock interaction with respect to the coordinates $x^\alpha$.

\section{Preliminaries}\label{sec preliminaries}

Spacetime is a four dimensional Lorentz manifold, that is, a manifold $M$ endowed with a metric tensor $g_{\mu\nu}$ of signature~$(-1,1,1,1)$, the so-called Lorentz metric. Around each point $p\,\in\,M$ there exists an open neighborhood $\mathcal{N}_p$ (called a patch) together with a homeomorphism~$x=(x^0,...,x^3):\,\mathcal{N}_p\rightarrow\R^4$, which defines coordinates on its image in $\R^4$. $\mathcal{N}_p$ together with the homeomorphism $x$ are called a chart. The collection of all such charts (covering the manifold) is called an \emph{atlas}. If the intersection of two coordinate patches is nonempty, ~$x\circ y^{-1}$ defines a mapping from an open set in $\R^4$ to another one (referred to as a coordinate transformation). If all coordinate transformations in the atlas are $C^k$ differentiable, the manifold $M$ is called a $C^k$-manifold and its atlas a $C^k$-atlas. 

We use the Einstein summation convention, that is, we always sum over repeated upper and lower indices, e.g.~$v^\mu w_\mu = v^0 w_0 +...+ v^3 w_3$ for~$\mu,\nu\in\{0,...,3\}$. Moreover, we use the type of index to indicate in which coordinates a tensor is expressed in, for instance, ${T^\mu}_\nu$ denotes a $(1,1)$-tensor in coordinates $x^\mu$ and ~${T^\alpha}_\beta$ denotes the same tensor in (different) coordinates $x^\alpha$. Under a change of coordinates, tensors transform via contraction with the \emph{Jacobian} $J^\mu_\alpha$ of the coordinate transformation, which is given by
\beq\label{Jacobian_Prelim}
J^\mu_\alpha= \frac{\partial x^\mu}{\partial x^\alpha} \, .
\eeq
For example, a vector transforms as $v^\mu=J^\mu_\alpha v^\alpha$ and a one-form as $v_\mu=J^\alpha_\mu v_\alpha$, where $J^\alpha_\mu$ denotes the (point-wise) inverse of \eqref{Jacobian_Prelim}. The metric tensor transforms according to
\beq \label{covariant transfo}
g_{\alpha\beta} = J^\mu_\alpha J^\nu_\beta g_{\mu\nu} \, ,
\eeq
which is crucial for our method here. $g^{\mu\nu}$ denotes the inverse of the metric, defined via
\beq \label{metric inverse}
g^{\mu\sigma}g_{\sigma\nu}={\delta^\mu}_\nu,
\eeq
for $\delta^\mu_\nu$ being equal to $1$ when $\mu=\nu$ and $0$ otherwise. By convention, we raise and lower tensor indices with the metric, for example, ${T^\mu}_\nu=  T^{\mu\sigma} g_{\nu\sigma} $.

In \eqref{Jacobian_Prelim}, the Jacobian is defined for a given change of coordinates. Vice versa, for a set of given functions $J^\mu_\beta$ to be a Jacobian of a coordinate transformation it is necessary and sufficient to satisfy (see appendix \ref{sec IC} for details),
\begin{eqnarray}
J^\mu_{\alpha,\beta} &=& J^\mu_{\beta,\alpha} \label{IC} \\
\det \left( J^\mu_\alpha \right) &\neq& 0, \label{nonzero det, prelim}
\end{eqnarray}
where ~$f_{,\alpha}= \frac{\partial f}{\partial x^\alpha}$ denotes partial differentiation with respect to coordinates $x^\alpha$. We henceforth refer to \eqref{IC} as the \emph{integrability condition} or as the \emph{curl equations}.

The \emph{Einstein equations}, which govern the gravitational field, read 
\beq\label{EFE}
G^{\mu\nu}=\kappa T^{\mu\nu} \,
\eeq
in units where $c=1$. $T^{\mu\nu}$ denotes the energy-momentum-tensor, describing the energy- and matter-content of spacetime, and 
\beq \label{Einstein tensor}
G^{\mu\nu}=R^{\mu\nu}-\frac12 R g^{\mu\nu} + \Lambda g^{\mu\nu}
\eeq
is the Einstein tensor \cite{Einstein}. $\Lambda$ denotes the cosmological constant and $\kappa= 8\pi \mathcal{G}$ incorporates Newton's gravitational constant $\mathcal{G}$. (Our method applies regardless the choice of $\Lambda \in \R$.) The Ricci tensor $R_{\mu\nu}$ is defined as the trace of the Riemann tensor, ~$R_{\mu\nu}=R^\sigma_{\mu\sigma\nu}$, and the scalar curvature is~$R={R^\sigma}_\sigma$. The metric tensor $g_{\mu\nu}$ enters the Einstein equations through the Christoffel symbols
\beq \label{Christoffel symbols}
\Gamma^\mu_{\nu \rho} = \frac12 g^{\mu\sigma}\left( g_{\nu\sigma,\rho} +g_{\rho\sigma,\nu} -g_{\nu \rho,\sigma} \right) ,
\eeq
since the Riemann curvature tensor is defined\footnote{The Riemann curvature tensor introduced in \cite{Weinberg} differs from the one used by us and in \cite{HawkingEllis} by a factor of $-1$ which, in \cite{Weinberg}, is compensated for by setting $\kappa = -8\pi \mathcal{G}$. MAPLE uses the sign convention in \cite{Weinberg} for the Riemann tensor, which is important to keep in mind when computing the Einstein tensor for \eqref{one} - \eqref{four} below with MAPLE.} as
\beq \nonumber
{R^\mu}_{\nu \rho \sigma} =  
\Gamma^\mu_{\ \nu \sigma,\rho}  - \Gamma^\mu_{\ \nu \rho,\sigma}
+\Gamma^\mu_{\ \lambda \rho}\Gamma^\lambda_{\ \nu \sigma}
- \Gamma^\mu_{\ \lambda \sigma}\Gamma^\lambda_{\ \nu \rho}  \, .
\eeq

By construction, due to the Bianchi identities of the Riemann curvature tensor, the Einstein tensor is divergence free,
\beq\label{div free}
\text{div}\, {G}=0,
\eeq
with respect to the covariant divergence of the metric $g_{\mu\nu}$. Through \eqref{div free}, the Einstein equations ensure conservation of energy in the matter source, that is,
\beq \label{Euler}
\text{div}\, {T}=0\, ,
\eeq
which was one of the guiding principles Einstein followed in the construction of the Einstein tensor \cite{Einstein}. In the case of a perfect fluid, that is, when the energy momentum tensor is given by
\beq
T^{\mu\nu}=(p+\rho)u^\mu u^\nu + p g^{\mu\nu} \, ,
\eeq
\eqref{Euler} are the general relativistic \emph{Euler} equations, with $\rho$ being the density, $p$ the pressure and $u^\mu$ the tangent vector of the fluid flow normalized such that $u^\nu u_\nu=-1$, (see for example \cite{HawkingEllis,SmollerTemple2}). \eqref{EFE} and \eqref{Euler} are the coupled \emph{Einstein Euler} equations, a system of 14 partial differential equations which closes once we prescribe a (barotropic) equation of state, $p=p(\rho)$, leaving 14 unknowns to solve for: $g_{\mu\nu}$, $\rho$ and $u^\nu$. In a \emph{locally inertial frame} around a point $p$, that is, in coordinates where the metric is the Minkowski metric up to second order metric-corrections in coordinate-distance to $p$, \eqref{Euler} reduce to the special relativistic Euler equations at $p$. For the Special Relativistic Euler equations, it is well-known that shock waves form from smooth initial data whenever the flow is sufficiently compressive,\cite{Christodoulou,Lax,Smoller}. This makes the study of shock waves inevitable for perfect fluid sources, arguing the result of this paper being fundamental to General Relativity.
 
Being discontinuous, shock wave solutions satisfy the conservation law \eqref{Euler} only in a weak sense, that is, 
\beq \label{Euler weak}
\int\limits_{M} T^{\mu\nu}\varphi_{,\nu} d\mu_M =0,
\eeq
for all test-functions $\varphi \: \in \: C^\infty_0(M)$, and where $d\mu_M$ denotes the volume element of spacetime. Across the hypersurface of discontinuity $\Sigma$, the so-called shock surface, the solution satisfies the Rankine Hugoniot conditions (RH conditions),
\beq\label{JC}
[T^{\mu\nu}] N_\nu =0,
\eeq
where $N^\nu$ is normal to the (time-like\footnote{Physical shock waves propagate slower than the speed of light.}) hypersurface $\Sigma$ and $[u]:=u_L-u_R$ for $u_L$ and $u_R$ denoting the left and right limit to $\Sigma$ of some function $u$ respectively. $[u]$ is referred to as the \emph{jump} in $u$ across $\Sigma$. One can bypass the weak formalism and work instead with the RH conditions due to the following basic fact: Suppose $T^{\mu\nu}$ is a strong solution everywhere away from $\Sigma$, then, $T^{\mu\nu}$ is a weak solution in a neighborhood of the $\Sigma$ if and only if the RH conditions \eqref{JC} hold everywhere on the surface, c.f. \cite{Smoller}.

We now discuss spherically symmetric spacetimes, examples of which are the Schwarz\-schild, the Oppenheimer Volkoff and the Friedmann Robertson Walker spacetimes, c.f. \cite{HawkingEllis,Weinberg}. In a spherically symmetric spacetime, assuming that one of the spaces of symmetry has constant scalar curvature equal to $1$, one can always introduce coordinates $\vartheta$ and $\varphi$ such that the metric is of the form
\beq\label{metric in SSY; general}
ds^2= -A dt^2 + B dr^2 + 2E dt dr + C d\Omega^2 \, ,
\eeq
where
\[ d\Omega^2 = d\varphi^2 + \sin^2\left(\vartheta\right)d\vartheta^2 \]
is the line element on the two-sphere and the metric coefficients $A$, $B$, $C$ and $E$ only depend on $t$ and $r$, \cite{Weinberg}. For our purposes, it suffices to define a spherically symmetric spacetime as a spacetime endowed with a metric which can be represented as in \eqref{metric in SSY; general}. It follows that a spherically symmetric metric is invariant under spatial rotation. The coordinate representation \eqref{metric in SSY; general} is preserved under coordinate transformations in the $(t,r)$-plane, that is, transformations which keep the angular variables, $\vartheta$ and $\varphi$, fixed. 

If $\partial_r C \neq 0$, one can simplify the metric further by introducing a new ``radial'' variable $r':=\sqrt{C}$ and removing the off-diagonal element through an appropriate coordinate transformation in the $(t,r')$-plane \cite{Weinberg}, denoting the resulting coordinates again by $t$ and $r$ the new metric reads
\beq\label{metric SSC}
ds^2= -A dt^2 + B dr^2 + r^2 d\Omega^2 \, .
\eeq
Coordinates where the metric is given by \eqref{metric SSC} are called Standard Schwarz\-schild Coordinates (SSC). In SSC, the metric has only two free components and the Einstein equations read
\begin{eqnarray}\label{EFEinSSC}
B_r + B \frac{B-1}{r} &=& \kappa AB^2r \, T^{00} \label{one} \\
B_t &=& -\kappa AB^2r \,T^{01} \label{two}\\
A_r - A \frac{B-1}{r} &=& \kappa AB^2r \,T^{11} \label{three}\\
B_{tt} - A_{rr} + \Phi &=& -2\kappa A B r^2 \,T^{22} \label{four}\, ,
\end{eqnarray}
with
\[\Phi := -\frac{BA_tB_t}{2AB} - \frac{B_t^2}{2B} - \frac{A_r}{r} + \frac{AB_r}{rB} + \frac{A_r^2}{2A} + \frac{A_r B_r}{2B} \, . \]

Regarding the metric regularity in the presence of shock waves, recall that a shock wave is a weak solution of the Euler equations \eqref{Euler}, discontinuous across a time-like hypersurface $\Sigma$ and smooth away from $\Sigma$. Now, by \eqref{one} - \eqref{three}, if the $T^{\mu\nu}$ are discontinuous, then the metric cannot be any smoother than Lipschitz continuous. In fact, in a spherically symmetric spacetime, as a consequence of the RH conditions, it is not possible that $T^{22}$ is discontinuous but all the remaining $T^{\mu\nu}$ are continuous. Therefore, any shock-discontinuity must be compensated for by a discontinuous first order metric derivative and the fourth Einstein equation \eqref{four} can only hold in a weak sense. In light of this, we henceforth assume that the gravitational metric in SSC is Lipschitz continuous, providing us with a consistent framework to address shock waves in General Relativity and agreeing in particular with the solution proven to exist by Groah and Temple, \cite{GroahTemple}. Moreover, Lipschitz continuity arises naturally in the problem of matching two spacetimes across an interface, c.f. \cite{Israel} or \cite{SmollerTemple}.

For the Einstein tensor to be defined point-wise almost everywhere, a $\oneone$ metric regularity is required, however, as we have just demonstrated, at the level of shock waves only Lipschitz continuity of the metric is guaranteed. For this low regularity the Einstein and Riemann curvature tensor can only be introduced in a weak (distributional) sense, (see \cite{SmollerTemple} for an definition of a weak shock solution of the Einstein equations). Nevertheless, in his $1966$ paper \cite{Israel} Israel proved that, at a \emph{single} shock surface, there exists coordinates where the metric is $\oneone$ regular if and only if the energy momentum tensor is bounded almost everywhere. Furthermore, if the metric is $C^{1,1}$ in some coordinates, then the second fundamental form is continuous across the shock surface and the Rankine Hugoniot jump conditions are satisfied. In fact, in a spherically symmetric spacetime, equivalence holds between the RH conditions and the existence of coordinates where the metric is $C^{1,1}$, as proven in \cite{SmollerTemple}. 

At the heart of Israel's method lies the choice of Gaussian Normal Coordinates with respect to the shock surface $\Sigma$, which we now explain in some more detail for a smooth shock surface in a $n$-dimensional Riemannian manifold.\footnote{This construction generalizes easily to Lorentz manifolds, as long that $\Sigma$ is non-null.} One first arranges by a smooth coordinate transformation that locally $\Sigma= \{p\in M :x^n(p)=0\}$ and then exchanges the $n$-th coordinate function by the arc-length-parameter of geodesics normal to $\Sigma$. In more detail, Gaussian Normal Coordinates are the mapping assigning the point $p\in M$ (sufficiently close to $\Sigma$) to the point
\beq\nonumber
 x^\alpha(p) = \left( s,x^{n-1}(q),...,x^1(q) \right) \  \in \R^n, 
\eeq 
where $s$ is the arc-length parameter of a geodesic curve $\gamma$ starting at the point $q\in\Sigma$ in the direction normal to $\Sigma$, with $\gamma(s)=p$, and $x^\alpha(q)=x^i(q)$ for all $\alpha=i \in \{1,..., n-1\}$. Computing now the Einstein tensor in such coordinates, one finds that each component of the resulting Einstein tensor contains only a single second order normal derivative, ~$\partial^2_{n}g_{\alpha\beta}$, while all other terms in the Einstein equations are in $L^\infty$ and thus ~$\partial^2_n g_{\alpha\beta} \,\in\, L^\infty$. Now, since all other second order metric derivatives are bounded by assumption, one arrives at the conclusion\footnote{Note that Lipschitz continuity of a function is equivalent for it to be in the Sobolev-space $W^{1,\infty}$, consisting of the functions with weak derivatives in $L^\infty$. These functions are differentiable almost everywhere \cite{Evans}.} $g_{\alpha\beta} \,\in\, \oneone$. Moreover, in Gaussian Normal coordinates the second fundamental form is given by $g_{\alpha\beta,n}$.

In Section \ref{Israel's Thm with new method}, we give a new constructive proof of Israel's result in spherically symmetric spacetimes, based on the method we introduce in Section \ref{intro to method}. It has remained an open problem whether or not such a theorem applies to the more complicated $C^{0,1}$-solutions containing shock wave interactions, for example, the SSC solutions proven to exist by Groah and Temple \cite{GroahTemple}. Our purpose here is to show that such solutions can be smoothed to $C^{1,1}$ in a neighborhood of a point of {\it regular shock wave interaction in SSC between shocks from different families}, a notion we now make precise.

\section{A Point of Regular Shock Wave Interaction in SSC bet\-ween Shocks from Different Characteristic Families}\label{sec: point regular interaction}

In this section we set up our basic framework and give the definition of a point of regular shock wave interaction in SSC, that is, a point $p$ where two radial shock waves enter or leave the point $p$ with distinct speeds. We henceforth restrict attention to radial shock waves, that is, shock surfaces $\Sigma$ that can (locally) be parameterized by
\beq \label{shock surface SSY}
\Sigma(t, \vartheta, \varphi) =(t,x(t), \vartheta, \varphi),
\eeq
and across which the stress-energy-momentum tensor, $T$, is discontinuous. Said differently, for a fixed time $t$ the shock surface is a two sphere (of symmetry) with radius $x(t)$. For simplicity, we assume $\Sigma$ to be a timelike surfaces, that is, the shock speed is smaller than the speed of light, $\dot{x}(t)^2 < \frac{A}{B}$. Nevertheless, Theorem \ref{TheoremMain} also applies to spacelike shock surfaces, since we can assume the metric components $A$ and $B$ in \eqref{metric SSC} to be both negative.  

For radial hypersurfaces in SSC, the angular variables play a passive role, and the essential issue regarding smoothing the metric components lies within the atlas of $\oneone$ coordinate transformations acting on the $(t,r)$-plane, i.e., angular coordinates are kept fixed. Therefore, it suffices to work with the so-called shock curve $\gamma$, that is, the shock surface $\Sigma$ restricted to the $(t,r)$-plane,
\beq \label{gamma}
\gamma(t)=(t,x(t)),
\eeq
with normal $1$-form
\beq \label{normal}
n_{\nu}=(\dot{x},-1).
\eeq
For radial shock surfaces (\ref{shock surface SSY}) in SSC, the Rankine Hugoniot (RH) conditions (\ref{JC}) take the simplified form
\begin{eqnarray}
\left[T^{00}\right]\dot{x}&=&\left[T^{01}\right], \label{RHwithN1} \\
\left[T^{10}\right]\dot{x}&=&\left[T^{11}\right]. \label{RHwithN2}
\end{eqnarray}

We now generalize the above framework to shock wave interactions (collisions) between shocks from different families. At the point of collision, shock waves typically change their speeds discontinuously, so that we cannot assume the shock surfaces to be smooth in $t$ at $t=0$. In light of this, it is useful to think of the incoming and outgoing branches of the two shock waves as four distinct shocks, two of which are defined for $t \leq 0$ and two for $t\geq 0$. In more detail, to establish our basic framework, suppose the timelike shock surfaces $\Sigma_i^\pm$ are parameterized in SSC by
\beq \label{radialsurface1+}
\Sigma_i^+(t,\theta,\phi) =(t,x^+_i(t),\theta,\phi), \ \ \ \ \text{for} \ \ t\geq0,
\eeq
and
\beq \label{radialsurface1-}
\Sigma_i^-(t,\theta,\phi) =(t,x^-_i(t),\theta,\phi), \ \ \ \ \text{for} \ \ t\leq0,
\eeq
where $i=1,2$. Assume in addition that $\Sigma^\pm_i$ intersect at $t=0$, that is, 
\beq \nonumber
x^\pm_1(0)=r_0=x^\pm_2(0),
\eeq
for some $r_0>0$. Restricted to the $(t,r)$-plane, the above shock surfaces are described by the shock curves
\beq \label{gammacurves}
\gamma_i^\pm(t)=(t,x^\pm_i(t)),
\eeq
with normal $1$-forms
\beq\label{normali}
(n^\pm_i)_{\nu}=(\dot{x}^\pm_i,-1).
\eeq
We assume the $\gamma_i^\pm$ are $C^3$ with all derivatives extending to $t=0$. Denoting with $[\cdot]_i^\pm$ the jump across the shock curve $\gamma_i^\pm$ the RH conditions read, in correspondence to \eqref{RHwithN1}-\eqref{RHwithN2},
\begin{eqnarray}
\left[T^{00}\right]_i^\pm\dot{x}_i^\pm &=&\left[T^{01}\right]^\pm_i, \label{RHwithN1; two} \\
\left[T^{10}\right]^\pm_i\dot{x}_i^\pm &=&\left[T^{11}\right]^\pm_i. \label{RHwithN2; two}
\end{eqnarray}

For the proof of Theorem \ref{TheoremMain}, it often suffices to restrict attention to the lower ($t<0$) or upper ($t>0$) part of a shock wave interaction that occurs at $t=0$. That is, it suffices to consider the lower or upper half plane in $\R^2$ separately, 
\beq\label{upper_lower-halfplane}
\R^2_{-}=\left\{(t,r): t < 0\right\}  \ \ \ \ \text{or} \ \ \ \R^2_{+}=\left\{(t,r): t > 0\right\},
\eeq
respectively. (We denote with $\overline{\R^2_{\pm}}$ the closure of $\R^2_\pm$.) Whenever it is clear that we restrict consideration to $\R^2_-$ or $\R^2_+$, we drop the superscript $\pm$ on the quantities introduced in \eqref{radialsurface1+} - \eqref{RHwithN2; two}.  

We are now prepared to give the definition of what we call a point of regular shock wave interaction in SSC between shocks from different families. By this we mean a point\footnote{The intersection of the shock surfaces is a two sphere, but abusing language we refer to it as a point, consistent with us suppressing the trivial angular dependence in our method.} $p$ where two shock waves collide, resulting in two outgoing shock waves, such that the metric is smooth away from the shock curves and Lipschitz continuous across each shock curve, allowing for a discontinuous $T^{\mu\nu}$ and the RH condition to hold. In case that $T^{\mu\nu}$ is given by a perfect fluid, this type of shock collision corresponds to an interaction of shock waves in different characteristic families, c.f. \cite{Smoller}. However, we only require the RH conditions to hold and do not explicitly assume a perfect fluid source.


\begin{Def}\label{shockinteract}
Let $r_0>0$, and assume $g_{\mu\nu}$ to be an SSC metric in $C^{0,1}\left(\mathcal{N} \right)$, where $\mathcal{N}\subset\R^2$ is a neighborhood of the point $p=(0,r_0)$ of intersection of the timelike shock curves $\gamma_i^\pm$, $i=1,2$, introduced in \eqref{gammacurves}. Let $\hat{\mathcal{N}}$ denote the open set consisting of all points in $\mathcal{N}$ not in the image of any $\gamma_i^\pm$. Then, we say that $p$ is a ``point of regular shock wave interaction in SSC between shocks from different families'' if:
\begin{enumerate}[(i)]
\item The pair $(g,T)$ is a strong solution of the SSC Einstein equations \eqref{one}-\eqref{four} in $\hat{\mathcal{N}}$, with $T^{\mu\nu}\in C^0(\hat{\mathcal{N}})$ and $g_{\mu\nu}\in C^2(\hat{\mathcal{N}})$.
\item The limits of $T^{\mu\nu}$ and of the metric derivatives $g_{\mu\nu,\sigma}$ exist on both sides of each shock curve $\gamma_i^\pm$, including the point $p$. 
\item The jumps in the metric derivatives $[g_{\mu\nu,\sigma}]^\pm_i(t)$ are $C^{3}$ functions for all $t\in(-\epsilon,0)$, respectively, for all $t\,\in\,(0,\epsilon)$.    
\item The (upper/lower)-limits
$$
\lim\limits_{t\rightarrow0}[g_{\mu\nu,\sigma}]_i^\pm(t)=[g_{\mu\nu,\sigma}]^\pm_i(0)
$$
exist and the (upper/lower)-limits for all derivatives of $[g_{\mu\nu,\sigma}]_i^\pm$ exist.
\item The stress tensor $T$ is bounded on $\mathcal{N}$ and satisfies the RH conditions $$[T^{\nu\sigma}]^\pm_i(n_i)_{\sigma}=0$$ at each point on $\gamma^\pm_i(t)$, $t\in(-\epsilon,0)$ or $t\,\in\,(0,\epsilon)$, and the limits of these jumps exist up to $p$ as $t\rightarrow 0$.
\end{enumerate}
\end{Def}

The structure assumed in Definition \ref{shockinteract} reflects the regularity of generic shock wave solutions of the coupled Einstein-Euler equations, for instance, the shock interactions simulated in \cite{VoglerTemple}, and our assumptions are consistent with the Groah-Temple existence-theory in \cite{GroahTemple}.  We expect Definition \ref{shockinteract} to be a most natural way of introducing shock wave interactions, since our assumptions are straightforward generalizations from either the flat case or the case of single relativistic shock waves, for instance, as in \cite{FreistuehlerRaoofi, SmollerTemple,SmollerTemple1,SmollerTemple2,SmollerTemple3,SmollerTemple4,SmollerTempleAstro}. In more detail, a typical shock wave solution of the compressible Euler equations is bounded (in $L^\infty$) and smooth away from the shock discontinuity, (one can actually assume $C^\infty$ as long that initial data is in $C^\infty$), consistent with the local existence results for symmetric hyperbolic first order PDE's \cite[Chapter 16.2]{Taylor}. Moreover, from the Euler equations holding in the weak sense across the shock surface, one can show with Gauss divergence theorem that limits of the fluid variables exist at the shock surface. Now, the regularity of $T^{\mu\nu}$ corresponds to the regularity of the fluid variables, so in Definition \ref{shockinteract} we can assume that $T^{\mu\nu}$ is smooth away from the shock surfaces and that limits to the shock surfaces exist. Since the fourth Einstein equations in SSC \eqref{four} contains second order derivatives of the metric, it is natural to assume $g_{\mu\nu}$ to be two degrees more regular than $T^{\mu\nu}$. Moreover, since the first three Einstein equations \eqref{one} - \eqref{three} only contain first order metric derivatives, it follows that limits of metric derivatives exist on the shock surfaces whenever these limits exist for $T^{\mu\nu}$. This is recorded in $(i)$, $(ii)$ and $(iv)$ of Definition \ref{shockinteract}. The $C^3$ regularity assumed in $(iii)$ is convenient for the PDE existence theory in Section \ref{Sec: Integrability Condition} and it should be possible to weaken this assumption. For our method, it should also be possible to weaken the assumption in $(i)$ from a $C^2$ to a $C^{1,1}$ metric regularity. However, since we can always assume more smoothness away from the shocks without essential loss of generality, we are content with our assumptions in $(i)$ and $(iii)$ and the slight loss of generality they might entail.

\section{Functions $\oone$ Across a Hypersurface}\label{sec: about oone across}

In this section we first define a function being \emph{$\oone$ across} a hypersurface. We then study the relation of a metric being $\oone$ across a hypersurface to the Rankine Hugoniot jump condition through the Einstein equations and derive a set of three equations central to our methods in Sections \ref{intro to method} - \ref{Sec: Proof_MainThm_Wrap-up}. Finally we derive a canonical form functions $\oone$ across a hypersurface can be represented in, which is the starting point for the method in Section \ref{Sec: Canonical Jacobian}.

\begin{Def}\label{Lipschitz across 1}\label{Lipschitz across; tensor}
Let $\Sigma$ be a smooth hypersurface in some open set $\mathcal{N} \,\subset\, \R^d$ with a normal vector-field nowhere lightlike. We call a function $f\,\in\, \oone(\mathcal{N})$ ``Lipschitz continuous across $\Sigma$'', (or $C^{0,1}$ across $\Sigma$), if $f$ is smooth in $\mathcal{N}\setminus\Sigma$ and limits of derivatives of $f$ to $\Sigma$ exist on each side of $\Sigma$ separately and are smooth functions. We call a metric $g_{\mu\nu}$ Lipschitz continuous across $\Sigma$ in coordinates $x^\mu$ if all metric components are $\oone$ across $\Sigma$.
\end{Def}

For us, ``smooth'' means enough continuous derivatives so that regularity is not an issue.\footnote{Throughout this paper, $f\in C^2(\mathcal{N}\setminus\Sigma)$ with left-/right-limits of first and second order derivatives existing suffices, with the exception of condition $(iii)$ in Definition \ref{shockinteract}, for which $f\in C^4(\mathcal{N}\setminus\Sigma)$ is sufficient.} The main point of the above definition is that we assume smoothness of $f$ away and tangential to the hypersurface $\Sigma$, but allow for the normal derivative of $f$ to be discontinuous, that is,
\beq \nonumber
[f_{,\sigma}]n^\sigma\neq0,
\eeq
where $n^\sigma$ is normal to $\Sigma$ with respect to some (Lorentz-) metric  $g_{\mu\nu}$ defined on $\mathcal{N}$. Moreover, the continuity of $f$ across $\Sigma$ and the existence of derivatives and their limits at each side of $\Sigma$ imply that derivatives of $f$ tangent to $\Sigma$ match-up continuously:

\begin{Lemma}\label{Lipschitz_across_lemma}
Suppose $f$ is $\oone$ across $\Sigma$ in the sense of Definition \ref{Lipschitz across 1} and let $[\cdot]$ denote the jump across $\Sigma$, then, for all $v^{\sigma}$ tangent to $\Sigma$,
\beq \label{Lipschitz across 1; eqn}
[f_{,\sigma}]v^{\sigma}=0.
\eeq
\end{Lemma}
\Proof
To prove \eqref{Lipschitz across 1; eqn}, we first recall that the continuity of $f$ across $\Sigma$ implies
\begin{eqnarray}\label{Lipschitz across 1; eqn1_proof}
0 &=& [f](q) = f_L(q) - f_R(q) , \ \ \ \ \ \forall \, q \in\, \Sigma,
\end{eqnarray}
where $f_{L/R}$ denotes the left and right limit of $f$ at $\Sigma$ respectively. Now, fix a point $q\in \Sigma$ and let $v$ be a vector tangent to $\Sigma$ at $q$. Denote with $c$ a $C^1$ curve in $\Sigma$ with $c(0)=q$ and $\frac{d c}{d s} (0)= v$ at $q$. Now, differentiating \eqref{Lipschitz across 1; eqn1_proof} along $c$, yields
\beq \label{Lipschitz across 1; eqn2_proof}
\frac{d}{ds} f_L\circ c(s) - \frac{d}{ds} f_R\circ c(s) = 0, \ \ \ \ \ \forall \, s.
\eeq
To complete the proof, we now show that 
\beq \label{Lipschitz across 1; eqn3_proof}
\frac{d}{ds} f_{L/R}\circ c (0)= \left( f_{,\sigma}\right)_{L/R} v^\sigma.
\eeq 
For this, let $\left(c_l\right)_{l\in \R}$ be a family of $C^1$-curves lying entirely on the left of $\Sigma$ and converging uniformly in $C^1$ to $c$ as $l$ approaches $0$. Using our incoming assumption that $f \in C^1(\mathcal{N}\setminus\Sigma)$ with limits of derivatives existing at each side of $\Sigma$ and that $f$ is continuously differentiable tangential to $\Sigma$,  we conclude that the difference quotient 
$$
\frac{ f \circ c_{l}(s+h) - f\circ c_{l}(s) }{h}
$$ 
is continuous in $h$ and $l$, which implies limits in $h$ and $l$ commute and thus
\begin{eqnarray}\nonumber
\frac{d}{ds} f_{L}\circ c 
&=& \frac{d}{ds} \left( \lim_{l\rightarrow0} f\circ c_{l}\right)  \cr
&=& \lim_{l\rightarrow0}\left( \frac{d}{ds} f\circ c_{l} \right) \cr
&=&  \lim_{l\rightarrow0} \left( f_{,\sigma} \frac{dc_{l}^\sigma}{ds}\right) \cr
&=&  \left( f_{,\sigma}\right)_{L} v^\sigma .
\end{eqnarray}
The above equation, and thus \eqref{Lipschitz across 1; eqn3_proof}, follows for the right-limit analogously. Now, \eqref{Lipschitz across 1; eqn2_proof} together with \eqref{Lipschitz across 1; eqn3_proof} yield \eqref{Lipschitz across 1; eqn} and complete the proof. 
\QED

In the following we clarify the implications of equation \eqref{Lipschitz across 1; eqn} together with the Einstein equations on the RH conditions \eqref{RHwithN1}, \eqref{RHwithN2}. For this, consider a spherically symmetric spacetime metric in SSC \eqref{metric SSC} and assume the first three Einstein equations (\ref{one})-(\ref{three}) hold and the stress tensor $T$ is discontinuous across a smooth radial shock surface, described in the $(t,r)$-plane by $\gamma(t)$ as in (\ref{shock surface SSY})-(\ref{normal}). To this end, condition (\ref{Lipschitz across 1; eqn}) applied to each metric component $g_{\mu\nu}$ in SSC, c.f. \eqref{metric SSC}, reads
\begin{eqnarray} \label{Lipschitz across; SSC}
\left[B_t\right]&=&-\dot{x} [B_r],\label{jumpone}\label{jumponeagain}\\
\left[A_t\right]&=& -\dot{x}[A_r]\label{jumptwo}\label{jumptwoagain}.
\end{eqnarray}
On the other hand, the first three Einstein equations in SSC (\ref{one})-(\ref{three}) imply
\begin{eqnarray}
 \left[B_r\right]&=&\kappa A B^2 r [T^{00}], \label{EFEinSSC1}\\
 \left[B_t\right]&=&-\kappa A B^2 r [T^{01}], \label{EFEinSSC2}\\
 \left[A_r\right]&=&\kappa A B^2 r [T^{11}].\label{EFEinSSC3}
\end{eqnarray}
Now, multiplying the first Einstein equation \eqref{EFEinSSC1} with $\dot{x}$, then using the first RH-condition \eqref{RHwithN1}, that is,
$$
\left[T^{00}\right]\dot{x}=\left[T^{01}\right],
$$
and finally applying the second Einstein equation \eqref{EFEinSSC2}, allows us the following computation:
\begin{eqnarray} \nonumber
[B_r] \dot{x} &=& \kappa A B^2 r [T^{00}] \dot{x} \cr
&=& \kappa A B^2 r [T^{01}] \cr
&=& - [B_t].
\end{eqnarray}
We conclude that \eqref{jumpone} is in fact equivalent to the first RH condition \eqref{RHwithN1}. The second condition, (\ref{jumptwo}), is independent of the RH-conditions, because $[A_t]$ does not appear in (\ref{EFEinSSC1})-(\ref{EFEinSSC3}). Now, multiplying the second Einstein equation \eqref{EFEinSSC2} with $\dot{x}$, then using the second RH-condition \eqref{RHwithN2}, that is,
$$
\left[T^{10}\right]\dot{x}=\left[T^{11}\right],
$$
and finally applying the third Einstein equation \eqref{EFEinSSC3} gives us
\begin{eqnarray}\nonumber
[B_t] \dot{x} &=& -\kappa A B^2 r [T^{01}] \dot{x} \cr 
&=& -\kappa A B^2 r [T^{11}] \cr 
&=& - [A_r].
\end{eqnarray}
The result, then, is that in addition to the assumption that the metric be $C^{0,1}$ across the shock surface in SSC, the RH conditions (\ref{RHwithN1}) and (\ref{RHwithN2}) together with the Einstein equations (\ref{EFEinSSC1})-(\ref{EFEinSSC3}), yield only one additional condition\footnote{This observation is consistent with Lemma 9, page 286, of \cite{SmollerTemple}, where only one jump condition need to be imposed to meet the full RH relations.} over and above (\ref{jumpone}) and (\ref{jumptwo}), namely,
\beq
[A_r]=-\dot{x}[B_t]\; \label{[Bt]shockspeed=[Ar]}.
\eeq
The RH conditions together with the Einstein equations will enter our method in Sections \ref{intro to method} -  \ref{Sec: Proof_MainThm_Wrap-up} only through equations \eqref{jumpone}, \eqref{jumptwoagain} and (\ref{[Bt]shockspeed=[Ar]}). We summarize the above important consideration in the following lemma:

\begin{Lemma}\label{RH_Lemma}
Assume the SSC metric, $g_{\mu\nu}$, is $\oone$ across a hypersurface $\Sigma$ in the sense of Definition \ref{Lipschitz across 1}. Assume the Einstein equations hold in the sense of \eqref{EFEinSSC1} - \eqref{EFEinSSC3}. Then \eqref{jumpone} - \eqref{jumptwo} hold and the RH conditions, \eqref{RHwithN1} - \eqref{RHwithN2}, are equivalent to \eqref{[Bt]shockspeed=[Ar]}.
\end{Lemma}

The following lemma provides a canonical form for any function Lipschitz continuous across a {\it single} shock curve $\gamma$ in the $(t,r)$-plane, under the assumption that the vector $n^{\mu}$, normal to $\gamma$, is obtained by raising the index in \eqref{normal} with respect to a  Lorentzian metric $C^{0,1}$ across $\gamma$. We later apply this lemma for the construction of Jacobians smoothing the metric to $C^{1,1}$.

\begin{Lemma}\label{characerization1} 
Given a function $f$ which is $\oone$ across a smooth curve $\gamma(t)=(t,x(t))$ in the sense of Definition \ref{Lipschitz across 1}, $t \in  (-\epsilon,\epsilon)$, in an open subset $\mathcal{N}$ of $\R^2$. Then there exists a function $\Phi \in  C^1(\mathcal{N})$ such that
\beq\label{characerization1, eqn}
f(t,r)= \varphi(t) \left|x(t)-r\right| +\Phi(t,r),
\eeq
where
\beq\label{kink in canonical form, single}
\varphi(t)=\frac12 \frac{[f_{,\mu}]n^{\mu}}{n^{\sigma}n_{\sigma}}\ \ \in\,C^1(-\epsilon,\epsilon),
\eeq
and $n_{\mu}(t)=(\dot{x}(t),-1)$ is the $1$-form normal to $v^\mu(t)=\dot{\gamma}^\mu(t)$ and indices are raised and lowered by a Lorentzian metric $g_{\mu\nu}$ which is $C^{0,1}$ across $\gamma$.
\end{Lemma}

\Proof
Suppose $\varphi$ is defined by \eqref{kink in canonical form, single}, then, by our assumptions, $\varphi$ is $C^1$ regular. To show the existence of $\Phi \, \in\, C^1(\mathcal{N})$ define $X(t,r)=x(t)-r$ and introduce
\beq\label{characerization1, eqn2}
\Phi=f- \varphi \left|X \right|,
\eeq
then \eqref{characerization1, eqn} holds and it remains to prove the $C^1$ regularity of $\Phi$. It suffices to prove
\[[\Phi_{,\mu}]n^\mu= 0= [\Phi_{,\mu}]v^\mu,\]
since $\Phi \, \in C^1(\mathcal{N} \setminus\gamma)$ follows immediately from \eqref{characerization1, eqn2} and the $C^1$ regularity of $f$ and $\varphi$ away from $\gamma$. By  Lemma \ref{Lipschitz_across_lemma}, $f$ satisfies equation \eqref{Lipschitz across 1; eqn} and using further that $\frac{d}{dX} |X| = H(X)$ and $[H(X)]=2$, we obtain that
\[ 
[\Phi_{,\mu}]v^\mu=-2\varphi X_{,\mu}v^\mu.
\]
But since $v^\mu(t)= \,^T \hspace{-0.3 em}(1,\dot{x}(t))$, we have 
$$
X_{,\mu}v^\mu=0
$$ 
and thus $[\Phi_{,\mu}]v^\mu=0$. Finally, the definition of $\varphi$ in \eqref{kink in canonical form, single} together with
\[X_{,\mu}n^\mu=n_{\mu}n^\mu\]
show that
\[[\Phi_{,\mu}]n^\mu=2\varphi\, n_\mu n^\mu - 2\varphi\, X_{,\mu}n^\mu =0,\]
which completes the proof. 
\QED

In words, the canonical form (\ref{characerization1, eqn}) separates off the $C^{0,1}$ kink of $f$ across $\gamma$ from its more regular $C^1$ behavior. In more detail, the absolute value function $\left|x(t)-r\right|$ locates the kink to the shock curve, with $\varphi$ giving the strength of the jump, while $\Phi$ encodes the remaining $C^1$ behavior of $f$. In Section \ref{Sec: Canonical Jacobian} we introduce a canonical form analogous to (\ref{characerization1, eqn}) for two shock curves, but such that it allows for the Jacobian to be in the weaker regularity class $\oone$, \emph{away} from the shock curves. To this end, suppose we are given timelike shock surfaces described in the $(t,r)$-plane by $\gamma_i(t)=(t,x_i(t))$, such that \eqref{radialsurface1+} - \eqref{normali} applies. To cover the generic case of shock wave interaction, we assume each $\gamma_i(t)$ is at least $C^2$ away from $t=0$, with the lower/upper-limit of their derivatives existing up to $t=0$. For our methods in Section \ref{Sec: Canonical Jacobian}, it suffices to consider the upper ($t>0$) or lower part ($t<0$) of a shock wave interaction at $t=0$ separately. In the following lemma we restrict without loss of generality to $\R^2_{+}$, the upper part of a shock wave interaction. 

\begin{Lemma}\label{characerization2} Let $\gamma_i(t)=(t,x_i(t))$ be two smooth curves defined on $I=(0,\epsilon)$, for some $\epsilon>0$, such that \eqref{radialsurface1+} - \eqref{normali} hold. Let $\mathcal{N}$ be an open neighborhood of $p=(0,r_0)$ in $\R^2$ and suppose $f$ is in $\oone(\mathcal{N}\cap\R^2_+)$, but such that $f$ is $C^2$ tangential to each $\gamma_i$, limits of derivatives of $f$ exist on each $\gamma_i$ and \eqref{Lipschitz across 1; eqn} holds on each $\gamma_i$.  Then there exists a function $\Phi \in \oone (\mathcal{N}\cap\R^2_{+})$, such that limits of its derivatives exist on each  $\gamma_i$ and match continuously, i.e.,
\beq\label{jumPhi} 
[\Phi_t]_i = 0 = [\Phi_r]_i,
\eeq
for $i=1,2$, and such that
\beq\label{canoicalfortwo}
f(t,r)= \sum_{i=1,2} \varphi_i(t) \left|x_i(t)-r\right| +\Phi(t,r),
\eeq
for all $(t,r)$ in $\mathcal{N}\cap\R^2_{+}$, where
\beq\label{kink in canonical form, two}
\varphi_i(t) = \frac12 \frac {[f_{,\mu}]_i (n_i)^{\mu}} {(n_i)^{\mu}(n_i)_\mu} \, \in C^{1}(I).
\eeq
here $(n_{i})_{\mu}(t)=(\dot{x}_i(t),-1)$ is the $1$-form normal to $v_i^\mu(t)=\dot{\gamma}_i^\mu(t)$, for $i=1,2$,  and indices are raised by a Lorentzian metric $C^{0,1}$ across each $\gamma_i$. 
\end{Lemma}

\Proof
Suppose $\varphi_i$ is given by \eqref{kink in canonical form, two}. The regularity of $\varphi$ follows immediately from our assumptions on $f$.  To show the existence of $\Phi \, \in\, C^{0,1}(\mathcal{N}\cap \R^2_+)$, define $X_i(t,r)=x_i(t)-r$ and
\beq\label{characerization1, eqn2_2case}
\Phi=f- \sum_{i=1,2}\varphi_{i} \left|X_i \right|,
\eeq
then $\Phi$ is in $C^{0,1}(\mathcal{N}\cap \R^2_+)$ and \eqref{canoicalfortwo} holds. Moreover, limits of derivatives of $\Phi$ exist on each $\gamma_i$, since limits of derivatives of $f$ are assumed to exist on each $\gamma_i$. It remains to prove \eqref{jumPhi}, for which it suffices to prove that
\[
[\Phi_{,\mu}]_i \, (n_i)^\mu \ = \ 0 \ = \ [\Phi_{,\mu}]_i \, (v_i)^\mu.
\]
By assumption, $f$ satisfies equation \eqref{Lipschitz across 1; eqn} with respect to each $\gamma_i$ and thus
\begin{eqnarray}\nonumber
[\Phi_{,\mu}]_i \, (v_i)^\mu 
&=& - \sum_{j=1,2} \varphi_j \, [H(X_j)]_i \,( X_j)_{,\mu}(v_i)^\mu \cr
&=& - 2\varphi_i \,(X_i)_{,\mu}(v_i)^\mu ,
\end{eqnarray}
where we used that $[H(X_j)]_i=2\delta_{ji}$ to obtain the last equality. Now, since $(v_i)^\mu(t)= \,^T \hspace{-0.3 em}(1,\dot{x}_i(t))$, we find that $X_{i,\mu} (v_i)^\mu=0$ and thus 
$$
[\Phi_{,\mu}]_i \, (v_i)^\mu=0.
$$ 
Finally, the definition of $\varphi_i$ in \eqref{kink in canonical form, two} together with $[H(X_j)]_i=2\delta_{ji}$ and
\[X_{i,\mu}(n_i)^\mu=(n_i)_{\mu}(n_i)^\mu\]
show that
\[[\Phi_{,\mu}]_i \, (n_i)^\mu \ = \ 2\varphi_i \, (n_i)_\mu (n_i)^\mu - 2\varphi_i \, X_{i,\mu} (n_i)^\mu \ = \ 0,\]
which completes the proof.
\QED

In Section \ref{Sec: Canonical Jacobian}, we use the canonical form \eqref{canoicalfortwo} to characterize all Jacobians in the $(t,r)$-plane capable to lift the metric regularity from $\oone$ to $\oneone$, unique up to addition of $C^1$ functions.

\section{The Smoothing Condition}\label{intro to method}

In this section we derive a \emph{point-wise} condition on the Jacobians of a coordinate transformation, necessary and sufficient for the Jacobians to lift the metric regularity from $C^{0,1}$ to $C^{1,1}$ in a neighborhood of a point on a single shock surface $\Sigma$. This condition is the starting point for our strategy and lies at the heart of the proofs in Sections \ref{Israel's Thm with new method} - \ref{Sec: Proof_MainThm_Wrap-up}.

We begin with the covariant transformation law
\beq\label{metrictrans}
g_{\alpha\beta}=J^{\mu}_{\alpha} J^{\nu}_{\beta} g_{\mu\nu},
\eeq
for the metric components at a point on a hypersurface $\Sigma$ for a general $C^{1,1}$ coordinate transformation $x^{\mu}\rightarrow x^{\alpha}$, where, as customary, the indices indicate the coordinate system. Let $J^{\mu}_{\alpha}$ denote the Jacobian of the transformation, that is,
\beq\label{Jacobian of coord}
J^{\mu}_{\alpha}=\frac{\partial x^{\mu}}{\partial x^{\alpha}}.
\eeq
Now, assume $J^\mu_\alpha$ and the metric components $g_{\mu\nu}$ are Lipschitz continuous across $\Sigma$, in the sense of Definition \ref{Lipschitz across 1}, with respect to coordinates $x^{\mu}$. Then, differentiating (\ref{metrictrans}) with respect to $\frac{\partial}{\partial x^{\gamma}}$ and taking the jump across $\Sigma$ we obtain
\beq\label{basicstart}
[g_{\alpha\beta,\gamma}]= J^\mu_\alpha J^\nu_\beta [g_{\mu\nu,\gamma}] + g_{\mu\nu} J^\mu_\alpha [J^\nu_{\beta,\gamma}]  +g_{\mu\nu} J^\nu_\beta [J^\mu_{\alpha,\gamma}] \, ,
\eeq
where $[f]=f_L-f_R$ denotes the jump in the quantity $f$ across the shock surface $\Sigma$, c.f. \eqref{JC}. Since both $g_{\mu\nu}$ and $J^{\mu}_{\alpha}$ are $\oone$ across $\Sigma$, the jumps are only on the derivative-terms. Now, $g_{\alpha\beta}$ is $C^{1}$ across $\Sigma$ if and only if
\beq\label{basicstart2}
[g_{\alpha\beta,\gamma}]=0
\eeq
for all $\alpha,\, \beta,\, \gamma\, \in\, \{0,...,3\}$, and (\ref{basicstart}) implies that \eqref{basicstart2} holds if and only if
\beq\label{smoothingcondt1}
[J^\mu_{\alpha,\gamma}] J^\nu_\beta g_{\mu\nu} + [J^\nu_{\beta,\gamma}]J^\mu_\alpha g_{\mu\nu} + J^\mu_\alpha J^\nu_\beta [g_{\mu\nu,\gamma}] =0.
\eeq
\eqref{smoothingcondt1} is a necessary and sufficient condition on the Jacobian for smoothing the metric from $\oone$ to $C^1$, in fact, even smoothing to $\oneone$ regularity as shown in Lemma \ref{smoothingcondt equiv oneone}. We refer to \eqref{smoothingcondt1} as the \emph{Smoothing Condition}.

\eqref{smoothingcondt1} is an inhomogeneous linear system with unknowns $[J^{\mu}_{\alpha,\gamma}]$, where we consider the $J^\mu_\alpha$-factors as given (free) parameters. A solution of \eqref{smoothingcondt1} alone does not ensure the existence of a Jacobian, since \eqref{smoothingcondt1} does not yet ensure the existence of $\oone$ functions $J^\mu_\alpha$ that take on the values $[J^{\mu}_{\alpha,\gamma}]$ and satisfy the integrability condition \eqref{IC}, necessary for integrating $J^\mu_\alpha$ to coordinates. It is therefore crucial to impose an appropriate integrability condition at the shock surface, namely
\beq\label{IC2}
[J^\mu_{\alpha,\beta}]=[J^\mu_{\beta,\alpha}].
\eeq

In the following we solve the linear system obtained from \eqref{smoothingcondt1} and \eqref{IC2} for $[J^\mu_{\alpha,\gamma}]$. We subsequently restrict to spherically symmetric spacetimes and assume that $g_{\mu\nu}$ denotes the metric in Standard Schwarzschild coordinates, \eqref{metric SSC}. (From now on, indices $\mu,\, \nu$ and $\sigma$ refer to SSC.) To implement our proof strategy of Theorem \ref{TheoremMain}, we write \eqref{smoothingcondt1} in its equivalent form in SSC,
\beq\label{smoothingcondt1_SSC}
[J^\mu_{\alpha,\sigma}] J^\nu_\beta g_{\mu\nu} + [J^\nu_{\beta,\sigma}]J^\mu_\alpha g_{\mu\nu} =  -J^\mu_\alpha J^\nu_\beta [g_{\mu\nu,\sigma}] ,
\eeq 
by using the chain rule and that $J^\sigma_\gamma$ is point-wise an invertible matrix. Similarly, we express \eqref{IC2} in its equivalent SSC-form, c.f. Appendix A,
\beq\label{IC2_SSC}
[J^\mu_{\alpha,\sigma}] J^\sigma_\beta - [J^\mu_{\beta,\sigma}] J^\sigma_\alpha =0.
\eeq

To simplify \eqref{smoothingcondt1_SSC}, suppose we are given a single radial shock surface $\Sigma$ in SSC locally parameterized by
\beq\label{radialsurface}
\Sigma(t,\theta,\phi) =(t,x(t),\theta,\phi),
\eeq
described in the $(t,r)$-plane by the corresponding shock curve
\beq
\gamma(t)=(t,x(t)).
\eeq
For such a hypersurface in SSC, the angular variables play a passive role, and the essential issue regarding smoothing the metric components by $C^{1,1}$ coordinate transformations, lies in the atlas of $(t,r)$-coordinate transformations. Thus we restrict to the atlas of $(t,r)$-coordinate transformations, which keep the SSC angular coordinates fixed, c.f. \eqref{metric SSC}. Then
\beq\label{transfo in tr plane, matrix}
\left(J^\mu_\alpha \right) = \left( \begin{array}{cccc} J^t_0 & J^t_1 & 0 & 0 \cr J^r_0 & J^r_1 & 0 & 0 \cr 0 & 0 & 1 & 0 \cr 0 & 0 & 0 & 1 \end{array} \right),
\eeq
with the coefficients just depending on the SSC $t$ and $r$, which implies $[J^\mu_{\alpha,\beta}] =0$ whenever $\mu\,\in\,\{\varphi,\,\vartheta\}$ or $\alpha\,\in\,\{2,3\}$ or $\beta\,\in\,\{2,3\}$. Now, \eqref{smoothingcondt1_SSC} and \eqref{IC2_SSC} form a linear inhomogeneous $8\times8$ system for the eight unknowns $[J^\mu_{\alpha,\sigma}]$, where $\mu,\sigma \in \{t,r\}$ and $\alpha \in \{0,1\}$. The following lemma states its unique solution. (To avoid confusion in later sections, it is convenient to denote its solution with $\mathcal{J}^\mu_{\alpha\sigma}$ instead of $[J^\mu_{\alpha,\sigma}]$.)

\begin{Lemma}\label{smoothcondition}
Consider a metric in SSC,
\beq\label{metric SSC2}
g_{\mu\nu} dx^\mu dx^\nu = -A(t,r)dt^2 + B(t,r)dr^2 + r^2 d\Omega^2\, ,
\eeq
let $\Sigma$ denote a single radial hypersurface, as in \eqref{radialsurface}, across which $g_{\mu\nu}$ is Lipschitz continuous and assume \eqref{transfo in tr plane, matrix}. Then,  defining 
$$
\mathcal{J}^\mu_{\alpha\sigma} := [J^\mu_{\alpha,\sigma}],
$$ 
the unique solution to the linear system \eqref{smoothingcondt1_SSC} and \eqref{IC2_SSC}  is given by
\begin{eqnarray}\label{smoothingcondt2}
&&\mathcal{J}^t_{0t} =-\frac12 \left( \frac{[A_t]}{A}J^t_0 + \frac{[A_r]}{A}J^r_0  \right); \ \ \ \ \ 
\mathcal{J}^t_{0r}=-\frac12 \left( \frac{[A_r]}{A}J^t_0 + \frac{[B_t]}{A}J^r_0  \right); \cr
&&\mathcal{J}^t_{1t} = -\frac12 \left( \frac{[A_t]}{A}J^t_1 + \frac{[A_r]}{A}J^r_1  \right); \ \ \ \ \ 
\mathcal{J}^t_{1r} = -\frac12 \left( \frac{[A_r]}{A}J^t_1 + \frac{[B_t]}{A}J^r_1  \right); \cr
&&\mathcal{J}^r_{0t} = -\frac12 \left( \frac{[A_r]}{B}J^t_0 + \frac{[B_t]}{B}J^r_0  \right); \ \ \ \ \ 
\mathcal{J}^r_{0r} = -\frac12 \left( \frac{[B_t]}{B}J^t_0 + \frac{[B_r]}{B}J^r_0  \right);   \cr
&&\mathcal{J}^r_{1t} = -\frac12 \left( \frac{[A_r]}{B}J^t_1 + \frac{[B_t]}{B}J^r_1  \right); \ \ \ \ \ 
\mathcal{J}^r_{1r} =-\frac12 \left( \frac{[B_t]}{B}J^t_1 + \frac{[B_r]}{B}J^r_1  \right) \, . 
\end{eqnarray}
\end{Lemma}
\Proof
Substituting \eqref{smoothingcondt2} into \eqref{smoothingcondt1_SSC} and \eqref{IC2_SSC}, a straightforward computation proves the existence. The uniqueness follows from the fact that the determinant of the linear system formed by \eqref{smoothingcondt1_SSC} and \eqref{IC2_SSC} has the non-zero value $- 16  \text{det} (J^\mu_\alpha)^3AB$. This completes the proof.
\QED

The above solution for coordinate transformations in the $(t,r)$-plane suffices for the construction of the Jacobians smoothing the metric in Section \ref{Sec: Canonical Jacobian} - \ref{Sec: Proof_MainThm_Wrap-up}. However, to prove that the existence of such a transformation implies the RH conditions, we have to study the solution of the smoothing condition for general coordinate transformations, that is, transformations which change angular variables as well. The solution is recorded in the following Lemma.

\begin{Lemma}\label{smoothcondition_full}
Consider a metric in SSC, \eqref{metric SSC2}, and let $\Sigma$ denote a single radial hypersurface, \eqref{radialsurface}, across which $g_{\mu\nu}$ is Lipschitz continuous. Then there exists a unique solution $\mathcal{J}^\mu_{\alpha\sigma} := [J^\mu_{\alpha,\sigma}]$ of the linear system formed by \eqref{smoothingcondt1_SSC} and \eqref{IC2_SSC}. The non-zero components of the solution are given by:
\begin{eqnarray}\label{smoothingcondt2_full}
&&\mathcal{J}^t_{0t} =-\frac12 \left( \frac{[A_t]}{A}J^t_0 + \frac{[A_r]}{A}J^r_0  \right); 
\ \ \ \ \ \mathcal{J}^t_{0r}=-\frac12 \left( \frac{[A_r]}{A}J^t_0 + \frac{[B_t]}{A}J^r_0  \right); \cr
&& \mathcal{J}^r_{0t} = -\frac12 \left( \frac{[A_r]}{B}J^t_0 + \frac{[B_t]}{B}J^r_0  \right); 
\ \ \ \ \ \mathcal{J}^r_{0r}= -\frac12 \left( \frac{[B_t]}{B}J^t_0 + \frac{[B_r]}{B}J^r_0  \right);   \cr
&&\mathcal{J}^t_{1t} = -\frac12 \left( \frac{[A_t]}{A}J^t_1 + \frac{[A_r]}{A}J^r_1  \right); 
\ \ \ \ \ \mathcal{J}^t_{1r}=-\frac12 \left( \frac{[A_r]}{A}J^t_1 + \frac{[B_t]}{A}J^r_1  \right); \cr
&&\mathcal{J}^r_{1t} = -\frac12 \left( \frac{[A_r]}{B}J^t_1 + \frac{[B_t]}{B}J^r_1  \right); 
\ \ \ \ \ \mathcal{J}^r_{1r}=-\frac12 \left( \frac{[B_t]}{B}J^t_1 + \frac{[B_r]}{B}J^r_1  \right) \, ; \cr
&&\mathcal{J}^t_{2t} = -\frac12 \left( \frac{[A_t]}{A}J^t_2 + \frac{[A_r]}{A} J^r_2  \right); 
\ \ \ \ \ \mathcal{J}^t_{2r}=-\frac12 \left( \frac{[A_r]}{A}J^t_2 + \frac{[B_t]}{A}J^r_2  \right) \, ; \cr
&& \mathcal{J}^r_{2t} = -\frac12 \left( \frac{[A_r]}{B}J^t_2 + \frac{[B_t]}{B}J^r_2  \right); 
\ \ \ \ \ \mathcal{J}^r_{2r}=-\frac12 \left( \frac{[B_t]}{B}J^t_2 + \frac{[B_r]}{B}J^r_2  \right) \, ; \cr
&&\mathcal{J}^t_{3t} =-\frac12 \left( \frac{[A_t]}{A}J^t_3 + \frac{[A_r]}{A}J^r_3  \right); 
\ \ \ \ \ \mathcal{J}^t_{3r} =-\frac12 \left( \frac{[A_r]}{A}J^t_3 + \frac{[B_t]}{A}J^r_3  \right); \cr
&&\mathcal{J}^r_{3t} = -\frac12 \left( \frac{[A_r]}{B}J^t_3 + \frac{[B_t]}{B}J^r_3  \right); 
\ \ \ \ \ \mathcal{J}^r_{3r}=-\frac12 \left( \frac{[B_t]}{B}J^t_3 + \frac{[B_r]}{B}J^r_3  \right) \, .  
\end{eqnarray}
\end{Lemma}
\Proof
The existence follows by inspections of \eqref{smoothingcondt2_full} and uniqueness follows from the fact that the determinant of the left hand side of the linear system formed by \eqref{smoothingcondt1_SSC} and \eqref{IC2_SSC} is non-zero, in fact, it is given by $2^{20} \det \left(g_{\mu\nu}\right)^{10} \det \left( J^\mu_\alpha \right)^{16}$. 
\QED 

\eqref{smoothingcondt2_full} is a necessary and sufficient condition for $[g_{\alpha\beta,\gamma}]=0$, since it solves \eqref{smoothingcondt1_SSC} and \eqref{IC2_SSC} uniquely.  In fact, assuming the $[J^\mu_{\alpha,\sigma}]$-terms come from an actual Jacobian of a coordinate transformation, c.f. \eqref{Jacobian of coord}, then \eqref{smoothingcondt2_full} is necessary and sufficient for raising the metric regularity to $\oneone$ in a neighborhood of a point on a single shock surface, as we prove in the following lemma. An analogous result holds for Jacobians restricted to the $(t,r)$-plane.

\begin{Lemma}\label{smoothingcondt equiv oneone}
Let $p$ be a point on a single smooth shock curve $\gamma$, and let $g_{\mu\nu}$ be the metric in SSC, $\oone$ across $\gamma$ in the sense of Definition \ref{Lipschitz across; tensor}. Let $J^\mu_\alpha$ be the Jacobian of a coordinate transformation defined on a neighborhood $\mathcal{N}$ of $p$ and assume $J^\mu_\alpha$ is $\oone$ across $\gamma$. Then the metric in the new coordinates, $g_{\alpha\beta}$, is in $\oneone(\mathcal{N})$ if and only if $J^\mu_\alpha$ satisfies \eqref{smoothingcondt2_full}.
\end{Lemma}
\Proof
We first prove that $g_{\alpha\beta} \,\in\, \oneone(\mathcal{N})$ implies \eqref{smoothingcondt1}. Suppose there exist coordinates $x^\alpha$ such that the metric in the new coordinates $g_{\alpha\beta}$ is $\oneone$ regular, then
\[
[g_{\alpha\beta,\gamma}]=0 \ \ \ \ \forall  \, \ \alpha,\, \beta,\, \gamma\, \in\, \{0,...,3\}.
\]
This directly implies \eqref{smoothingcondt1} and since $J^\mu_\alpha$ are the Jacobians of an actual coordinate transformation they satisfy the integrability condition \eqref{IC2} as well. By Lemma \ref{smoothcondition_full} the jumps in the derivatives of the Jacobian $[J^\mu_{\alpha,\gamma}]$ then satisfy \eqref{smoothingcondt2_full}.

We now prove the opposite direction. Suppose the Jacobians $J^\mu_\alpha$ satisfy \eqref{smoothingcondt2_full}, by Lemma \ref{smoothcondition_full}, they then meet the smoothing condition \eqref{smoothingcondt1}, which implies all metric derivatives $g_{\alpha\beta,\gamma}$ to match continuously across the shock curve $\gamma$, that is, $[g_{\alpha\beta,\gamma}]=0$ for all $\alpha,\, \beta,\, \gamma\, \in\, \{0,...,3\}$. Since $g_{\mu\nu}$ and $J^\mu_\alpha$ are assumed to be smooth away from $\gamma$, it follows that $g_{\alpha\beta} \,\in\, C^1(\mathcal{N})$. Moreover, one-sided limits of all first and second order metric derivatives exist at $\gamma$, which already implies $g_{\alpha\beta} \,\in\, \oneone(\mathcal{N})$. Namely, since the first order derivatives match continuously across $\gamma$, it is straight forward to verify that the second order derivatives away from $\gamma$ agree with the weak second order derivatives almost everywhere\footnote{Without the continuous matching of lower order derivatives, partial integration would lead to boundary terms, so that the weak derivative would generally differ from the derivatives away from $\gamma$.}. Now, given $q_1,\, q_2 \in \mathcal{N}$, define $c$ to be a $C^1$ curve lying in $\mathcal{N}$ such that $c(0)=q_1$ and $c(1)=q_2$, we obtain the Lipschitz bound
\begin{eqnarray} \label{smoothingcondt equiv oneone, eqn1}
\left|g_{\alpha\beta,\gamma}(q_2)-g_{\alpha\beta,\gamma}(q_1) \right| 
&\leq & \left| \int^1_0 \frac{d}{ds} g_{\alpha\beta,\gamma} \circ c(s) ds \right| \cr 
&\leq & \sum_{\delta= 0}^4 \left( \sup_{q\in\mathcal{N}\setminus\gamma} \left| g_{\alpha\beta ,\gamma\delta}(q) \right|\right)  \left| \int^1_0 \frac{d c}{ds} ds \right| \cr
&\leq & \sum_{\delta= 0}^4 \left( \sup_{q\in\mathcal{N}\setminus\gamma} \left| g_{\alpha\beta ,\gamma\delta}(q) \right|\right)  \left| q_2 -q_1  \right|.
\end{eqnarray}
\QED

It is instructive at this point to discuss why a $C^{1,1}$ atlas is generic for addressing the metric regularity across shock surfaces: For raising the metric regularity to $C^1$ across a shock curve, that is, arranging for $[g_{\alpha\beta,\gamma}]=0$, \eqref{smoothingcondt2_full} requires the Jacobian to ``mirror'' the $\oone$ metric regularity compensating for all discontinuous first order metric derivatives. If the coordinate transformation is $C^2$, the jumps in $J^{\mu}_{\alpha,\beta}$ vanish, and (\ref{smoothingcondt1}) reduces to
\beq\nonumber
0= J^\mu_\alpha J^\nu_\beta  [g_{\mu\nu,\gamma}].
\eeq
But this contradicts the Jacobians being non-singular. We conclude that it is precisely the lack of covariance in \eqref{smoothingcondt1} for $C^{1,1}$ transformations, providing the necessary degrees of freedom, (namely $[J^{\mu}_{\alpha,\gamma}]$), allowing for a $\oone$ metric to be smoothed across a single shock surface. This illustrates that there is no hope of lifting the metric regularity within a $C^2$ atlas. Also, lowering the atlas regularity to H\"older continuous Jacobians, we again lack free parameters to meet \eqref{smoothingcondt1} point-wise, since derivatives of H\"older continuous functions are not bounded.  From this, we conclude that a $\oneone$ atlas is generic to address shock waves in GR.

Before we close this section, we need to address a technical issue for proving the reverse implication of Theorem \ref{TheoremMain}: So far we addressed Jacobians which are smooth along $\Sigma$, which suffices for the construction of Jacobian smoothing the metric in Section \ref{Israel's Thm with new method} - \ref{Sec: Proof_MainThm_Wrap-up}, however, in order to specify all such Jacobian it is important to show that the above regularity already exhausts the choice of all possible $C^{0,1}$ Jacobians.  Namely, in general, Lipschitz continuous Jacobians could also have discontinuous derivatives \emph{along} $\Sigma$, which might not match up continuously across $\Sigma$. In order to address such Jacobians, we assume that limits of Jacobian derivatives along curves transversal to $\Sigma$ exist. For such Jacobians one can then define the jump across $\Sigma$ as
\beq\label{def jump via curve}
[u]_c=\lim\limits_{s\searrow 0}u\circ c(s) - \lim\limits_{s\nearrow 0}u\circ c(s)
\eeq
for some continuous curve $c(\cdot)$ transversal to $\Sigma$ such that $c(0)\,\in\,\Sigma$, and for $u\equiv J^\mu_{\alpha,\sigma}$ being assumed regular enough for the above limits to exist. We expect that $C^{0,1}$ regular Jacobians for which the above limits fail to exist are not capable of raising the metric regularity, since the smoothing condition cannot be made sense of. We therefore deem such Jacobians irrelevant for the purpose of metric smoothing. The following Lemma proves that Jacobians smoothing the metric regularity have a well-defined jump and that $[J^\mu_{\alpha,\sigma}]$ meet \eqref{Lipschitz across 1; eqn}.

\begin{Lemma}\label{smoothingcondt => J is oone across}
Let $g_{\mu\nu}$ be a SSC-metric $\oone$ across a single smooth shock curve $\gamma$. Let $J^\mu_\alpha$ be $\oone$ regular functions which meet \eqref{smoothingcondt1_SSC} and \eqref{IC2_SSC} with limits assumed to exist in the sense of \eqref{def jump via curve} for any smooth curve $c$ transversal to $\gamma$. Then the value $[J^\mu_{\alpha,\sigma}]_c$ is independent of the choice of the curve $c$, the $J^\mu_\alpha$ are $C^1$ along $\gamma$, and \eqref{Lipschitz across 1; eqn} holds, that is,
$$
[J^\mu_{\alpha,\sigma}] \dot{\gamma}^\sigma =0.
$$
\end{Lemma}
\Proof
Assume $[J^\mu_{\alpha,\sigma}]_c$ satisfy \eqref{smoothingcondt1_SSC} and \eqref{IC2_SSC} for any curve $c(\cdot)$ transversal to $\gamma$, where $[\cdot]$ is defined by \eqref{def jump via curve}. Lemma \ref{smoothcondition_full} also applies for the jump being defined via \eqref{def jump via curve}, thus $[J^\mu_{\alpha,\sigma}]_c$ is given by \eqref{smoothingcondt2_full}. Now, the right hand side of \eqref{smoothingcondt2_full} depends on the undifferentiated Jacobian and $[g_{\mu\nu,\sigma}]_c$ only, and since the metric is assumed to be smooth along the shock curves, we conclude that $[g_{\mu\nu,\sigma}]_c = [g_{\mu\nu,\sigma}]$  for any curve $c(\cdot)$.  Thus $[J^\mu_{\alpha,\sigma}]_c= [J^\mu_{\alpha,\sigma}]$ is independent of the choice of $c$.

To prove the regularity along $\gamma$ and that $[J^\mu_{\alpha,\sigma}]$ satisfy \eqref{Lipschitz across 1; eqn}, assume for contradiction that $\left(J^\mu_{\alpha,\sigma} \right)_{L/R}$ is discontinuous at $t_0$ and assume without loss of generality that $\gamma(t) = (t,0)$. 
Now, consider two smooth curves, $s\mapsto c_1(s)$ and $s \mapsto c_2(s)$, which intersect $\gamma$ in $\gamma(t_0)=c_1(0)=c_2(0)$, such that the image of $c_1$ lies entirely below the line $\{t=t_0\}$, while the image of $c_2$ lies below  $\{t=t_0\}$ for $s<0$ and above $\{t=t_0\}$ for $s>0$. Then $[J^\mu_{\alpha,t}]_{c_1}=0$ and $[J^\mu_{\alpha,t}]_{c_2} \neq 0$, which contradicts $[J^\mu_{\alpha,\sigma}]_c$ being independent of the choice of the curve $c$. Thus $J^\mu_\alpha$ is $C^1$ along $\gamma$ which also implies \eqref{Lipschitz across 1; eqn}, by Lemma \ref{Lipschitz_across_lemma}. 
\QED

To close this section, we discuss how our method proceeds from here on. Lemma \ref{smoothcondition} establishes the remarkable result that there is no algebraic obstruction to lifting the metric regularity. Namely, the smoothing condition \eqref{smoothingcondt1_SSC} has the solution \eqref{smoothingcondt2_full}, and its solvability is \emph{neither} connected to the RH conditions \eqref{RHwithN1}-\eqref{RHwithN2} nor the Einstein equations. However, by \cite{SmollerTemple,Israel},  the RH conditions must be imposed for coordinate transformations smoothing the metric to exist. The point here is that to prove the existence of coordinate transformations lifting the regularity of SSC metrics to $C^{1,1}$ at $p\in\Sigma$, one must prove that there exists a set of functions $J^{\mu}_{\alpha}$ defined in a neighborhood of $p$, such that (\ref{smoothingcondt2_full}) holds at $p$, and such that the integrability condition (\ref{IC}), necessary for $J^{\mu}_{\alpha}$ to be the Jacobian of a coordinate transformation, holds in a neighborhood of $p$. The RH conditions are the necessary and sufficient condition for the existence of such functions.

\section{The Single-Shock-Case: A New Proof of Israel's Theorem}\label{Israel's Thm with new method}\label{Israel}

We have shown in Lemma \ref{smoothingcondt equiv oneone} that \eqref{smoothingcondt2_full} is a necessary and sufficient condition on $[J^{\mu}_{\alpha,\sigma}]$ for lifting the SSC metric regularity to $\oneone$ in a neighborhood of a shock curve, provided the value $[J^{\mu}_{\alpha,\sigma}]$ comes from an actual Jacobian $J^{\mu}_{\alpha}$. We now address the issue of how to construct such Jacobians in a neighborhood of a \emph{single} shock curve, that is, we study how to construct a set of functions $J^{\mu}_{\alpha}$, acting on the $(t,r)$-plane and satisfying the smoothing condition \eqref{smoothingcondt2} on the shock curve, such that the integrability condition (\ref{IC}) holds in a neighborhood of the curve. The goal of this section is to prove the following version of Israel's Theorem, which is a special case of Theorem \ref{TheoremMain}, but, in order to give an easy and rigorous introduction to our method, it is instructive to prove this special case first.

\begin{Thm}\label{SingleShockThm}{\rm (Israel's Theorem)}
Suppose $g_{\mu\nu}$ is an SSC metric that is $\oone$ across a radial shock surface $\Sigma$ in the sense of Definition \ref{Lipschitz across 1}, such that it solves the Einstein equations \eqref{one} - \eqref{four} strongly away from $\Sigma$ for a $T^{\mu\nu}$ which is continuous away from $\Sigma$. Let $p$ be a point on $\Sigma$. Then the following are equivalent:
\begin{enumerate}[(i)]
\item There exists a $\oneone$ coordinate transformation of the $(t,r)$-plane, defined in some neighborhood $\mathcal{N}$ of $p$, such that the transformed metric components 
are $\oneone$ functions of the new coordinates.
\item The RH conditions, \eqref{RHwithN1} - \eqref{RHwithN2}, hold on $\Sigma \cap \mathcal{N}'$ for some $\mathcal{N}' \supset \mathcal{N}$.
\end{enumerate}
Furthermore, the above equivalence also holds for the full atlas of $C^{1,1}$ coordinate transformations, not restricted to the $(t,r)$-plane.
\end{Thm}

The first step in the proof of Theorem \ref{SingleShockThm} is to construct function $J^\mu_\alpha$ that satisfy the smoothing condition \eqref{smoothingcondt2_full} on the shock curve, that is, the condition guaranteeing $[g_{\alpha\beta,\gamma}]=0$. The next lemma shows that the RH conditions are necessary and sufficient for such functions to exist. In fact, to prove sufficiency, it suffices to restrict attention to coordinate transformations in the $(t,r)$-plane, that is, to functions $J^\mu_\alpha$ of the form \eqref{transfo in tr plane, matrix}.

\begin{Lemma}\label{canonicalformforJ1}
Let $\mathcal{N}$ be a neighborhood of a point $p$, for $p$ lying on a single shock curve $\gamma$ across which the SSC metric $g_{\mu\nu}$ is Lipschitz continuous in the sense of Definition \ref{Lipschitz across 1}, and let $g_{\mu\nu}$ be defined on $\mathcal{N}$. Then, there exists functions $J^\mu_\alpha \in \oone(\mathcal{N})$ of the form \eqref{transfo in tr plane, matrix}, which satisfy the smoothing condition \eqref{smoothingcondt2} on $\gamma\cap\mathcal{N}$, if and only if the RH conditions \eqref{[Bt]shockspeed=[Ar]} hold on $\gamma\cap\mathcal{N}$. Furthermore, any such function $J^\mu_\alpha$ is of the ``canonical form'' 
\begin{eqnarray}\label{Jacobian1}
J^\mu_\alpha (t,r) &=& \varphi^\mu_\alpha (t) \left| x(t)-r \right| + \Phi^\mu_\alpha(t,r)
\end{eqnarray}
with 
\beq \label{Jacobian1_coeff}
\varphi^\mu_\alpha (t) =-\frac12 \mathcal{J}^\mu_{\alpha\, r}(t),
\eeq 
where $\mathcal{J}^\mu_{\alpha r}$ is given in \eqref{smoothingcondt2}, $\mu \in \{t,r\}$, $\alpha\in \{0,1\}$, and ~$\Phi^\mu_\alpha\in \oone(\mathcal{N})$ satisfy
\beq\label{nojumps1}
[\partial_r \Phi^\mu_\alpha]=0=[\partial_t \Phi^\mu_\alpha].
\eeq 
Explicitly, the Jacobian coefficients are given by 
\begin{eqnarray} \label{Jacobian1_coeff_expl}
\varphi^t_0(t)&=& \frac{[A_r]\phi(t) + [B_t]\omega(t)}{4A\circ \gamma(t)} \cr
\varphi^t_1(t)&=&\frac{[A_r]\nu(t) + [B_t]\zeta(t)}{4A\circ \gamma(t)} \cr
\varphi^r_0(t)&=&\frac{[B_t]\phi(t) + [B_r]\omega(t)}{4B\circ \gamma(t)}\cr
\varphi^r_1(t)&=&\frac{[B_t]\nu(t) + [B_r]\zeta(t)}{4B\circ \gamma(t)}\, ,
\end{eqnarray}
where
\beq\label{restrictedR2fct}
\phi=\Phi^t_0\circ\gamma, \ \ \omega= \Phi^r_0\circ\gamma, \ \ \nu= \Phi^t_1\circ\gamma, \ \ \zeta= \Phi^r_1\circ\gamma\, .
\eeq
Furthermore, the above equivalence also holds for the full atlas of $C^{0,1}$ coordinate transformations, not restricted to the $(t,r)$-plane, with corresponding canonical form, \eqref{Jacobian1} - \eqref{Jacobian1_coeff}, for $\mathcal{J}^\mu_{\alpha r}$ given by \eqref{smoothingcondt2_full}.
\end{Lemma}

\Proof
Suppose there exists a set of $\oone$ functions $J^\mu_\alpha$ satisfying \eqref{smoothingcondt2_full}, then Lemma \ref{smoothingcondt => J is oone across} implies     
\beq\label{ansatz is lipschitz across}
[J^{\mu}_{\alpha,t}]=-\dot{x}[J^{\mu}_{\alpha,r}]
\eeq
for all $\mu\in\left\{t,r\right\}$ and $\alpha\in\left\{0,1\right\}$. Combining \eqref{ansatz is lipschitz across} for the special case $\mu=t$ and $\alpha=0$ with the right hand side in \eqref{smoothingcondt2_full} leads to
\[
-\frac12 \left( \frac{[A_t]}{A}J^t_0 + \frac{[A_r]}{A}J^r_0  \right) = \frac{\dot{x}}{2} \left( \frac{[A_r]}{A} J^t_0 + \frac{[B_t]}{A} J^r_0  \right).
\]
Using now the jump relations for the metric tensor, \eqref{jumponeagain} - \eqref{jumptwoagain}, which hold by the assumed smoothness of $g_{\mu\nu}$ along $\Sigma$, c.f. Lemma \ref{RH_Lemma}, we conclude 
\beq \label{non-trivial RH} 
[A_r]=-\dot{x}[B_t]
\eeq 
holds or $J^t_0= 0$. However, combining \eqref{ansatz is lipschitz across} with the right hand side in \eqref{smoothingcondt2_full} for the remaining cases ($\mu\neq t$ and $\alpha\neq 0$), we find that \eqref{non-trivial RH} indeed holds, since otherwise $J^\mu_\alpha= 0$ for all $\mu\in\{t,r\}$ and for all $\alpha\in\{0,...,3\}$ and such Jacobians would lead to a singular metric tensor. Equation \eqref{non-trivial RH} is the non-trivial RH condition \eqref{[Bt]shockspeed=[Ar]} and it follows by Lemma \ref{RH_Lemma} that the RH conditions are satisfied.

For proving the opposite direction, as a consequence of Lemma \ref{smoothcondition}, it suffices to show that all $t$- and $r$-derivatives of the functions $J^{\mu}_{\alpha}$, defined in \eqref{Jacobian1}, satisfy (\ref{smoothingcondt2}) for all $\mu\in\left\{t,r\right\}$ and $\alpha\in\left\{0,1\right\}$. Observing that (\ref{restrictedR2fct}) implies the identities
\beq\nonumber
\phi=J^t_0\circ\gamma, \  \ \ \nu=J^t_1\circ\gamma, \  \ \ \omega=J^r_0\circ\gamma, \ \ \ \zeta=J^r_1\circ\gamma \ ,
\eeq
and using the $C^1$ matching of the functions $\Phi^\mu_\alpha$, \eqref{nojumps1}, as well as the RH conditions in the form  (\ref{jumponeagain}), (\ref{jumptwoagain}) and (\ref{[Bt]shockspeed=[Ar]}), it follows immediately that the Jacobian ansatz \eqref{Jacobian1} satisfies (\ref{smoothingcondt2}). This proves the existence of functions $J^\mu_\alpha$ satisfying the smoothing condition \eqref{smoothingcondt2}. Clearly, these functions then also meet the full smoothing condition, \eqref{smoothingcondt2_full}. Finally,  applying Lemma \ref{characerization1}, it follows that all functions satisfying \eqref{smoothingcondt2} assume the canonical form \eqref{Jacobian1}.
\QED

To complete the proof of Israel's Theorem, we need to make sure that the functions $J^{\mu}_{\alpha}$, defined in \eqref{Jacobian1}, are integrable to coordinates, by showing that they solve the integrability condition, \eqref{IC}, that is,
$$
J^\mu_{\alpha,\beta} = J^\mu_{\beta,\alpha} .
$$ 
The only way for the $J^\mu_\alpha$ to meet \eqref{IC} is through the free functions $\Phi^\mu_\alpha$. To accomplish this, we consider $\Phi^t_1$ and $\Phi^r_1$ as given $C^2$ functions and introduce $U:= (\Phi^t_0, \Phi^r_0)$ as the unknowns of the PDE resulting from \eqref{IC}. The goal then is to prove we can solve the integrability condition for $U \in C^1(\mathcal{N})\cap C^{2}(\mathcal{N}\setminus \gamma)$. The first step is to write \eqref{IC} as a meaningful PDE in $U$. To begin with, using that the integrability conditions \eqref{IC} are equivalent to 
\beq \nonumber
J^\mu_{\alpha,\sigma} J^\sigma_\beta - J^\mu_{\beta,\sigma} J^\sigma_\alpha = 0,
\eeq
according to Lemma \ref{Appendix_Lemma2}. Assuming further $\Phi^t_1$ is such that $J^t_1\neq0$, we obtain the equivalent equation
\beq\nonumber
J^\mu_{0,t}  + J^\mu_{0,r} \frac{ J^r_1}{J^t_1}  \  = \ \frac{ J^\mu_{1,t}}{J^t_1} J^t_0 + \frac{J^\mu_{1,r}}{J^t_1} J^r_0 ,
\eeq
which in matrix notation is given by
\beq \nonumber
\frac{\partial}{\partial t}\left(\begin{array}{c} J^t_0 \cr J^r_0  \end{array}\right)
+\, \frac{ J^r_1}{J^t_1} \ \frac{\partial}{\partial r} \left(\begin{array}{c} J^t_0 \cr J^r_0  \end{array}\right) \,
= \,  \frac{1}{J^t_1} \left(\begin{array}{cc}  J^t_{1,t} &  J^t_{1,r}  \cr  J^r_{1,t}   &  J^r_{1,r}  \end{array}\right)  \left(\begin{array}{c} J^t_0 \cr J^r_0  \end{array}\right) \, .
\eeq
Now, it is straightforward to verify that \eqref{IC}, for $J^\mu_\alpha$ defined in \eqref{Jacobian1} and $U:= (\Phi^t_0, \Phi^r_0)$, is equivalent to
\beq \label{Israel_IC_PDE}
\partial_t U + c\ \partial_r U \, -\, \mathcal{M}\, U  =\, \Big(  |X| \mathcal{M}  - H(X) \left(\dot{x} - c \right)  
\Big) \left(\begin{array}{c} \varphi^t_{0} \cr \varphi^r_{0} \end{array} \right)   
- |X| \left(\begin{array}{c} \dot{\varphi}^t_{0} \cr \dot{\varphi}^r_{0} \end{array} \right)  ,
\eeq
where the coefficients are given by
\beq \nonumber 
c = \frac{J^r_1}{J^t_1}
\eeq
and  
\beq \nonumber 
\mathcal{M} =   \frac{1}{J^t_1} \left( \begin{array}{cc} J^t_{1,t}  &  J^t_{1,r} \cr J^r_{1,t} & J^r_{1,r}   \end{array} \right).
\eeq

\eqref{Israel_IC_PDE} is a system of \emph{non-local} PDEs, since the right hand side of \eqref{Israel_IC_PDE} contains the Jacobian coefficients $\varphi^t_0$ and $\varphi^r_0$ which depend on $U\circ \gamma$ itself, and standard existence theory cannot be applied in general. However, the resolution to this problem is to prescribe initial data on the shock curve, since the right hand side of \eqref{Israel_IC_PDE} then turns into a given source term. Then \eqref{Israel_IC_PDE} is, in fact, a linear strictly hyperbolic system of first order differential equation, to which standard existence theory can be applied, c.f. \cite{John}, provided we choose
\beq \label{shock not characteristic}
\zeta\neq \dot{x} \nu,
\eeq
which ensures that the shock curve is non-characteristic. Thus \eqref{Israel_IC_PDE} can be solved along characteristic curves, yielding a solution $U \in C^{0,1}(\mathcal{N})$ which is smooth away from $\gamma$, and the only obstacle to a solution $U$ with the \emph{necessary} $C^1$ regularity, \eqref{nojumps1}, is the presence of the (discontinuous) Heaviside functions $H(X)$ in \eqref{Israel_IC_PDE}. However, the coefficients of $H(X)$ in \eqref{Israel_IC_PDE} vanish on the shock curve precisely when the RH jump conditions hold, as stated in the next lemma, which then yields the desired $C^1$ regularity across $\gamma$.

\begin{Lemma} \label{techlemma2}
Assume the assumptions of Theorem \ref{SingleShockThm} and denote with $f$ and $h$ the coefficient functions of the Heaviside function $H(X)$ in the first and second component of \eqref{Israel_IC_PDE}, respectively. Then,
\beq\label{f=0=h}
f\circ\gamma = 0 = h\circ\gamma
\eeq
if and only if the RH conditions, \eqref{RHwithN1} - \eqref{RHwithN2}, hold on $\gamma$. 
\end{Lemma}
\Proof
To derive an explicit expression for the coefficients to $H(X)$ in \eqref{Israel_IC_PDE}, note that the matrix $\mathcal{M}$ contains Heaviside functions as well. Then, collecting all terms containing $H(X)$ and using  $X\circ\gamma=0$ and \eqref{restrictedR2fct}, we find 
\begin{eqnarray} \label{f and h on gamma}
f\circ \gamma &=& \varphi^t_0 \, \dot{x}\, \nu -\varphi^t_1\, \dot{x}\, \phi +\varphi^t_1\, \omega -\varphi^t_0 \,\zeta \, ,\cr
h\circ \gamma &=& \varphi^r_0\, \dot{x}\, \nu - \varphi^r_1\, \dot{x}\, \phi +\varphi^r_1\, \omega - \varphi^r_0 \, \zeta.
\end{eqnarray}
Now, replace $\varphi^t_0$ and $\varphi^t_1$ by their definition, \eqref{Jacobian1_coeff_expl}, then a straightforward computation shows that
\[f\circ \gamma = 0\]
is equivalent to
\beq \label{techlemma2, eqn1}
\left( [A_r] + \dot{x} [B_t] \right) \left( \phi \zeta - \nu \omega \right) = 0.
\eeq
Now, using
\beq \label{techlemma2, eqn2}
\left( \phi \zeta - \nu \omega \right) = \det \left( J^\mu_\alpha \circ \gamma \right) \neq 0,
\eeq
we conclude that $f\circ \gamma =0$ if and only if $[A_r] + \dot{x} [B_t] =0$, that is, \eqref{[Bt]shockspeed=[Ar]}, holds, which is equivalent to the second RH condition, \eqref{RHwithN2}. Similarly, replacing $\varphi^r_0$ and $\varphi^r_1$ in \eqref{f and h on gamma} by their definition, \eqref{Jacobian1_coeff_expl}, a straightforward computation shows the equivalence of $h\circ \gamma =0$ and
\beq \label{techlemma2, eqn3}
\left( [B_t] + \dot{x} [B_r] \right) \left( \phi \zeta - \nu \omega \right) =0.
\eeq
Now, using again \eqref{techlemma2, eqn2}, it follows that $h\circ \gamma =0$ if and only if \eqref{jumptwo} holds, which is equivalent to the first RH condition \eqref{RHwithN1}. This completes the proof.
\QED

\noindent \emph{Proof of Theorem \ref{SingleShockThm}:} 
We begin by showing that (i) implies (ii). Assume there exist coordinates $x^\alpha$ such that $g_{\alpha\beta}= J^\mu_\alpha J^\nu_\beta g_{\mu\nu}$ is in $\oneone$. The $\oneone$ regularity of $g_{\alpha\beta}$ implies that the Jacobian $J^\mu_\alpha$ satisfies \eqref{smoothingcondt2_full} on $\gamma$. Now, Lemma \ref{canonicalformforJ1} implies that the RH condition hold. 

We now show that (ii) implies (i). Suppose the RH conditions, \eqref{RHwithN1}-\eqref{RHwithN2}, and thus \eqref{[Bt]shockspeed=[Ar]} hold on $\gamma \cap \mathcal{N}'$. By Lemma \ref{canonicalformforJ1}, the functions $J^\mu_\alpha$ defined in \eqref{Jacobian1} satisfy the smoothing condition \eqref{smoothingcondt2} and thus lift the metric regularity from $\oone$ to $\oneone$. To ensure the existence of coordinates $x^\alpha$, such that $J^\mu_\alpha=\frac{\partial x^\mu}{\partial x^\alpha}$, we need to prove the existence of a $C^1$ regular function $U=(\Phi^t_0,\Phi^r_0)$ that satisfy the integrability condition \eqref{Israel_IC_PDE}, c.f. Lemma \eqref{equiv of IC and exi of integrable function}. However, to make sure that the $C^2$ metric regularity is preserved away from the shock curve, we must prove that $U$ is $C^2$ away from $\gamma$.  For this, we choose $\Phi^t_1$ and $\Phi^r_1$ to be smooth non-zero functions such that \eqref{shock not characteristic} holds, which is the condition that the shock curve is not characteristic. Now, imposing smooth initial data on the shock curve, \eqref{Israel_IC_PDE} is a strictly hyperbolic system of linear first order PDE's. The standard existence theory in \cite{John}, Chapter 2.5, then yields a solution $U\in C^{0,1}(\mathcal{N})$, which is smooth away from the shock curve, for some neighborhood $\mathcal{N} \subset \mathcal{N}'$ of $\gamma$.  Moreover, we pick the initial data such that $\det J \neq 0$ everywhere on the shock curve and, by continuity of the solution, $\det  J $ is thus non-vanishing in some neighborhood of $\gamma$, which we again denote by $\mathcal{N}$.

Now, for the Jacobian \eqref{Jacobian1} to satisfy the smoothing condition \eqref{smoothingcondt2}, it remains to prove a $C^1$ regularity of $U$ across the shock curve, that is, \eqref{nojumps1}. For this, take the jump of \eqref{Israel_IC_PDE} across $\gamma$ and use the vanishing of $f$ and $h$ in \eqref{f=0=h} and the continuity of $U$ to conclude that
\begin{eqnarray}\label{nojumps along characteristics}
[\partial_t U]\, \nu + [\partial_r U]\, \zeta &=&0.
\end{eqnarray}
In addition, $U$ being $\oone$ across $\gamma$ implies \eqref{Lipschitz across 1; eqn}, that is,
\begin{eqnarray}\label{nojumps across shock}
[\partial_t U]\, 1 + [\partial_r U]\, \dot{x} &=& 0.
\end{eqnarray} 
Now, since $(\nu,\zeta)$ is tangent to the characteristic curves of \eqref{Israel_IC_PDE}, and since \eqref{shock not characteristic} ensures that the shock curve is non-characteristic, that is, $(\nu,\zeta)\neq (1,\dot{x})$ at $\gamma$, \eqref{nojumps along characteristics} and \eqref{nojumps across shock} are independent conditions, which then yields the desired $C^1$ regularity across $\gamma$, \eqref{nojumps1}. We conclude that the $J^\mu_\alpha$ constructed meet the smoothing condition \eqref{smoothingcondt2} and are integrable to coordinates, c.f. Lemma \eqref{equiv of IC and exi of integrable function}. By Lemma \ref{smoothingcondt equiv oneone}, we conclude that the metric in the resulting coordinates, $g_{\alpha\beta}$, is $C^{1,1}$ regular. This completes the proof of Theorem \ref{Israel}.
\hfill $\Box$

\section{The Shock-Collision-Case: The Canonical Jacobian} \label{Sec: Canonical Jacobian}

In Section \ref{Israel's Thm with new method}, we have shown how to construct Jacobians smoothing the metric tensor across a single shock surface. We now proceed proving Theorem \ref{TheoremMain}, which concerns the metric smoothing in a neighborhood of a point of shock wave interaction. In principal, we follow the constructive proof of Theorem \ref{SingleShockThm}: We first extend our Jacobian ansatz \eqref{Jacobian1} to the case of two interacting shock waves and then show that this set of functions can be integrated to coordinates. However, in contrast to the single shock case addressed in the previous section, we have to pursue the construction on the upper and lower half-plane, $\R^2_\pm$, separately and then show that the resulting functions can be ``glued'' together in a way appropriate to smooth the metric.         In more detail, we have to pursue three major steps. The first one is to apply the canonical form \eqref{canoicalfortwo} of Lemma \ref{characerization2} to construct the Jacobians $J^\mu_\alpha$ on $\R^2_\pm$ separately, such that $J^\mu_\alpha$ meet the smoothing condition \eqref{smoothingcondt2} across each of the shock curves, which is the subject of this section. The second step is to show that there exists a choice of free functions $\Phi^\mu_\alpha$, such that the $J^\mu_\alpha$ constructed in this section solve the integrability condition on $\R^2_\pm$ separately, which is achieved in Section \ref{Sec: Integrability Condition}. Finally, in Section \ref{Sec: Matching Conditions}, we demonstrate how to match the $J^\mu_\alpha$, constructed on the upper and lower half-plane, across the $\{t=0\}$-interface in a way preserving the $C^{1,1}$ metric regularity, by first deriving so-called ``matching conditions'' and then proving these conditions are met by the $J^\mu_\alpha$ constructed in Section \ref{Sec: Canonical Jacobian} - \ref{Sec: Integrability Condition}. 

The subsequent proposition gives the canonical form of Jacobians that meet the smoothing condition \eqref{smoothingcondt1_SSC} across each shock curve in either $\R^2_+$ or $\R^2_-$, and act on the $(t,r)$-plane only, that is, they are of the form \eqref{transfo in tr plane, matrix}. Without loss of generality we formulate the proposition for $\R^2_+$. An additional obstacle in the case of shock interactions is that, in contrast to the single shock case, the restriction $J^\mu_\alpha \circ \gamma_i$ does not only depend on the (free) functions $\Phi^\mu_\alpha$ but also on the Jacobian coefficients from the other shock curve, that is, 
\beq \label{errorsource}
J^\mu_\alpha \circ \gamma_1 = (\varphi_2)^\mu_{\alpha}\, \big|x_1(\cdot) - x_2(\cdot)\big| + \Phi^\mu_{\alpha} \circ \gamma_1 .
\eeq
Thus, since the smoothing conditions \eqref{smoothingcondt2} depend on $J^\mu_\alpha \circ \gamma_i$ itself, we have to make sure that the smoothing condition indeed lead to an (implicit) definition of the Jacobian coefficients $(\varphi_i)^\mu_{\alpha}$, in the sense that the coefficients only depend on the free functions $\Phi^\mu_\alpha$. 

\begin{Prp}\label{canonicalformforJ}
Let $p$ be a point of regular shock wave interaction in SSC between shocks from different families, in the sense of Definition \ref{shockinteract} with (i) - (iv) being met, with corresponding SSC metric, $g_{\mu\nu}$, defined on $\mathcal{N}\cap \overline{\R^2_+}$. Then the following is equivalent:
\begin{enumerate}[(i)]
\item There exists a set of functions $J^\mu_\alpha \in \oone \left(\mathcal{N}\cap \overline{\R^2_+} \right)$ of the form \eqref{transfo in tr plane, matrix} which satisfies the smoothing condition \eqref{smoothingcondt2} on $\gamma_i\cap\mathcal{N} \cap \overline{\R^2_+}$, for $i=1,2$.
\item The RH condition \eqref{[Bt]shockspeed=[Ar]} holds on each shock curve $\gamma_i\cap\mathcal{N} \cap \overline{\R^2_+}$, for $i=1,2$, as in Definition \ref{shockinteract}, (v).
\end{enumerate}
Furthermore, any such set of functions $J^\mu_\alpha$ satisfying \eqref{smoothingcondt2} is of the ``canonical form'' 
\beq\label{Jacobian2}
J^\mu_\alpha(t,r)= \sum_{i=1,2} (\varphi_i)^\mu_{\alpha}(t) \left| x_i(t)-r \right| + \Phi^\mu_\alpha(t,r),
\eeq
where $\Phi^\mu_\alpha \in \oone \left(\mathcal{N}\cap \overline{\R^2_+} \right)$ have matching derivatives across each shock curve $\gamma_i(t)$, for $t>0$, that is,
\beq\label{nojumps2}
[\Phi^\mu_{\alpha,\, r}]_i \ = \ 0 \ = \ [\Phi^\mu_{\alpha,\, t}]_i  \ \ \ \ \forall \, \mu \, \in \, \{t,r\},  \ \forall \, \alpha \, \in\, \{0,1\},
\eeq
and where $(\varphi_i)^\mu_\alpha$ is defined implicitly through
\beq\label{Jacobiancoeff2_impl}
(\varphi_i)^\mu_\alpha =  -\frac12 (\mathcal{J}_i)^\mu_{\alpha r}  
\eeq
with $(\mathcal{J}_i)^\mu_{\alpha r}$ denoting the values $\mathcal{J}^\mu_{\alpha r}$ in \eqref{smoothingcondt2} with respect to $\gamma_i$. Explicitly, the values for $(\varphi_i)^\mu_\alpha$ are given by
\begin{eqnarray}
(\varphi_i)^t_{0} = -\frac{B_i}{A_i} \, \dot{x}_i \, (\varphi_i)^r_{0}, \label{Jacobiancoeff2_expl_t-0} \\
(\varphi_i)^t_{1} = -\frac{B_i}{A_i}\, \dot{x}_i \, (\varphi_i)^r_{1}, \label{Jacobiancoeff2_expl_t-1} 
\end{eqnarray}
\begin{align}
(\varphi_i)^r_{0}   =  
\frac{ \frac1{4B_i} \Big(  [B_t]_i \, \Phi^t_{0}|_i   +   [B_r]_i \, \Phi^r_{0}|_i   \Big)  +  \frac{1}{4 B_j}  \Big(  [B_t]_j \, \Phi^t_{0}|_j   +   [B_r]_j\,   \Phi^r_{0}|_j   \Big) \mathcal{B}_{ij} }      {1 -   \mathcal{B}_{ij} \mathcal{B}_{ji} } \label{Jacobiancoeff2_expl_r-0} \\
(\varphi_i)^r_{1} = 
\frac{\frac1{4B_i} \Big(  [B_t]_i \, \Phi^t_{1}|_i  +   [B_r]_i \, \Phi^r_{1}|_i   \Big)  +  \frac{1}{4 B_j}  \Big(  [B_t]_j \, \Phi^t_{1}|_j + [B_r]_j\,   \Phi^r_{1}|_j   \Big) \mathcal{B}_{ij} }       { 1 -   \mathcal{B}_{ij} \mathcal{B}_{ji}  } \, , \label{Jacobiancoeff2_expl_r-1} 
\end{align}
with $j\neq i$ in \eqref{Jacobiancoeff2_expl_r-0} and \eqref{Jacobiancoeff2_expl_r-1}, and where we define $A_i = A \circ \gamma_i$, $B_i = B \circ \gamma_i$,
\beq \label{restrictedR2fct, interaction}
\Phi^\mu_{\alpha}|_i= \Phi^\mu_\alpha \circ \gamma_i \, 
\eeq
and 
\beq\label{Jacobiancoeff2_expl_B}
\mathcal{B}_{ij} = \frac{|x_1(\cdot)-x_2(\cdot)|}{4B_i}  \left(  [B_r]_i   - \frac{B_j}{A_j} \dot{x}_j \, [B_t]_i  \right).
\eeq 
Furthermore, the above equivalence also holds for the full atlas of $C^{1,1}$ coordinate transformations, not restricted to the $(t,r)$-plane, with corresponding canonical form, \eqref{Jacobian2} - \eqref{Jacobiancoeff2_impl}, for $(\mathcal{J}_i)^\mu_{\alpha r}$ given by \eqref{smoothingcondt2_full}.
\end{Prp}

The error in \cite{ReintjesTemple}, was to falsely omit the term $(\varphi_j)^\mu_{\alpha}\, \big|x_1(\cdot) - x_2(\cdot)\big|$ in \eqref{errorsource}, which we correct here. The effect on the Jacobian coefficients \eqref{Jacobiancoeff2_expl_r-0} - \eqref{Jacobiancoeff2_expl_r-1} is precisely the appearance of the non-zero function $\mathcal{B}_{ij}$, and \eqref{Jacobiancoeff2_expl_r-0} - \eqref{Jacobiancoeff2_expl_r-1} reduce to the (incorrect) formulas in \cite{ReintjesTemple} upon setting $\mathcal{B}_{ij} = 0$.

\Proof
We first prove that (i) implies (ii). For this, suppose that there exist $\oone$ functions $J^\mu_\alpha$, defined on $\mathcal{N}\cap \overline{\R^2_+}$, that meet the smoothing condition \eqref{smoothingcondt2_full} on each $\gamma_i(t)$, for $t>0$. By Lemma \ref{smoothingcondt => J is oone across}, $J^\mu_\alpha$ satisfies  \eqref{Lipschitz across 1; eqn}, that is,
\beq\label{ansatz is lipschitz across, interaction}
[J^{\mu}_{\alpha,t}]_i=-\dot{x}_i[J^{\mu}_{\alpha,r}]_i .
\eeq
Substituting into \eqref{ansatz is lipschitz across, interaction} the expressions for $[J^\mu_{\alpha,\nu}]_i$ from the smoothing condition \eqref{smoothingcondt2_full} with respect to $\gamma_i$, and using that $g_{\mu\nu}$ is $\oone$ across $\gamma_i$, in the sense of \eqref{jumpone}-\eqref{jumptwo}, we find that 
$$
\left( [A_r]_i + \dot{x}_i [B_t]_i \right) \, J^\mu_\alpha \circ \gamma_i \, = \, 0 \ \ \ \ \forall \, \mu \, \in \, \{t,r\},  \ \forall \, \alpha \, \in\, \{0,1,2,3\}.
$$
Since not all of the above $J^\mu_\alpha$ can vanish simultaneously, while maintaining a non-vanishing determinant, we conclude that \eqref{ansatz is lipschitz across, interaction} implies \eqref{[Bt]shockspeed=[Ar]} on $\gamma_i$. By Lemma \ref{RH_Lemma}, \eqref{jumpone} together with \eqref{[Bt]shockspeed=[Ar]} imply the RH conditions on $\gamma_i$, which proves that (i) implies (ii).

We now prove that (ii) implies (i). Suppose that the RH conditions hold on each $\gamma_i$, $i=1,2$, then, by Lemma \ref{RH_Lemma} and since $g_{\mu\nu}$ is $\oone$ across each $\gamma_i$, \eqref{jumpone}, \eqref{jumptwo} and \eqref{[Bt]shockspeed=[Ar]} hold on each shock curve. We first need to show that $J^\mu_\alpha$, as defined in \eqref{Jacobian2} - \eqref{Jacobiancoeff2_impl}, satisfies in fact the smoothing condition, \eqref{smoothingcondt2}. However, this does not yet imply the existence of functions $J^\mu_\alpha$ which meet \eqref{smoothingcondt2}, since the coefficients $(\varphi_i)^\mu_\alpha$ in \eqref{Jacobiancoeff2_impl} depend, through \eqref{smoothingcondt2}, on the $J^\mu_\alpha$ themselves. We thus have to prove that \eqref{Jacobiancoeff2_impl} indeed gives us an implicit definition of the coefficients $(\varphi_i)^\mu_{\alpha}$. We do this, by showing that \eqref{Jacobiancoeff2_impl} is indeed equal to the explicit expressions \eqref{Jacobiancoeff2_expl_t-0} - \eqref{Jacobiancoeff2_expl_r-1} for $(\varphi_i)^\mu_\alpha$. Then, since the $(\varphi_i)^\mu_\alpha$ depend on the metric and the (free) functions $\Phi^\mu_\alpha$ alone, the claimed existence follows. 
 
To begin, we prove that the $J^\mu_\alpha$ defined in \eqref{Jacobian2}  satisfy the smoothing condition. For this, differentiate \eqref{Jacobian2} with respect to $r$ and take the jump across $\gamma_i$ of the resulting expression, then, using \eqref{nojumps2}, this leads to
\begin{eqnarray}\label{canonicalforJ_pf_eqn3}
[J^\mu_{\alpha,r}]_i(t) &=& -\sum_{j=1,2} (\varphi_j)^\mu_{\alpha}(t) [H(x_j(t)-r)]_i  \cr
&=& -2(\varphi_i)^\mu_{\alpha}(t),
\end{eqnarray}
where we used for the last equality the identity for the Heaviside function
\beq \label{Heaviside_jump}
[H(x_j(t)-r)]_i = 2\delta_{ji}.
\eeq  
Now, since $(\varphi_i)^\mu_{\alpha}$ is defined in \eqref{Jacobiancoeff2_impl} in terms of the value $(\mathcal{J}_i)^\mu_{\alpha r}$ of the smoothing conditions \eqref{smoothingcondt2},  we conclude that \eqref{canonicalforJ_pf_eqn3} implies the smoothing conditions to hold for the $r$-derivative of $J^\mu_\alpha$. Now, take the $t$-derivative of \eqref{Jacobian2} and the jump across $\gamma_i$ of the resulting expression. Using again \eqref{nojumps2} and \eqref{Heaviside_jump}, we obtain
\begin{eqnarray}\label{canonicalforJ_pf_eqn4}
[J^\mu_{\alpha,t}]_i &=& 2(\varphi_i)^\mu_{\alpha} \, \dot{x}_i \cr
&=& -\dot{x}_i (\mathcal{J}_i)^\mu_{\alpha r} ,
\end{eqnarray}
where we used the implicit definition of $(\varphi_i)^\mu_{\alpha}$, \eqref{Jacobiancoeff2_impl}, for the last equality. Now, inserting the values for $(\mathcal{J}_i)^\mu_{\alpha r}$ from \eqref{smoothingcondt2} and using the RH-condition, \eqref{jumpone}, \eqref{jumptwo} and \eqref{[Bt]shockspeed=[Ar]}, to eliminate $\dot{x}_i$ on the right hand side of \eqref{canonicalforJ_pf_eqn4} and comparing the resulting expression with the expressions for $(\mathcal{J}_i)^\mu_{\alpha t}$ from the smoothing condition, \eqref{smoothingcondt2}, finally proves
\beq \label{RH_smoothingcondt}
 -\dot{x}_i (\mathcal{J}_i)^\mu_{\alpha r} = (\mathcal{J}_i)^\mu_{\alpha t},
\eeq
which, by \eqref{canonicalforJ_pf_eqn4}, implies that $J^\mu_\alpha$ defined in \eqref{Jacobian2} satisfies the smoothing condition.

To complete the existence proof, it remains to prove the explicit expression for $(\varphi_i)^\mu_\alpha$. To begin with, we first derive the identities \eqref{Jacobiancoeff2_expl_t-0} and \eqref{Jacobiancoeff2_expl_t-1}. For this, consider the smoothing conditions \eqref{smoothingcondt2} and use the RH condition in the form \eqref{jumpone}, \eqref{jumptwo} and \eqref{[Bt]shockspeed=[Ar]}, to derive the following straighforward relations,
\begin{eqnarray}
(\mathcal{J}_i)^t_{0r} &=& -\frac{B_i}{A_i} \, \dot{x}_i \, (\mathcal{J}_i)^r_{0 r}, \label{coeff_rel_1} \\
(\mathcal{J}_i)^t_{1 r} &=& -\frac{B_i}{A_i}\, \dot{x}_i \, (\mathcal{J}_i)^r_{1 r}.  \label{coeff_rel_2}
\end{eqnarray}
\eqref{Jacobiancoeff2_expl_t-0} and \eqref{Jacobiancoeff2_expl_t-1} then already follow from the above identities together with the implicit expression of the Jacobian coefficients, \eqref{Jacobiancoeff2_impl}.

It remains to prove \eqref{Jacobiancoeff2_expl_r-0} and \eqref{Jacobiancoeff2_expl_r-1}. For this, observe that the restriction of the Jacobian as introduced by \eqref{Jacobian2} is given by
\beq\label{Jacobian_restriction}
J^\mu_\alpha \circ \gamma_i \, = \, (\varphi_j)^\mu_{\alpha} \, f + \Phi^\mu_{\alpha}|_i \; , \ \ \ \ \ \ \ \ \text{for} \ j\neq i,
\eeq
with $\Phi^\mu_{\alpha}|_i$ introduced in \eqref{restrictedR2fct, interaction} and where we define 
$$
f(t)=\left| x_1(t) - x_2(t) \right|.
$$
Now, starting with the implicit expression for $(\varphi_i)^r_{0}$ in \eqref{Jacobiancoeff2_impl}, and substituting first \eqref{smoothingcondt2} and then \eqref{Jacobian_restriction}, we obtain
\beq \nonumber 
(\varphi_i)^r_{0} = \frac1{4B_i} \Big( [B_t]_i \left((\varphi_j)^t_{0} \, f + \Phi^t_{0}|_i \right)   +  [B_r]_i \left((\varphi_j)^r_{0} \, f + \Phi^r_{0}|_i \right)  \Big) \ \ \ \ \  \text{for} \ \ j\neq i.
\eeq
Applying \eqref{Jacobiancoeff2_expl_t-0} to eliminate $(\varphi_j)^t_{0}$, the above equation becomes
\beq\label{Jac_coeff_pf_eqn2}
(\varphi_i)^r_{0}   
=   \frac1{4B_i} \Big(  [B_t]_i \, \Phi^t_{0}|_i   +   [B_r]_i\,   \Phi^r_{0}|_i   \Big)
+  \mathcal{B}_{ij} \, (\varphi_j)^r_{0}\, , 
\eeq
for $ j\neq i$, where $\mathcal{B}_{ij}$ is defined by \eqref{Jacobiancoeff2_expl_B}, that is,
$$
\mathcal{B}_{ij} = \frac{f}{4B_i}  \left(  [B_r]_i   - \frac{B_j}{A_j} \dot{x}_j \, [B_t]_i  \right)\,  .
$$ 
Exchanging $i$ and $j$ in \eqref{Jac_coeff_pf_eqn2}, we get
\beq\label{Jac_coeff_pf_eqn3}
(\varphi_j)^r_{0}   
=    \frac1{4B_j} \Big(  [B_t]_j \, \Phi^t_{0}|_j   +   [B_r]_j\,   \Phi^r_{0}|_j   \Big)
+  \mathcal{B}_{ji} \, (\varphi_i)^r_{0}  .
\eeq
Substituting \eqref{Jac_coeff_pf_eqn3} into \eqref{Jac_coeff_pf_eqn2} and solving the resulting expression for $(\varphi_i)^r_{0}$, we finally obtain 
\beq\label{Jac_coeff_pf_eqn4}
(\varphi_i)^r_{0}   =   
\frac{ \frac1{4B_i} \Big(  [B_t]_i \Phi^t_{0}|_i + [B_r]_i \Phi^r_{0}|_i \Big)  +  \frac{1}{4 B_j}  \Big(  [B_t]_j \, \Phi^t_{0}|_j + [B_r]_j\, \Phi^r_{0}|_j \Big) \mathcal{B}_{ij} }   {1 -   \mathcal{B}_{ij} \mathcal{B}_{ji} } .
\eeq
Note that $ 1 -   \mathcal{B}_{ij} \mathcal{B}_{ji} \neq 0$ for $t$ sufficiently close to $0$, since $f(t)=|x_1(t)-x_2(t)|$ vanishes at $t=0$ and thus $\mathcal{B}_{ij}$ converges to $0$ as $t$ approaches $0$, c.f. \eqref{Jacobiancoeff2_expl_B}. This proves \eqref{Jacobiancoeff2_expl_r-0}.

To prove \eqref{Jacobiancoeff2_expl_r-1}, we substitute 
\eqref{smoothingcondt2} and \eqref{Jacobian_restriction} into the implicit expression for $(\varphi_i)^r_{1}$ in \eqref{Jacobiancoeff2_impl}, which yields
\beq\label{Jac_coeff_pf_eqn5}
(\varphi_i)^r_{1} = \frac1{4B_i} \Big( [B_t]_i \left((\varphi_j)^t_{1} \, f + \Phi^t_{1}|_i \right)   +  [B_r]_i \left((\varphi_j)^r_{1} \, f + \Phi^r_{1}|_i \right)  \Big), \ \ \ \ \ j\neq i.
\eeq
Applying \eqref{Jacobiancoeff2_expl_t-1} to eliminate $(\varphi_j)^t_{1}$, \eqref{Jac_coeff_pf_eqn5} becomes
\beq\label{Jac_coeff_pf_eqn6}
(\varphi_i)^r_{1}   =   \frac1{4B_i} \Big(  [B_t]_i \, \Phi^t_{1}|_i   +   [B_r]_i\,   \Phi^r_{1}|_i   \Big)
+  \mathcal{B}_{ij} \, (\varphi_j)^r_{1}\, , 
\eeq
for $ j\neq i$ and $\mathcal{B}_{ij}$ defined in \eqref{Jacobiancoeff2_expl_B}. Exchanging $i$ and $j$ in \eqref{Jac_coeff_pf_eqn6}, gives
\beq\label{Jac_coeff_pf_eqn7}
(\varphi_j)^r_{1}   =    \frac1{4B_j} \Big(  [B_t]_j \, \Phi^t_{1}|_j   +   [B_r]_j\,   \Phi^r_{1}|_j   \Big)
+  \mathcal{B}_{ji} \, (\varphi_i)^r_{1}  .
\eeq
Substituting \eqref{Jac_coeff_pf_eqn7} into \eqref{Jac_coeff_pf_eqn6} and solving the resulting expression for $(\varphi_i)^r_{1}$, we finally obtain 
\beq\label{Jac_coeff_pf_eqn8}
(\varphi_i)^r_{1} = 
\frac{\frac1{4B_i} \Big(  [B_t]_i  \Phi^t_{1}|_i  +   [B_r]_i \Phi^r_{1}|_i \Big)  +  \frac{1}{4 B_j}  \Big(  [B_t]_j \, \Phi^t_{1}|_j + [B_r]_j\, \Phi^r_{1}|_j   \Big) \mathcal{B}_{ij} }       { \left( 1 -   \mathcal{B}_{ij} \mathcal{B}_{ji} \right) },
\eeq
which proves \eqref{Jacobiancoeff2_expl_r-1}.

We finally prove that any set of Lipschitz continuous functions $J^\mu_\alpha$ which satisfies the smoothing condition is of the canonical form \eqref{Jacobian2}. By Lemma \ref{smoothingcondt => J is oone across}, any such function $J^\mu_\alpha$ is $\oone$ across each of the shock curves, $\gamma_i(t)$ for $t>0$, and satisfies in particular \eqref{Lipschitz across 1; eqn}. Thus, applying Lemma \ref{characerization2} for fixed $\mu$ and $\alpha$, there exists a function $\Phi^\mu_\alpha \, \in\, \oone(\mathcal{N}\cap\R^2_+)$ which satisfies \eqref{nojumps2} and which meets the canonical form \eqref{Jacobian2} for the coefficient 
\beq\label{canonicalforJ_pf_eqn1}
(\varphi_i)^\mu_\alpha 
\ =\ \frac12\frac{[J^\mu_{\alpha,\sigma}]_i\, (n_i)^\sigma}{(n_i)_\sigma (n_i)^\sigma}
\ =\ \frac12\frac{(\mathcal{J}_i)^\mu_{\alpha \sigma}\, (n_i)^\sigma}{(n_i)_\sigma (n_i)^\sigma},
\eeq
where we used in the last equality that the $J^\mu_\alpha$ are assumed to satisfy the smoothing condition. Using that the normal 1-form for $\gamma_i$ is given by $(n_i)_\sigma = (\dot{x}_i,-1)$ and thus $(n_i)^\sigma= \ ^T(-\frac{\dot{x}_i}{A},-\frac1{B})$ and $(n_i)_\sigma (n_i)^\sigma =  \frac{A- B\dot{x}^2}{AB}$, we find that \eqref{canonicalforJ_pf_eqn1} yields
\beq\label{canonicalforJ_pf_eqn2}
(\varphi_i)^\mu_\alpha  = -\frac12  \frac{ B\dot{x}_i (\mathcal{J}_i)^\mu_{\alpha t}\, + A  (\mathcal{J}_i)^\mu_{\alpha r} }{A - B\dot{x}^2} .
\eeq
Finally using that $J^\mu_\alpha$ satisfies \eqref{Lipschitz across 1; eqn} in the form $(\mathcal{J}_i)^\mu_{\alpha t}=-\dot{x}_i(\mathcal{J}_i)^\mu_{\alpha r}$, we find that $(\varphi_i)^\mu_{\alpha}$ in \eqref{canonicalforJ_pf_eqn2} agrees with \eqref{Jacobiancoeff2_impl}, that is, $(\varphi_i)^\mu_\alpha =  -\frac12 (\mathcal{J}_i)^\mu_{\alpha r}$. In summary, we conclude that $J^\mu_\alpha$ assumes the canonical form \eqref{Jacobian2}, which completes the proof.
\QED

\section{The Shock-Collision-Case: The Integrability Condition}\label{Sec: Integrability Condition}

From Proposition \ref{canonicalformforJ}, we conclude that choosing four arbitrary $C^{1,1}$ functions $\Phi^\mu_\alpha$, (with a non-vanishing determinant), the construction \eqref{Jacobian2} - \eqref{Jacobiancoeff2_expl_B} yields four $\oone$ functions $J^\mu_\alpha$ which satisfy the smoothing condition on $\gamma_1^+$ and $\gamma_2^+$. However, for $J^{\mu}_{\alpha}$ to be proper Jacobians, integrable to a coordinate system, we must use the free functions $\Phi^\mu_\alpha$ to meet the integrability condition \eqref{IC}, c.f. Lemma \ref{equiv of IC and exi of integrable function}. For this, we rewrite \eqref{IC} as a system of first order differential equation in the unknowns $U:= \, ^T(\Phi^t_0, \Phi^r_0)$ and we consider $\Phi^t_1$ and $\Phi^r_1$ as arbitrary given smooth functions, (at least in $C^3$). The resulting equations are a special case of a non-local system of partial differential equations for which we prove the existence of solutions in Section \ref{Sec: Non-local PDE}.

To begin with, we write the integrability conditions, 
$$
J^\mu_{\alpha,\beta}=J^\mu_{\beta,\alpha},
$$ 
in their equivalent form 
\beq\label{IC_SSC_again_1}
J^\mu_{\alpha,\sigma} J^\sigma_\beta = J^\mu_{\beta,\sigma} J^\sigma_\alpha ,
\eeq
c.f. Lemma \ref{Appendix_Lemma2}. Without loss of generality, suppose $\alpha=0$ and $\beta=1$, then,  assuming $\Phi^t_1$ is such that $J^t_1\neq0$, we write \eqref{IC_SSC_again_1} in its equivalent form
\beq\nonumber
J^\mu_{0,t}  +\frac{ J^r_1}{J^t_1} \,  J^\mu_{0,r}  \  = \ \frac{ J^\mu_{1,t}}{J^t_1} J^t_0 + \frac{J^\mu_{1,r}}{J^t_1} J^r_0 ,
\eeq
which in matrix notation is given by
\beq\label{IC_SSC_again_2}
\frac{\partial}{\partial t}\left(\begin{array}{c} J^t_0 \cr J^r_0  \end{array}\right)
+\, \frac{ J^r_1}{J^t_1} \ \frac{\partial}{\partial r} \left(\begin{array}{c} J^t_0 \cr J^r_0  \end{array}\right) \,
= \,  \frac{1}{J^t_1} \left(\begin{array}{cc}  J^t_{1,t} &  J^t_{1,r}  \cr  J^r_{1,t}   &  J^r_{1,r}  \end{array}\right)  \left(\begin{array}{c} J^t_0 \cr J^r_0  \end{array}\right) \, .
\eeq
Now, it is straightforward to verify that \eqref{IC} for $J^\mu_\alpha$ defined in \eqref{Jacobian2} assumes the form
\beq\label{IC_2shocks}
\partial_t U + c\, \partial_r U \,=\, F(U),
\eeq
where 
\beq\label{IC_2shocks_def_b}
c = \frac{J^r_1}{J^t_1},
\eeq
and  
\beq \label{IC_2shocks_def_F}
F = \mathcal{M}\, U    +    \sum_{i=1,2} \left\{ \Big(  |X_i| \mathcal{M}  - H(X_i) \left(\dot{x}_i - c \right)  
\Big) \left(\begin{array}{c} (\varphi_i)^t_{0} \cr (\varphi_i)^r_{0} \end{array} \right)   
- |X_i| \left(\begin{array}{c} (\dot{\varphi}_i)^t_{0} \cr (\dot{\varphi}_i)^r_{0} \end{array} \right)    \right\}
\eeq
with
\beq\label{IC_2shocks_def_C}
\mathcal{M} =   \frac{1}{J^t_1} \left( \begin{array}{cc} J^t_{1,t}  &  J^t_{1,r} \cr J^r_{1,t} & J^r_{1,r}   \end{array} \right).
\eeq
The goal of this section is the following proposition, which proves existence of a $C^{1,1}$ solution in the region  
\beq \label{N+}
\mathcal{N}_+=\mathcal{N}\cap\overline{\mathbb{R}^2_+},
\eeq
where $\mathcal{N}$ is some neighborhood of $p=(0,r_0)$ in the $(t,r)$-plane. An analogous result holds for  the set $\mathcal{N}_-=\mathcal{N}\cap\overline{\mathbb{R}^2_-}$.

\begin{Prp} \label{soln_IC_Prop}
Assume $C^2$ regular initial data $U_0(r)$ and assume $\Phi^t_1$ and $\Phi^r_1$ are given $C^3$ functions. Then, there exist a neighborhood $\mathcal{N}$ of $p$ and there exist a $C^{1,1}$ regular function $U= \, ^T(\Phi^t_0, \Phi^r_0)$ which solves the integrability condition, \eqref{IC_2shocks} - \eqref{IC_2shocks_def_C}, in the region $\mathcal{N}_+ = \mathcal{N}\cap\overline{\mathbb{R}^2_+}$, such that $U(0,r)=U_0(r)$ for all $(0,r)\in\mathcal{N}_+$.
\end{Prp}

The difficulty of proving existence of solutions to \eqref{IC_2shocks} - \eqref{IC_2shocks_def_C} is that $F(U)$ contains terms which depend on $U\circ \gamma_i$ and its derivatives. The presence of these terms turn \eqref{IC_2shocks} - \eqref{IC_2shocks_def_C} into a system of \emph{non-local} PDE's, for which one cannot apply standard existence theory of hyperbolic PDE's. To overcome these difficulties, we study in Section \ref{Sec: Non-local PDE} a system of PDE's generalizing \eqref{IC_2shocks} - \eqref{IC_2shocks_def_C} in a way suitable for isolating the non-local structure, and we prove existence of $\oone$ solutions for these non-local PDE's. In Section \ref{Sec: Bootstrapping}, we use this existence result and the explicit structure of the integrability condition \eqref{IC_2shocks} for a bootstrapping argument which yields the $C^{1,1}$ regularity claimed in Proposition \ref{soln_IC_Prop}. Note that, to pursue the construction of the Jacobian smoothing the metric tensor from $\oone$ to $C^{1,1}$, the $C^{1,1}$ regularity of $U:= \, ^T(\Phi^t_0, \Phi^r_0)$ is fundamental, c.f. Proposition \ref{canonicalformforJ}.

\subsection{Existence Theory of a Non-Local First Order PDE} \label{Sec: Non-local PDE}

To introduce the system of non-local PDE's generalizing the integrability condition \eqref{IC_2shocks}, we first introduce a suitable function space. We start with conditions assumed in the set $\mathcal{N}_+$, introduced in \eqref{N+}, and seek solutions in a smaller set of similar shape. Moreover, let $\mathcal{N}_+^c$ denote the union of the three open sets in the upper half plane obtained by taking the complement of the three curves  $\{(t,r)|t=0\}$, $\gamma_1$ and $\gamma_2$, in $\mathcal{N}_+$.  We denote these three open sets in $\R^2_+$ , which are the connected components of $\mathcal{N}_+^c$, by $\mathcal{N}_L$, $\mathcal{N}_M$ and $\mathcal{N}_R$.  Here $\mathcal{N}_L$ denotes the left most region, between $t=0$ and $\gamma_1$, $\mathcal{N}_M$ is the middle region between $\gamma_1$ and $\gamma_2$, and $\mathcal{N}_R$ denotes the right most region between $\gamma_1$ and $t=0$.  
 
We now work with functions that are Lipschitz continuous across the shocks $\gamma_i$, $i=1,2$, and are in $C^l(\bar{\mathcal{N}}_p)$ for each $p=L,M,R$ and for $l\geq2$,\footnote{To prove the main Theorem of this section, $l\geq1$ suffices, but for the issue of metric smoothing addressed in this paper the case $l\geq2$ is relevant.} where $\bar{\mathcal{N}}_p$ denotes the closure of $\mathcal{N}_p$.  That is,  $C^l(\bar{\mathcal{N}}_p)$ is the space of functions whose derivatives up to order $l$ exist and are bounded in  $\mathcal{N}_p$, and have continuous extensions to the boundary curves.    Now define 
\begin{eqnarray}\label{0.0}
 C^{0,1}_l(\mathcal{N}_+)=C^{0,1}(\mathcal{N}_+)\cap_{p=L,M,R} C^{l}(\bar{\mathcal{N}}_p).
\end{eqnarray}
In particular, the restriction of a function in $C^{0,1}_l(\mathcal{N}_+)$ to one of the boundary curves $t=0$, or $\gamma_i$, $i=1,2$, is a $C^l$ function whose first $l$ derivatives are limits of the $C^l$ derivatives in the open sets on each side, and hence are bounded by their norms, (c.f. \eqref{Lipschitz across 1; eqn3_proof} and the subsequent argument in the proof of Lemma \ref{Lipschitz_across_lemma}). 

We consider now the following initial value problem which is basic to our analysis:
\begin{eqnarray}
&\partial_t U + c\, \partial_r U = F(U) \label{Cauchyproblem_PDE} \\
&U(0,r)=U_0(r),\label{Cauchyproblem_initial}
\end{eqnarray}
where $U(t,r)\in\mathbb{R}^2$, $c=c(t,r)$ is a given scalar real valued function in $C^{0,1}_l(\mathcal{N}_+)$, and
\begin{eqnarray}\label{3}
F(U)=\mathcal{M}U+\sum_{i=1,2}\left(\mathcal{A}_i\cdot (U\circ\gamma_i)+\mathcal{B}_i\left|X_i\right|\cdot\frac{d}{dt}\left(U\circ\gamma_i\right)\right)
\end{eqnarray}
with
\begin{eqnarray}\label{4}
\mathcal{M}=\sum_{i=1,2}M_iH(X_i)+M,
\end{eqnarray}
\begin{eqnarray}\label{5}
\mathcal{A}_i =\sum_{j=1,2}\mathcal{A}_{ij}H(X_j)+A_i,
\end{eqnarray}
and $M,M_i,A_i,\mathcal{A}_{ij},\mathcal{B}_i$ are given $2\times2$ matrix valued functions of $(t,r)$ in $C^{0,1}_l(\mathcal{N}_+)$, $H$ denotes the Heaviside step function, defined by $H(x)=-1$ for $x<0$ and $H(x)=1$ for $x>0$, and 
\begin{eqnarray}\label{6}
X_i(t,r)=x_i(t)-r .
\end{eqnarray}

Our goal now is to prove that solutions $U(t,r) \in C^{0,1}({\mathcal{N}}_+)$ of \eqref{Cauchyproblem_PDE} - (\ref{Cauchyproblem_initial}) exist  for smooth initial data $U_0$, such that these solutions are more regular away from the shock curves, as specified in \eqref{hat-space} below. For this we introduce the following iteration scheme which we define inductively. For the induction, start with $U^0=0$, and define $U^k$ for $k\geq1$ in terms of $U^{k-1}$ by 
\begin{eqnarray}\label{IC_iteration_1}
&& \partial_t U^k + c\, \partial_r U^k =\, F^{k-1}, \cr
&& U^k(0,r)=U_0(r),
\end{eqnarray}
where we set $F^{k-1} = F(U^{k-1})$.

Observe that the PDEs in (\ref{Cauchyproblem_PDE}) are coupled through the right hand side, but the iteration scheme \eqref{IC_iteration_1} is an uncoupled system of equations for $U^k$ once $U^{k-1}$ (and hence $F^{k-1}$) is a given function, because $c(t,r)$ is scalar.  The problem then is how to prove convergence of the above scheme, for which we have to overcome two major difficulties: The first one is the low regularity in $F^{k-1}$, containing Heaviside and absolute value functions, which can be handled by a careful analysis. The second difficulty is the presence of leading order derivatives in $F^{k-1}$, which we overcome thanks to the factor $|X_i|$ in each term in $F^{k-1}$ containing a derivative of $U^{k-1}$, since $|X_i|=|x_i(t)-r|$ can be bounded by $t$ up to some constant, which then suffices to prove \emph{local} existence for $t$ sufficiently small. 

The Cauchy problem for the $k$-th iterate in \eqref{IC_iteration_1} is well-posed and the solution can be computed explicitly through the method of characteristics. To see this, we start by introducing the characteristic curves $t\mapsto (t,\mathcal{X}(t,r'))$ for \eqref{IC_iteration_1}, where $\mathcal{X}$ is defined to be the solution of the ODE
\begin{eqnarray}\label{ODE_characteristics}
\frac{d}{d t} \mathcal{X}(t,r') &=& c\left(t,\mathcal{X}(t,r')\right), \\\nonumber
 \mathcal{X}(0,r')&=&r',
\end{eqnarray}
in which $r'$ is treated as a parameter. Note, the solution $\mathcal{X}(t,r')$ of \eqref{ODE_characteristics} gives the $r$-position at time $t$ of the characteristic emanating from  $(0,r')$.  

The next Lemma proves that if $c\in C^{0,1}_l(\mathcal{N}_+)$, then a solution $r=\mathcal{X}(t,r')$ of \eqref{ODE_characteristics} and its inverse $r'=\mathcal{Y}(t,r)$ exist and both functions are $C^{1,1}$ regular, but with a loss of $C^l$ regularity across the unique characteristic emanating from $r'=r_0$, which we subsequently denote by $\gamma_0$.\footnote{In $(t,r')$-coordinates $\gamma_0$ coincides with the line $r'=r_0$.} We make the basic assumption that the characteristic speed $c\in C^{0,1}_l(\mathcal{N}_+)$ is bounded away from both shock speeds, 
$$
c(t,r) \neq \dot{x}_i(t),
$$ 
for $i=1,2$. For convenience, and without loss of generality, we now assume that $c$ lies between the two shock speeds throughout $\mathcal{N}_+$, 
\begin{eqnarray}\label{shockspeedassumption}
\dot{x}_1(t)<c(t,r)<\dot{x}_2(t).
\end{eqnarray} 
\eqref{shockspeedassumption} implies that characteristics intersect one and only one of the shock curves within finite time, which allows us to define $\tau(r')$ to be the time at which the characteristic starting at $(0,r')$ intersects a shock curve:  $\gamma_1$ if $r'<r_0$, $\gamma_2$ if $r'>r_0$ and with $\tau(r_0)=0$. In fact, the function $\tau$ exists for $r'$ sufficiently close to $r_0$, because the shock curves $\gamma_1$ and $\gamma_2$ both emanate from $(0,r_0)$.  Up until the time of intersection of the characteristics with a $\gamma_i$, the function $c$ is $C^l$ by assumption, and $\mathcal{X}$ must be a $C^l$ function of $(t,r')$ for all $t<\tau(r')$ and all $r'\neq r_0$, by continuous dependence of solutions of (\ref{ODE_characteristics}) on its parameters.  It follows that the function $r' \mapsto \tau(r')$ is defined in a neighborhood of $(0,r_0)$ and is $C^l$ regular for $r'\neq r_0$, because the $\gamma_i$ are smooth away from $r'= r_0$. We are now prepared to state and prove the following lemma about the existence and regularity of the characteristic curves.

\begin{Lemma}\label{ODE_char_Lemma_exist}
The solution $\mathcal{X}(t,r')$ of the ODE (\ref{ODE_characteristics}) exists in $\mathcal{N}_+=\mathcal{N} \cap \overline{\R^2_+}$ for a sufficiently small neighborhood $\mathcal{N}$ of $(0,r_0)$, it is $C^{1,1}$ throughout $\mathcal{N}_+$, is $C^l$ at each point $(t,r')\in\mathcal{N}_+$ where $t\neq\tau(r')$ and $r'\neq r_0$, and its derivatives up to order $l$ extend continuously to each side of the curves $t\neq\tau(r')$ and $r'\neq r_0$. Moreover, $\mathcal{X}$ has an inverse $\mathcal{Y}(t,r)\in C^{1,1}(\mathcal{N}_+)$ which solves
\beq \label{characteristics_inverse}
r = \mathcal{X}(t,\mathcal{Y}(t,r)).
\eeq
$\mathcal{Y}$ is $C^l$ away from the shock curves and $\gamma_0$, with its derivatives (up to order $l$) extending continuously to each side of the boundary curves $\gamma_i$, $i=0,1,2$.  
\end{Lemma}

\Proof 
Since $c$ is Lipschitz continuous, it follows from the standard existence and uniqueness theory for ODE's, (the Picard-Lindel\"off Theorem), that there exists a unique solution $\mathcal{X}(t,r')$ of (\ref{ODE_characteristics}), and $\mathcal{X}$ will be $\oneone$ regular in $t$ and $\oone$ regular in $r'$ in some neighborhood $\mathcal{N}_+$ of $(0,r_0)$. Moreover, regular dependence of $\mathcal{X}$ on $c(\cdot,\mathcal{X})$, by the basic theory of ODE's, implies that $\mathcal{X}$ is $C^l$ regular in the regions where $t < \tau(r')$, since $c$ is $C^l$ regular away from the shock curves. To prove the $C^l$ regularity in the region where $t> \tau(r')$ and $r'\neq r_0$, compute $\mathcal{X}$ at the line $t=\tau(r')$.  Considering the resulting $C^l$ regular function as initial data at $t=\tau(r')$ and applying again the ODE-theorem of regular dependence, we find that $\mathcal{X}$ is $C^l$ regular in the region where $t>\tau(r')$ and $r'\neq r_0$. By the same theorem we find that $\mathcal{X}$ and derivatives up to order $l$ extend smoothly to each side of the curves $t=\tau(r')$ and $r'=r_0$. 

It remains to prove the $C^{1,1}$ regularity of $\mathcal{X}$ with respect to differentiation in $r'$. For this, differentiate the ODE \eqref{ODE_characteristics} with respect to $r'$. Solving the resulting linear equation leads to
\beq \nonumber
\partial_{r'} \mathcal{X}(t,r') = \exp \left( \int\limits_0^t \partial_{\mathcal{X}} c\left(s,\mathcal{X}(s,r')\right) ds \right),
\eeq
and for $t>\tau(r')$ we write the above equation as
\beq \nonumber
\partial_{r'} \mathcal{X}(t,r') 
= \exp \left( \int\limits_0^{\tau(r')} \partial_{\mathcal{X}} c\left(s,\mathcal{X}(s,r')\right) ds +\int\limits_{\tau(r')}^t \partial_{\mathcal{X}} c\left(s,\mathcal{X}(s,r')\right) ds \right).
\eeq
From the above two equations and the continuity of $\tau(\cdot)$, it is immediate that the jumps of $\partial_{r'} \mathcal{X}$ across the shock curves $t=\tau(r')$ as well as the line $r'=r_0$ vanish and thus $\partial_{r'} \mathcal{X} \in C^0(\mathcal{N}_+)$. Now, since $\mathcal{X}$ is $C^l$ in the region where $t\neq \tau(r')$ and $r'\neq r_0$ with derivatives of order up to $l$ extend to the boundary lines $t= \tau(r')$ and $r'= r_0$, and since derivatives of $\mathcal{X}$ are continuous in all of $\mathcal{N}_+$, we conclude that second order weak derivatives are in $L^\infty(\mathcal{N}_+)$ and therefore $\mathcal{X} \in C^{1,1}(\mathcal{N}_+)$, c.f. \cite{Evans}.\footnote{In the above reasoning, we use that for a function $f$ which is $C^1$ away from some hypersurface, its weak derivative agrees with its strong derivative almost everywhere, provided $f$ is continuous across the hypersurface. Note that, if $f$ fails to be continuous, partial integration leads to boundary terms so that its weak derivative differs from its strong derivative and is generally not a function in some $L^p$ space but a true distribution, as can easily be verified by direct computation from the definition of weak derivative in, e.g., \cite{Evans}.} (Observe that $\mathcal{X}$ is in general not $C^l$ regular across the line $r'=r_0$, since $\tau(r')$ is only Lipschitz continuous at $r'=r_0$.)

We now show the existence of a function $\mathcal{Y}$ which satisfies \eqref{characteristics_inverse}. For this, define the $C^1$ function
\beq \nonumber
f(t,r,r') = \mathcal{X}(t,r') - r,
\eeq 
then, by the choice of initial data in \eqref{ODE_characteristics}, it follows that
\begin{eqnarray}\nonumber
f(0,r_0,r_0) = 0 , \cr
\partial_{r'}f(0,r_0,r_0) = 1 .
\end{eqnarray}
Thus, by the Implicit Function Theorem, there exist a neighborhood $\mathcal{N}$ of $p=(0,r_0)$ and a function $\mathcal{Y} \in C^1(\mathcal{N}_+)$ which satisfies
\beq \nonumber
f(t,r,\mathcal{Y}(t,r)) = 0 
\eeq
for all $(t,r) \in \mathcal{N}_+$. This proves the existence of $\mathcal{Y}$ satisfying \eqref{characteristics_inverse}.

We now prove the claimed regularity of $\mathcal{Y}$. Using that $\mathcal{Y}$ is $C^1$ and differentiating \eqref{characteristics_inverse} with respect to $\partial_r$ and $\partial_t$, we obtain
\begin{eqnarray} \nonumber
1 &=& \partial_\mathcal{Y}\mathcal{X}(t,\mathcal{Y}) \partial_r\mathcal{Y}(t,r) \cr
0 &=&  \partial_t \mathcal{X}(t,\mathcal{Y}) + \partial_\mathcal{Y} \mathcal{X}(t,\mathcal{Y}) \partial_t \mathcal{Y} \cr
&=&  c\left(t,\mathcal{X}(t,\mathcal{Y})\right) + \partial_\mathcal{Y} \mathcal{X}(t,\mathcal{Y}) \partial_t \mathcal{Y} ,
\end{eqnarray}
where we use \eqref{ODE_characteristics} for the last equality. By the ODE \eqref{ODE_characteristics}, $\partial_t \mathcal{X}$ and $\partial_{r'} \mathcal{X}$ are non-vanishing for $r'$ close to $r_0$ and $t\geq 0$ sufficiently small. Now, solving the above system for $\partial_t \mathcal{Y}$ and $\partial_r \mathcal{Y}$, we obtain 
\beq \label{derivatives_Y}
\partial_t\mathcal{Y} = -\frac{c\left(\cdot,\mathcal{X}(\cdot,\mathcal{Y})\right)}{\partial_\mathcal{Y}\mathcal{X}(\cdot,\mathcal{Y})} \ \ \ \ \text{and} \ \ \ \partial_r\mathcal{Y} = \frac{1}{\partial_\mathcal{Y}\mathcal{X}(\cdot,\mathcal{Y})}.
\eeq 
Since the right hand sides in \eqref{derivatives_Y} are Lipschitz continuous, we conclude that $\mathcal{Y} \in C^{1,1}(\mathcal{N}_+)$. Moreover, since $\mathcal{Y}$ maps the region away from the shock curves and $\gamma_0$ to the regions of $C^l$ regularity of $\mathcal{X}$, (that is, where $t\neq \tau(r')$ and $r'\neq r_0$), and since $\mathcal{X}$ maps the regions of its $C^l$ regularity to the region of $C^l$ regularity of $c$, we find from \eqref{derivatives_Y} that $\mathcal{Y}$ is $C^l$ away from the shock curves and $\gamma_0$, with its derivatives up to order $l$ being continuously extendable to the boundary curves. This proves the claimed regularity of $\mathcal{Y}$ and concludes the proof.
\QED

We now compute the explicit solution of the iterative scheme \eqref{IC_iteration_1} using the method of characteristics. Treating $r'$ as fixed along the characteristic curves, our iteration scheme \eqref{IC_iteration_1} takes  the form
\begin{eqnarray}\label{IC_iteration_2}
&& \frac{d}{d t} U^k\left(t,\mathcal{X}(t,r')\right)  \,=\, F^{k-1}\left(t,\mathcal{X}(t,r')\right), \cr
&& U^k(0,r)=U_0(r).
\end{eqnarray}
Integrating \eqref{IC_iteration_2} gives
\beq\label{IC_iteration_solns}
U^k(t,\mathcal{X}(t,r')) = U_0(r') +  \int_0^t F^{k-1}\left(s,\mathcal{X}(s,r')\right) ds .
\eeq
Substituting $r=\mathcal{X}(t,r')$ and $r'=\mathcal{Y}(t,r)$ into (\ref{IC_iteration_solns}) we obtain the following exact expression for the solution $U^k$ of (\ref{IC_iteration_1}):
\beq\label{IC_iteration_soln}
U^k(t,r) =  U_0(\mathcal{Y}(t,r)) +  \int_0^t F^{k-1}\left(s,\mathcal{X}\big(s,\mathcal{Y}(t,r)\big)\right) ds .
\eeq
Using (\ref{derivatives_Y}), one can verify directly that (\ref{IC_iteration_soln}) solves (\ref{IC_iteration_1}).      

Our main problem in deducing the regularity of $U^k$ is the problem of determining the regularity of $\int_0^t F^{k-1}\left(s,\mathcal{X}(s,r')\right) ds$.  One difficulty is that $\mathcal{X}$ is in $C^l$ only away from $\gamma_i$ and the line $r'=r_0$, according to Lemma \ref{ODE_char_Lemma_exist}. Thus, we must account for the fact that both $\mathcal{X}$ and $\mathcal{Y}$ introduce an additional lower $C^{1,1}$ regularity across $\gamma_0$, the unique characteristic emanating from the initial point $(0,r_0)$.          Furthermore, by \eqref{3}-\eqref{6}, the coefficients of the $U^{k-1}$ in $F^{k-1}$ involve functions that are either in $C^{0,1}_l$ or else Heaviside functions times functions in $C^{0,1}_l$ and a careful analysis (done in the subsequent lemma) shows that the integral over $F^{k-1}$ in \eqref{IC_iteration_soln} generally suffers from a lack of regularity across $\gamma_0$, similar to the one identified for $\mathcal{X}$. This does not affect the convergence proof, but we must account for it in the function spaces used to prove convergence as we do in the following.  

To begin with, we define a modification $\hat{C}_{l}^{0,1}(\mathcal{N}_+)$ of $C_{l}^{0,1}(\mathcal{N}_+)$. We need only modify the space $C^l(\bar{\mathcal{N}}_M)$ to allow for a loss of $C^1$ regularity across the single characteristic emanating from $(0,r_0)$. For this purpose, define $\mathcal{N}_{M}^L$ as the open region in $\mathcal{N}_{M}$ to the left of $\gamma_0$, and $\mathcal{N}_{M}^R$ as the open region in $\mathcal{N}_{M}$ to the right of $\gamma_0$. 
Now, our modification $\hat{C}^{0,1}_l(\mathcal{N}_+)$ of $C^{0,1}_l(\mathcal{N}_+)$ is given by
\beq \label{hat-space}
 \hat{C}^{0,1}_l(\mathcal{N}_+) = C^{0,1}(\mathcal{N}_+) \cap C^{l}(\bar{\mathcal{N}}_L)
 \cap C^l(\bar{\mathcal{N}}_M^L) \cap C^l(\bar{\mathcal{N}}_M^R) \cap C^{l}(\bar{\mathcal{N}}_R).
\eeq
In correspondence with our definition above, c.f. \eqref{0.0}, $C^l(\bar{\mathcal{N}}_M^L)$ and $C^l(\bar{\mathcal{N}}_M^R)$ denote the space of functions whose derivatives exist up to order $l$, are bounded, and are continuously extendable to the boundaries. 

We now introduce a norm on $\hat{C}^{0,1}_l(\mathcal{N}_+)$. For this, define the $C^l$ norm on the closure of $\mathcal{N}_p$ as
$$
\|f\|_{C^{l}(\bar{\mathcal{N}}_p)}=\sum_{m=0,...,l} \|\partial^m f\|_{C^0(\bar{\mathcal{N}}_p)},
$$
where $\mathcal{N}_p$ denotes $\mathcal{N}_L$, $\mathcal{N}^L_M$, $\mathcal{N}^R_M$ or $\mathcal{N}_R$, and $\|\cdot\|_{C^0(\bar{\mathcal{N}}_p)}$ denotes the supremums-norm, c.f. \cite{Evans}. Here $\partial^m$ denotes either an $m$-th order derivative in the interior, or a one-sided $m$-th order derivative on the boundary of $\mathcal{N}_p$. The natural norm that makes $\hat{C}^{0,1}_l(\mathcal{N}_+)$ into a Banach space is then given by
\beq \label{norm2}
 \|\cdot\|_{\hat{C}^{0,1}_l(\mathcal{N}_+)}=\|\cdot\|_{C^{0,1}(\mathcal{N}_+)}+\| \cdot \|_{\hat{C}^l(\mathcal{N}_+)},
\eeq
where we define, for $l\geq0$,
\beq \label{norm2_addition}
\| \cdot \|_{\hat{C}^l(\mathcal{N}_+)} = \| \cdot \|_{C^{l}(\bar{\mathcal{N}}_L)} + \| \cdot \|_{C^{l}(\bar{\mathcal{N}}_M^{L})} + \| \cdot\|_{C^{l}(\bar{\mathcal{N}}_M^{R})} + \| \cdot \|_{C^{l}(\bar{\mathcal{N}}_R)}. 
\eeq

Essentially, functions bounded in the norm \eqref{norm2} are $C^l$ away from the shocks and the characteristic $\gamma_0$, Lipschitz continuous across them, such that the restriction of the function to the shock curves and $\gamma_0$ are $C^l$ functions that inherit bounds on derivatives from the bounds on each side.\footnote{Note that $\gamma_0$ is a $C^l$ curve, as can be seen as follows: By Lemma \ref{ODE_char_Lemma_exist}, $\mathcal{X}$ is $C^l$ regular away from the shock curves and $\gamma_0$, with all its derivatives extending to the boundaries, including $\gamma_0$, so that $\mathcal{X}$ is $C^l$ tangential to $\gamma_0$. But all derivatives of $\mathcal{X}$ tangential to $\gamma_0$ are in fact the derivatives of $\gamma_0$, from which we conclude that $\gamma_0$ itself is in $C^l$.}  In particular, for $i=0,1,2$, we immediately obtain the estimate
\begin{eqnarray}\nonumber 
\|f\circ\gamma_i\|_{C^l}\leq \|f\|_{\hat{C}^{0,1}_l(\mathcal{N}_+)} .
\end{eqnarray}
It follows that functions in $\hat{C}^{0,1}_l(\mathcal{N}_+)$ have the property that their restriction to one of the boundary curves $t=0$, or $\gamma_i$, $i=1,2$, or $\gamma_0$, is a $C^l$ function whose first $l$ derivatives are limits of the $C^l$ derivatives in the open sets on each side, and hence are bounded by their norms. In fact, functions in $\hat{C}^{0,1}_l(\mathcal{N}_+)$ agree with functions in $C^{0,1}_l(\mathcal{N}_+)$ except that they may be only $C^{0,1}$ across $\gamma_0$.

Finally, we note that the regularity of the map $r=\mathcal{X}(t,r')$ implies that if $f\in\hat{C}^{0,1}_l(\mathcal{N}_+)$, and $f(t,r)$ is replaced by $\hat{f}(t,r')=f(t,\mathcal{X}(t,r')),$ then $\hat{f}$ is in a space equivalent to $\hat{C}^{0,1}_l(\mathcal{N}_+)$, obtained by substituting for the sets $\mathcal{N}_p$ in $(t,r)$-coordinates, their images $\mathcal{N}_p'$ in $(t,r')$-coordinates, that is,
$$
\mathcal{N}_p' = \left\{(t,r')\in \R^2_+ \,|\, (t,\mathcal{X}(t,r')) \in \mathcal{N}_p \right\}.
$$
We denote this space by $\hat{C}^{0,1}_l(\mathcal{N}_+')$. The space $\hat{C}^{0,1}_l(\mathcal{N}_+')$ equipped with the norm 
\begin{eqnarray}\label{normprime}
\|\hat{f}\|_{\hat{C}^{0,1}_l(\mathcal{N}_+')}\equiv\|f\|_{\hat{C}^{0,1}_l(\mathcal{N}_+)},
\end{eqnarray}
is again a Banach space. Note, by Lemma \ref{ODE_char_Lemma_exist}, we have $\mathcal{X} \in \hat{C}^{0,1}_l (\mathcal{N}_+') \cap C^{1,1}(\mathcal{N}_+')$ and $\mathcal{Y} \in \hat{C}^{0,1}_l (\mathcal{N}_+) \cap C^{1,1}(\mathcal{N}_+)$. For simplicity, we subsequently often drop the dependence of the domain, $\mathcal{N}_+$ or $\mathcal{N}_+'$, in the norms considered and write instead          
\beq \nonumber
\|\cdot\|_{C^0} = \|\cdot\|_{C^0({\mathcal{N}}_+')} ,  \ \ \ \ 
\| \cdot \|_{\hat{C}^l} = \| \cdot \|_{\hat{C}^l(\mathcal{N}_+')}, \ \ \text{and} \ \  \|\cdot\|_{\hat{C}^{0,1}_l} =  \|\cdot\|_{\hat{C}^{0,1}_l(\mathcal{N}_+')}.
\eeq 

We subsequently assume, without loss of generality, that the set $\mathcal{N}_+'$ is of the form      
\begin{eqnarray}\label{TxR-space}
\mathcal{N}_+' =[0,T)\times(r_0-R,r_0+R),
\end{eqnarray}
for some $R>0$ with $R\leq T$. One can indeed assume \eqref{TxR-space} without loss of generality, since one can always restrict the domain of definition of the $U^k$ to arrange for \eqref{TxR-space}. Note that $T$ is also the upper temporal boundary of $\mathcal{N}_+$. We now state our main lemma for the regularity of $U^k$ and the convergence of our iterative scheme.

\begin{Lemma}\label{RegularityLemma}
Consider functions of the forms
\begin{eqnarray}
\psi(t,r') &=& \int_{0}^{t}H\big(x_i(s)-\mathcal{X}(s,r')\big)f(s,\mathcal{X}(s,r'))ds,\label{RegularityLemma_eqn1}  \\
\phi(t,r') &=& \int_{0}^{t}f(s,\mathcal{X}(s,r'))ds,\label{RegularityLemma_eqn2}
\end{eqnarray}
where $H(x)$ is the Heaviside function, and we assume that
$$
f\in \hat{C}^{0,1}_l(\mathcal{N}_+) 
$$
for some $l\geq 2$, then 
\beq \nonumber
\phi, \,  \psi \ \in \ \hat{C}^{0,1}_l(\mathcal{N}_+').
\eeq
Furthermore, $\partial_t\phi$ and $\partial_t\psi$ are continuous across $\gamma_0$, but $\partial_{r'} \phi$ and $\partial_{r'} \psi$ are in general discontinuous. Denoting with $[\cdot]_0$ the jump across $\gamma_0$, we have
\begin{align} 
 [\partial_{r'} \phi]_0(t) &=& \int_0^t \big[\partial_{r'}f\circ\mathcal{X}\big]_0(s) ds,  \label{phi_derivative-jump} \hspace{3cm} \\
[\partial_{r'} \psi]_0(t) &=& 2 f(0,r_0) \left(\partial_{r'} \tau\right)_i - \int_0^t \big[\partial_{r'}f\circ\mathcal{X}\big]_0(s) ds , \label{psi_derivative-jump}
\end{align} 
where $\left(\partial_{r'} \tau\right)_i$ denotes the left-limit of $\partial_{r'} \tau(r')$ as $r'$ approaches $r_0$ for $i=1$ in \eqref{RegularityLemma_eqn1} and the right-limit for $i=2$. Assuming further $\|f\circ\mathcal{X}\|_{\hat{C}^l(\mathcal{N}_+)} <~\infty$, for $\left\|\cdot\right\|_{\hat{C}^l}$ defined by \eqref{norm2_addition} and $(f\circ\mathcal{X})(t,r')\equiv  f(t,\mathcal{X}(t,r'))$, then, for $1\leq k \leq l$, we obtain the estimates           
\begin{eqnarray}
\left\|\psi\right\|_{\hat{C}^k({\mathcal{N}}'_+)}
&\leq & \alpha\, \|f\circ\mathcal{X}\|_{\hat{C}^{k-1}({\mathcal{N}}_+')} + \beta\, T\, \|f\circ\mathcal{X}\|_{\hat{C}^k({\mathcal{N}}_+')} , \label{RegularityLemma_eqn3} \\
\left\|\phi\right\|_{\hat{C}^k({\mathcal{N}}'_+)}
&\leq &\alpha\, \|f\circ\mathcal{X}\|_{\hat{C}^{k-1}({\mathcal{N}}_+')} + \beta\, T\, \|f\circ\mathcal{X}\|_{\hat{C}^k({\mathcal{N}}_+')} , \label{RegularityLemma_eqn3b}
\end{eqnarray}
where $\alpha,\,\beta$ are constants depending on $\mathcal{X}$ and $\tau$,  and $T$ is determined by $\mathcal{N}_+'$ through \eqref{TxR-space}. 
\end{Lemma}

\Proof  
It suffices here to prove $\psi\in \hat{C}^{0,1}_l(\mathcal{N}_+')$ together with the bound \eqref{RegularityLemma_eqn3} and \eqref{psi_derivative-jump} only, as the case for $\phi\in \hat{C}^{0,1}_l(\mathcal{N}_+')$ with \eqref{RegularityLemma_eqn3b} is a straightforward consequence and \eqref{phi_derivative-jump} follows analogously to \eqref{psi_derivative-jump}. Without loss of generality, we assume $i=1$ in \eqref{RegularityLemma_eqn1}.

To begin, we derive formulas for $\psi$ in the different regions, formulas from which the respective regularity easily follows. Let us consider the case $r'>r_0$ first, for which the Heaviside function is equal to $-1$, since $x_1(s)-\mathcal{X}(s,r')<0$ for all $s\geq 0$. Thus, we obtain for $(t,r') \in \mathcal{N}_R'$ that
\beq \label{RegularityLemma_eqn4a}
\psi\Big|_{\mathcal{N}_R'} (t,r') = -\int_{0}^{t}f(s,\mathcal{X}(s,r'))ds,
\eeq
while for $(t,r') \in \mathcal{N}_M^{\prime R}$ we get
\beq \label{RegularityLemma_eqn4b}
\psi\Big|_{\mathcal{N}_M^{\prime R}} (t,r') = -\int_{0}^{\tau(r')}f(s,\mathcal{X}(s,r'))ds - \int_{\tau(r')}^{t}f(s,\mathcal{X}(s,r'))ds.
\eeq
In the case $r'<r_0$, since $x_1(s)-\mathcal{X}(s,r')>0$ whenever $s<\tau(r')$, we conclude for $(t,r') \in \mathcal{N}_L'$ that
\beq \label{RegularityLemma_eqn4c}
\psi\Big|_{\mathcal{N}_L'} (t,r') = \int_{0}^{t} f(s,\mathcal{X}(s,r'))ds.
\eeq
It is only in $\mathcal{N}_M^{\prime L}$ that we integrate across the discontinuity of the Heaviside function, and using that the Heaviside function is identical $1$ whenever $s<\tau(r')$ and $-1$ for $s>\tau(r')$, we finally obtain
\beq \label{RegularityLemma_eqn4d}
\psi\Big|_{\mathcal{N}_M^{\prime L}} (t,r') = \int_{0}^{\tau(r')}f(s,\mathcal{X}(s,r'))ds-\int_{\tau(r')}^{t}f(s,\mathcal{X}(s,r'))ds.
\eeq

Now, from \eqref{RegularityLemma_eqn4a} and \eqref{RegularityLemma_eqn4c}, since the region of integration lies in the regions where $f(\cdot,\mathcal{X}(\cdot,\cdot))$ is $C^l$ regular, that is, in $\mathcal{N}_L'$ and $\mathcal{N}_R'$, it follows immediately that $\psi \in C^l(\bar{\mathcal{N}}_R')$ and $\psi \in C^l(\bar{\mathcal{N}}_L')$. To prove the $C^l$ regularity in $\bar{\mathcal{N}}_M^{\prime L}$ and $\bar{\mathcal{N}}_M^{\prime R}$, first observe that in \eqref{RegularityLemma_eqn4b} and \eqref{RegularityLemma_eqn4d} we have $r'\neq r_0$ which implies that $\tau$ is $C^l$ regular, moreover, left- and right-limits of $\tau$ and of all its derivatives exist at $r'=r_0$. In addition, the regions of integration of both integrals in \eqref{RegularityLemma_eqn4b} and of both integrals in \eqref{RegularityLemma_eqn4d} lie in the regions where $f(\cdot,\mathcal{X}(\cdot,\cdot))$ is $C^l$ regular. Using in addition that derivatives of $f$ extends continuously to the shocks and $\gamma_0$ and that the restriction of $f$ to any shock curve or $\gamma_0$ is in $C^l$, it follows that $\psi \in C^l(\bar{\mathcal{N}}_M^{\prime L})$ and $\psi \in C^l(\bar{\mathcal{N}}_M^{\prime R})$.

We next prove $\psi \in C^{0,1}(\mathcal{N}'_+)$, for which we first prove continuity. Taking the limit $t\nearrow \tau(r')$ in \eqref{RegularityLemma_eqn4c} and $t\searrow \tau(r')$ in \eqref{RegularityLemma_eqn4d}, we find that $\psi$ matches continuously across $\gamma_1$. Computing the respective limits in \eqref{RegularityLemma_eqn4a} - \eqref{RegularityLemma_eqn4b} gives continuity across $\gamma_2$. Taking the limit $r'\nearrow r_0$ in \eqref{RegularityLemma_eqn4d} and $r'\searrow r_0$ in \eqref{RegularityLemma_eqn4b}, and using that $\tau(r_0)=0$, we find that $\psi$ matches continuously across the characteristic $\gamma_0$, which then proves $\psi \in C^{0}(\mathcal{N}'_+)$. Now, since in addition first order derivatives exist in $\mathcal{N}_L'$, $\mathcal{N}^{\prime L}_M$, $\mathcal{N}^{\prime R}_M$ and $\mathcal{N}_R'$, and extend continuously to the boundaries, it follows that $\psi$ is Lipschitz continuous on $\mathcal{N}_+'$, (c.f. the footnote in the proof of Lemma \ref{ODE_char_Lemma_exist}).

We proceed by proving $[\partial_t \psi]_0=0$ and \eqref{psi_derivative-jump}. Taking the $t$- and $r'$-derivatives of \eqref{RegularityLemma_eqn4b} and \eqref{RegularityLemma_eqn4d}, we get
\begin{eqnarray}
\partial_{t} \psi(t,r') &=& - f\big(t,\mathcal{X}(t,r')\big) , \cr
\partial_{r'} \psi(t,r') &=& \left(\partial_{r'}{\tau(r')} \pm \partial_{r'}{\tau(r')}\right) f\big(\tau(r'),\mathcal{X}(\tau(r'),r')\big),  \label{RegularityLemma_eqn5} \cr
&\pm & \int_{0}^{\tau(r')}\partial_{r'}f(s,\mathcal{X}(s,r'))ds-\int_{\tau(r')}^{t}\partial_{r'}f(s,\mathcal{X}(s,r'))ds, 
\end{eqnarray}
where, in the expression for $\partial_{r'} \psi$, the $+$ sign corresponds to differentiation of \eqref{RegularityLemma_eqn4d} and the $-$ sign comes from \eqref{RegularityLemma_eqn4b}.  Taking the left/right-limit of \eqref{RegularityLemma_eqn5} to $r_0$, it is immediate that $\partial_t \psi$ is continuous across $\gamma_0$. To prove \eqref{psi_derivative-jump}, take the right-limit of $\partial_{r'} \psi$ in \eqref{RegularityLemma_eqn5}, that is, compute $r' \searrow r_0$ for the expression containing the $-$ sign, and use $\tau(r_0)=0$, which leads to
\beq\nonumber
\left(\partial_{r'} \psi\right)_R (t) = - \int_0^t (\partial_{r'}f)_R(s,\mathcal{X}(s,r_0)) ds, 
\eeq
where $(\cdot)_{L/R}$ denotes the left/right-limit at $r_0$ respectively. Similarly, taking the left-limit of $\partial_{r'} \psi$ in \eqref{RegularityLemma_eqn5} gives
\beq\nonumber
\left(\partial_{r'} \psi \right)_L (t) = 2 f(0,r_0) \left(\frac{\partial \tau}{\partial r'}\right)_L (r_0) - \int_0^t (\partial_{r'}f)_L(s,\mathcal{X}(s,r_0)) ds.
\eeq
Now, computing $[\partial_{r'} \psi]_0 = \left(\partial_{r'} \psi \right)_L  - \left(\partial_{r'} \psi\right)_R $ immediately gives \eqref{psi_derivative-jump}. 

We now prove the estimate \eqref{RegularityLemma_eqn3}. Consider first the regions $\mathcal{N}_L'$ and $\mathcal{N}_R'$, where it is immediate from \eqref{RegularityLemma_eqn4a} and \eqref{RegularityLemma_eqn4c} that we can bound all terms in $\left\|\psi\right\|_{{C}^k(\bar{\mathcal{N}}'_p)}$ containing a $t$-derivative by $\|f\circ\mathcal{X}\|_{C^{k-1}(\bar{\mathcal{N}}'_p)}$, while $T\, \|f\circ\mathcal{X}\|_{C^{k}(\bar{\mathcal{N}}'_p)}$ bounds the term $\partial^l_{r'} \psi$. This proves, for $p=L,R$, the estimate 
\beq \label{RegularityLemma_eqn6}
\left\|\psi\right\|_{{C}^k(\bar{\mathcal{N}}'_p)}
\leq  \alpha\, \|f\circ \mathcal{X}\|_{\hat{C}^{k-1}({\mathcal{N}}'_+)} + \beta\, T\, \|f\circ \mathcal{X}\|_{\hat{C}^k({\mathcal{N}}'_+)} ,
\eeq
for $1\leq k \leq l$ and some combinatorial constants $\alpha$ and $\beta$ only depending on $k$. It remains to derive suitable estimates on $\left\|\psi\right\|_{{C}^k}$ in the intermediate regions, $\mathcal{N}_M^{\prime L}$ and $\mathcal{N}_M^{\prime R}$. For this, observe that the expression for the $t$-derivative in \eqref{RegularityLemma_eqn5} implies that all derivatives of $\partial_{t} \psi$, up to order $k-1$, can be bounded by $\alpha \|f\circ\mathcal{X}\|_{C^{k-1}(\mathcal{N}_M^{\prime p})}$, $p=L,R$. Furthermore, this implies that all derivatives of $\partial_{r'} \psi$ of order up to $k$ are bounded by $\alpha \|f\circ\mathcal{X}\|_{C^{k-1}(\mathcal{N}_M^{\prime p})}$, as long as they include at least one time derivative, since partial derivatives commute. It remains to estimate $\partial^k_{r'} \psi$, for $1\leq k \leq l$. We begin with $k=1$. By the above equation we have
\begin{eqnarray}\nonumber
|\partial_{r'} \psi(t,r')| &\leq & 2\left| \partial_{r'}{\tau(r')}\right| \, \left| f\big(\tau(r'),\mathcal{X}(\tau(r'),r')\big) \right|  \cr 
&+ &  \int_{0}^{\tau(r')} | \partial_{r'}f(s,\mathcal{X}(s,r')) | ds + \int_{\tau(r')}^{t}|\partial_{r'}f(s,\mathcal{X}(s,r'))|ds,
\end{eqnarray}
and thus, for $p=L,R$, 
\beq\nonumber
\|\partial_{r'} \psi \|_{C^0(\mathcal{N}_M^{\prime p})} 
\leq  \alpha \| f\circ\mathcal{X}\|_{C^0(\mathcal{N}_M^{\prime p})} + \beta \, T\, \| f\circ\mathcal{X} \|_{\hat{C}^1(\mathcal{N}_+)} ,
\eeq
where $\alpha$ depends only on the $C^0$-norm of $\partial_{r'}{\tau(r')}$ for $r'\neq r_0$. Proceeding in a similar way, we find that $\partial^k_{r'} \psi$ for $k \leq l$ is bounded by
\beq\nonumber
 \| \partial^k_{r'} \psi \|_{C^0(\mathcal{N}_M^{\prime p})} 
 \leq    \alpha\,  \| f\circ\mathcal{X} \|_{\hat{C}^{k-1}(\mathcal{N}_+')} +  T\, \beta\, \| f\circ\mathcal{X} \|_{\hat{C}^{k}(\mathcal{N}_+')},
\eeq
where $\alpha$ and $\beta$ depends only on $\tau$ and $\mathcal{X}$.  Combining the above estimates with \eqref{RegularityLemma_eqn6} finally proves \eqref{RegularityLemma_eqn3}. The estimate \eqref{RegularityLemma_eqn3b} follows analogously.
\QED

We now prove the existence of a solution to the non-local Cauchy problem \eqref{Cauchyproblem_PDE} - \eqref{Cauchyproblem_initial}, by proving the convergence of the iteration scheme \eqref{IC_iteration_1}. In more detail, consider the iteration scheme \eqref{IC_iteration_1} with its  iterates $U^k(t,r)$, defined in \eqref{IC_iteration_solns}, which solve \eqref{IC_iteration_1} for $C^l$ initial data $U_0$ defined in a neighborhood of $r=r_0$, for $l\geq2$. We prove there exists a neighborhood $\mathcal{N}$ of $(0,r_0)$ in $\R^2$, such that the $U^k$ lie in $\hat{C}^{0,1}_l(\mathcal{N}_+)$, for $\mathcal{N}_+ = \mathcal{N} \cap \overline{\R^2_+}$. We then prove there exists a smaller neighborhood $\tilde{\mathcal{N}} \subset \mathcal{N}$ of $(0,r_0)$, such that a subsequence of $\left( U^k \right)_{k\in\mathbb{N}}$ converges to a function $U$ in ${C}^{0,1}({\tilde{\mathcal{N}}}_+)$, for $\tilde{\mathcal{N}}_+ = \tilde{\mathcal{N}} \cap \overline{\R^2_+}$, which is $C^{1,1}$ away from $\gamma_0$, $\gamma_1$ and $\gamma_2$, and solves the non-local Cauchy problem, \eqref{Cauchyproblem_PDE} - \eqref{Cauchyproblem_initial}, with initial data $U_0$. We formulate and prove the theorem in the special case $l=2$, the case relevant for the metric smoothing. 

\begin{Thm} \label{Thmiteration}
For any $C^2$ function $U_0$, defined in an interval containing $r=r_0$, there exists a function $U$ in ${C}^{0,1}({\mathcal{N}}_+)$ which solves the Cauchy Problem \eqref{Cauchyproblem_PDE} - \eqref{Cauchyproblem_initial} almost everywhere in $\mathcal{N}_+$, for initial data $U_0$, where $\mathcal{N}_+ = \mathcal{N} \cap \overline{\R^2_+}$ for some neighborhood $\mathcal{N}$ of $(0,r_0)$ in $\R^2$. Moreover, $U|_{\bar{\mathcal{N}}_p} \in {C}^{1,1}(\bar{{\mathcal{N}}}_p)$, i.e., derivatives of $U$ are in ${C}^{0,1}(\bar{{\mathcal{N}}}_p)$, for $\mathcal{N}_p$ being any of the regions $\mathcal{N}_L$, $\mathcal{N}_M^L$, $\mathcal{N}_M^R$ and $\mathcal{N}_R$.     Furthermore, $\partial_t U$ is continuous across $\gamma_0$ and the jump in $\partial_{r} U$ across $\gamma_0$ satisfies          
\beq\label{jump_r-deriv_U}
\big\| [\partial_{r} U]_0 \big\|_{C^0([0,T])} \leq \mathcal{C} \sum_{i=1,2}\left| \left( M_i(0,r_0) + \sum_{j=1,2} \mathcal{A}_{ji}(0,r_0) \right) U_0(r_0) \right| ,
\eeq   
for some constant $\mathcal{C}>0$ which is finite for $T>0$ sufficiently small, where $T$ denotes the upper temporal boundary of $\mathcal{N}_+$, c.f. \eqref{TxR-space}.  
\end{Thm}

\Proof
The proof is comprised of the following six steps. In Step 1, we show that $U^k \in \hat{C}^{0,1}_l(\mathcal{N}_+')$ for all $k \in \mathbb{N}$, as long that the first iterate is of this regularity. In Step 2, we derive the estimate 
\beq \label{Thm_PDE_eqn-step2}
\big\| U^k \big\|_{\hat{C}^{0,1}_{1}({\mathcal{N}}_+')} 
\leq  \alpha\, \big\| U_0\big\|_{C^{1}} + T\,\beta\, \left( \big\| U^{k-1}\big\|_{\hat{C}^{0,1}_{1}({\mathcal{N}}_+')} + \big\| U^{k-2} \big\|_{\hat{C}^{0,1}_{1}({\mathcal{N}}_+')} \right) ,
\eeq
for constants $\alpha$ and $\beta$. Note that the factor $T$ on the right hand side can only be achieved because the additional derivative term $\dot{U}^k$ is weighted by $|X_i(t,r)|$, (and $|X_i(t,r)|$ is bounded by $\beta T$), a peculiar and important structure of the non-local PDE \eqref{Cauchyproblem_PDE}. In Step 3, we built on estimate \eqref{Thm_PDE_eqn-step2} to derive an analogous estimate with respect to $\|\cdot\|_{\hat{C}^{0,1}_{2}({\mathcal{N}}_+')}$ in which the previous iterates are weighted by a factor $T$. From this estimate, in Step 4, we derive a $k$-independent upper bound on $\|U^k\|_{\hat{C}^{0,1}_{2}({\mathcal{N}}_+')}$ assuming that $T$ is sufficiently small. To obtain this bound, it is crucial to get the factor $T$ on the right hand side in \eqref{Thm_PDE_eqn-step2}. As explained in Step 5, suitable convergence of a subsequence in $C^{1,1}$ now follows by the Arzel\`a-Ascoli Theorem. In Step 6, we derive \eqref{jump_r-deriv_U} which then concludes the proof. \vspace{.2cm}

\noindent \textbf{Step 1:} We first prove the regularity of the iterates, $U^k$. For simplicity, we work in $(t,r')$-coordinates and assume $\mathcal{N}_+'$ is of the form \eqref{TxR-space} and  we often write $U^k$ instead of $U^k(\cdot,\mathcal{X}(\cdot,\cdot))$. To begin with, recall the expression for the $k$-th iterate,  in \eqref{IC_iteration_soln},
$$
U^k(t,\mathcal{X}(t,r')) = U_0(r') +  \int_0^t F^{k-1}\left(s,\mathcal{X}(s,r')\right) ds,
$$
where $F^{k-1}=F(U^{k-1})$, and by \eqref{3},
\begin{eqnarray}\nonumber
F(U)=\mathcal{M}U+\sum_{i=1,2}\left(\mathcal{A}_i\cdot (U\circ\gamma_i)+\mathcal{B}_i\left|X_i\right|\dot{U}_i \right),
\end{eqnarray}
where we define $\dot{U}_i = \frac{d}{dt}\left(U\circ\gamma_i\right)$. Recall further that $\mathcal{A}_i$ and $\mathcal{M}$ contain Heaviside functions with coefficients in $C^{0,1}_l$, while $\mathcal{B}_i$ is in $C^{0,1}_l$, c.f. \eqref{4} - \eqref{6}. Assume for the moment that $U^{k-1} \in \hat{C}^{0,1}_l(\mathcal{N}_+')$ for some $k\in \N$. Then, since $\mathcal{X}$ is in $\hat{C}^{0,1}_l(\mathcal{N}_+')$ for some neighborhood $\mathcal{N}_+'$, and since
$$
\int_0^t \Big( F^{k-1} - \sum_{i=1,2}\mathcal{B}_i\left|X_i\right|\dot{U}^{k-1}_i \Big) ds
$$ 
is a sum of terms of the form \eqref{RegularityLemma_eqn1} - \eqref{RegularityLemma_eqn2}, Lemma \ref{RegularityLemma} implies that the previous integral is in $\hat{C}^{0,1}_l(\mathcal{N}_+')$. In addition, the integral over $\mathcal{B}_i\left|X_i\right|\dot{U}^{k-1}_i$ compensates the derivative $\dot{U}^{k-1}_i\equiv \frac{d}{dt}U^{k-1}\circ \gamma_i$, so that applying Lemma \ref{RegularityLemma} for derivatives higher than first order in~$t$ (i.e., higher derivatives in $t$ and $r'$ containing at least one derivative in $t$) gives
$$
\int_0^t  \mathcal{B}_i\left|X_i\right|\dot{U}^{k-1}_i \: ds \ \in \ \hat{C}^{0,1}_l(\mathcal{N}_+') .
$$ 
Combining these two regularity results, we conclude that $U^{k-1}\in \hat{C}^{0,1}_l(\mathcal{N}_+')$ implies $U^k\in \hat{C}^{0,1}_l(\mathcal{N}_+')$.  Therefore, choosing $U^0=0$ as the first iterate, we conclude that $$U^k \in \hat{C}^{0,1}_l(\mathcal{N}_+') \ \ \ \ \forall \ k \in \mathbb{N}.$$ 

\noindent \textbf{Step 2:} We proceed by deriving a $k$-independent upper bound on the $\hat{C}^{0,1}_l$-norm of $U^k$ for all $k\geq0$. Once this is achieved, the Arzela Ascoli Theorem yields convergence of our iterative scheme. For this, let $\alpha>0$ and $\beta>0$ denote universal constants (changing from estimate to estimate) which only depend on $\mathcal{X}$, $\tau$,  $\mathcal{M}$, $\mathcal{A}_i$, $\mathcal{B}_i$ and $x_i$, for $i=1,2$, and the volume of $\mathcal{N}_+$. To begin, we derive estimates of the supremums-norm of $U^k$ over ${\mathcal{N}}'_+$.  From the expression \eqref{IC_iteration_soln} for the iterate $U^k$, using that the region of integration is given by $0 \leq t \leq T$, (where $T$ is the upper time boundary of $\mathcal{N}'_+$, c.f. \eqref{TxR-space}), we conclude with the estimate\footnote{$\|\cdot\|_{C^0}$ applied to a Heaviside function should be understood in an almost everywhere sense, which corresponds to the $L^\infty$ norm, c.f. \cite{Evans}.}
\begin{eqnarray}\label{iteration_estimate_sup}
\| U^k \|_{C^0({\mathcal{N}}_+')} 
&\leq & \| U_0\|_{C^0} + T\, \|F^{k-1}\|_{C^0} \cr
&\leq & \| U_0\|_{C^0} + T\, \beta\, \left( \| U^{k-1} \|_{C^{0}} + \sum_{i=1,2} \| \dot{U}_i^{k-1} \|_{C^{0}} \right)\cr
&\leq & \| U_0\|_{C^0({\mathcal{N}}_+')} + T\, \beta\, \| U^{k-1} \|_{\hat{C}^{1}({\mathcal{N}}_+')}.
\end{eqnarray}
(Recall that $\|\cdot\|_{\hat{C}^{l}({\mathcal{N}}_+')}$ is defined as the $C^l$-norm over the regions in between the shock curves, $\gamma_0$ and the line $t=0$, including the values of derivatives on these boundary curves, c.f. \eqref{norm2_addition}.) To derive higher derivative estimates, separate the dependence of $F^{k-1}$ on $U^{k-1}$ and on the tangential derivatives $\dot{U}^{k-1}_i $, ($i=1,2$), by writing
\beq\label{Thm_PDE_eqn1a}
F^{k-1}=f_0(U^{k-1})  + f^i_1(U^{k-1})H(X_i)  + f^i_2\, \dot{U}^{k-1}_i,
\eeq
where from \eqref{3} - \eqref{5}, omitting the dependence on $\mathcal{X}$, we have
\begin{eqnarray}\nonumber 
f_0(U^{k-1}) &=& M\cdot U^{k-1} + \sum_{i=1,2} A_i\cdot  U^{k-1} \circ \gamma_i   \cr
f^i_1(U^{k-1})\, H(X_i) &=&  \sum_{i=1,2} \left( M_i \cdot U^{k-1} + \sum_{j=1,2}\mathcal{A}_{ji}\cdot  U^{k-1} \circ \gamma_j \right) H(X_i)   \cr
f^i_2\ \dot{U}^{k-1}_i &=& \sum_{i=1,2}\mathcal{B}_i\left|X_i\right|\,  \dot{U}^{k-1}_i  \, .
\end{eqnarray}
Observe that $f_0(U^{k-1}),\, f_1^i(U^{k-1})$ and $f^i_2$ are in $\hat{C}^{0,1}_l(\mathcal{N}_+')$ and linear in $U^{k-1}$. From the explicit expression for $U^{k}$, \eqref{IC_iteration_soln}, we now obtain that 
\begin{eqnarray}\label{Thm_PDE_eqn1}
\| U^k \|_{\hat{C}^{1}} 
&\leq &  \| U_0\|_{{C}^{1}} +  \left\| \int_0^t F^{k-1}\left(s,\mathcal{X}(s,r')\right) ds \right\|_{\hat{C}^{1}} \cr
&\leq &  \| U_0\|_{{C}^{1}} +  \left\| \int_0^t f_0(U^{k-1}) ds \right\|_{\hat{C}^{1}} \cr 
&+&   \left\| \int_0^t f^i_1(U^{k-1})H(X_i) ds \right\|_{\hat{C}^{1}} +  \left\| \int_0^t f^i_2\, \dot{U}^{k-1}_i ds \right\|_{\hat{C}^{1}}  .  
\end{eqnarray}
The second term is of the form (\ref{RegularityLemma_eqn2}) and the third term is of the form (\ref{RegularityLemma_eqn1}). Thus, \eqref{RegularityLemma_eqn3b} of Lemma \ref{RegularityLemma} implies for the second term that
\begin{eqnarray}\label{Thm_PDE_eqn1b}
\left\| \int_0^t f_0(U^{k-1}) ds \right\|_{\hat{C}^{1}}  
&\leq & \alpha \, \left\| f_0(U^{k-1}) \right\|_{{C}^{0}}  +  T\, \beta\, \left\| f_0(U^{k-1}) \right\|_{\hat{C}^{1}} \cr
&\leq & \alpha\, \left\| U^{k-1} \right\|_{C^{0}}  +  T\, \beta\,  \left\| U^{k-1}\right\|_{\hat{C}^{1}},
\end{eqnarray}
where we absorb $\left\| f_0\right\|_{C^{0}}$ and $\left\| f_0\right\|_{\hat{C}^{1}} $ into the universal constants $\alpha$ and $\beta$ in the last line. Similarly, \eqref{RegularityLemma_eqn3} of Lemma \ref{RegularityLemma} implies for the third term in \eqref{Thm_PDE_eqn1} the estimate
\begin{eqnarray}\label{Thm_PDE_eqn1c}
\left\| \int_0^t f^i_1(U^{k-1})H(X_i) ds \right\|_{\hat{C}^{1}} 
&\leq & \alpha \, \left\| f^i_1(U^{k-1}) \right\|_{C^{0}}  +  T\, \beta\, \left\| f^i_1(U^{k-1}) \right\|_{\hat{C}^{1}} \cr
&\leq & \alpha\, \left\| U^{k-1} \right\|_{C^{0}}  +  T\, \beta\,  \left\| U^{k-1}\right\|_{\hat{C}^{1}}.
\end{eqnarray}
Using that $\dot{U}^{k-1}_i$ is independent of $r'$ and that the integrand is independent of $t$, we estimate the fourth term in \eqref{Thm_PDE_eqn1} as
\begin{eqnarray}\label{Thm_PDE_eqn2}
\left\| \int_0^t f^i_2\, \dot{U}^{k-1}_i ds \right\|_{\hat{C}^{1}}  
&\leq &  \left( T \left\| f^i_2\right\|_{C^{0}}    + T \left\| \partial_{r'}f^i_2\right\|_{C^{0}} + \left\| f^i_2\right\|_{C^{0}} \right) \left\|\dot{U}^{k-1}_i \right\|_{C^{0}} \cr
 &\leq &  \beta T \left\|{U}^{k-1} \right\|_{\hat{C}^{1}},
\end{eqnarray}
where we used for the last inequality that
\beq \label{Thm_PDE_eqn2b}
\left\| f^i_2\right\|_{C^{0}}  \leq \left\| \mathcal{B}_i \right\|_{C^0} \left\| X_i \right\|_{C^0}  \leq \beta\, T, 
\eeq
since $X_i(t,r) =x_i(t) - r$ is bounded by $\beta T$, as is shown in the following estimate
\begin{eqnarray}\label{bound_X}
|X_i(t,r)| &=& |x_i(t)-r| \cr 
&\leq & |x_i(t)-x_i(0)| + |r_0-r| \cr
&\leq & \beta |t| + R \cr
&\leq & \beta\, T, 
\end{eqnarray} 
for which we used that $x_i(0)=r_0$, $R\leq T$ (c.f. \eqref{TxR-space}) and $\beta$ is taken to be some constant bounding the Lipschitz constant of $x_i$. Substituting the estimates \eqref{Thm_PDE_eqn1b} - \eqref{Thm_PDE_eqn2} into \eqref{Thm_PDE_eqn1} leads to
\begin{eqnarray}\nonumber
\| U^k \|_{\hat{C}^{1}} 
&\leq & \| U_0\|_{C^{1}} +\alpha\, \left\| U^{k-1} \right\|_{C^{0}} + T\,\beta\, \left\| U^{k-1}\right\|_{\hat{C}^{1}} ,  
\end{eqnarray}
and substituting \eqref{iteration_estimate_sup} for $\left\| U^{k-1} \right\|_{C^{0}}$ in the above inequality finally gives
\beq\label{Thm_PDE_eqn3}
\| U^k \|_{\hat{C}^{1}({\mathcal{N}}_+')} 
\leq  \alpha\, \| U_0\|_{C^{1}} + T\,\beta\, \left( \left\| U^{k-1}\right\|_{\hat{C}^{1}({\mathcal{N}}_+')} + \left\| U^{k-2} \right\|_{\hat{C}^{1}({\mathcal{N}}_+')} \right) .
\eeq

Estimate \eqref{Thm_PDE_eqn3} implies the sought after estimate \eqref{Thm_PDE_eqn-step2} of Step 2. Namely, the $C^{0,1}$-norm of a function $f$, defined on a bounded region with a boundary that is piece-wise $C^1$, can always be bounded by the $C^0$-norm of the weak derivative of $f$, applied in an almost everywhere sense, c.f. \cite{Evans}. Thus $\|\cdot\|_{C^{0,1}(\mathcal{N}_+')}$ and $\|\cdot\|_{\hat{C}^1(\mathcal{N}_+')}$ are equivalent norms on $\hat{C}^{0,1}_l(\mathcal{N}_+')$, for $l=1$. (Note also that estimate \eqref{Thm_PDE_eqn-step2} is sufficient to conclude convergence of a subsequence in $C^{0,1}(\mathcal{N}_+')$ as a result of the Arcela Ascoli Theorem.)  Nevertheless, to get the crucial $C^{1,1}$ regularity away from the boundary curves $t=0$ and $\gamma_i$, for $i=0,1,2$, we extend the above estimates to $\|\cdot\|_{\hat{C}^{0,1}_l}$, for $l=2$, in the next step. \vspace{.2cm}

\noindent \textbf{Step 3:}  From the explicit expression for $U^{k}$, \eqref{IC_iteration_soln}, with $F^{k-1}$ decomposed as in \eqref{Thm_PDE_eqn1a}, we obtain that 
\begin{eqnarray}\label{Thm_PDE_eqn4}
\| U^k \|_{\hat{C}^{2}} 
&\leq &   \| U_0\|_{C^{2}}  +  \left\| \int_0^t f^i_2\, \dot{U}^{k-1}_i ds \right\|_{\hat{C}^{2}} \cr  
&+&  \left\| \int_0^t f_0(U^{k-1}) ds \right\|_{\hat{C}^{2}} +  \left\| \int_0^t f^i_1(U^{k-1})H(X_i) ds \right\|_{\hat{C}^{2}}.  
\end{eqnarray}
The third and fourth term are each a sum of terms of the form (\ref{RegularityLemma_eqn1})-(\ref{RegularityLemma_eqn2}), so that \eqref{RegularityLemma_eqn3} - \eqref{RegularityLemma_eqn3b} of Lemma \ref{RegularityLemma} imply that   
\begin{eqnarray}\label{Thm_PDE_eqn4a}
&& \left\| \int_0^t f_0(U^{k-1}) ds \right\|_{\hat{C}^{2}} +  \left\| \int_0^t f^i_1(U^{k-1})H(X_i) ds \right\|_{\hat{C}^{2}} \cr  
&\leq & \alpha\, \left\| U^{k-1} \right\|_{\hat{C}^{1}}  +  T\, \beta\,  \left\| U^{k-1}\right\|_{\hat{C}^{2}},
\end{eqnarray}
where, similarly to the $\hat{C}^1$-estimates, we absorb again the $\hat{C}^1$- and $\hat{C}^2$-norms of $f_0$ and $f^i_1$ into $\alpha$ and $\beta$. Using that $\dot{U}^{k-1}_i$ is independent of $r'$ and that the integrand is independent of $t$, we estimate the second term in \eqref{Thm_PDE_eqn4} as
\begin{eqnarray}\nonumber
\left\| \int_0^t f^i_2\, \dot{U}^{k-1}_i \right\|_{\hat{C}^{2}}  
&\leq & \left\| \int_0^t f^i_2\, \dot{U}^{k-1}_i  \right\|_{\hat{C}^{1}} + \left\|\int_0^t \partial^2_{r'} f^i_2\, \dot{U}^{k-1}_i \right\|_{\hat{C}^0} \cr 
&+& \left\|\partial_{r'} f^i_2\, \dot{U}^{k-1}_i \right\|_{\hat{C}^0} + \left\|\partial_{t} \left( f^i_2 \right)  \dot{U}^{k-1}_i\right\|_{\hat{C}^0} + \left\|f^i_2\, \ddot{U}^{k-1}_i \right\|_{\hat{C}^0}  .
\end{eqnarray}
The first term is bounded by \eqref{Thm_PDE_eqn2}, the second one by
\begin{eqnarray}\nonumber
\left\|\int_0^t \partial^2_{r'} f^i_2\, \dot{U}^{k-1}_i \right\|_{\hat{C}^0} 
&\leq & T\, \left\|f^i_2\right\|_{\hat{C}^2} \left\|\dot{U}^{k-1}_i \right\|_{C^0} \cr
&\leq & T\, \beta\, \left\|{U}^{k-1}\right\|_{\hat{C}^1},
\end{eqnarray}
while the third and fourth term are bounded by
\begin{eqnarray}\nonumber
\left\|\partial_{r'} f^i_2\, \dot{U}^{k-1}_i \right\|_{\hat{C}^0} + \left\|\partial_{t} \left( f^i_2 \right)  \dot{U}^{k-1}_i\right\|_{\hat{C}^0} 
&\leq & \left\| f^i_2 \right\|_{\hat{C}^1}  \left\|\dot{U}^{k-1}_i\right\|_{C^0}, \cr
&\leq & \alpha  \left\|{U}^{k-1}\right\|_{\hat{C}^1}
\end{eqnarray}
and the fifth term has the bound  
\begin{eqnarray}\nonumber
\left\|f^i_2\, \ddot{U}^{k-1}_i \right\|_{\hat{C}^0} 
&\leq & \left\| f^i_2 \right\|_{\hat{C}^0} \left\|\ddot{U}^{k-1}_i \right\|_{C^0}  \cr
&\leq & T\, \beta\,  \left\|U^{k-1} \right\|_{\hat{C}^2} ,
\end{eqnarray}
where we use \eqref{Thm_PDE_eqn2b} to obtain the last inequality. Combining the above estimates and bounding all $\hat{C}^1$-norms by the $\hat{C}^2$-norm whenever the term contains a factor $T$, we obtain the following bound on the second term in \eqref{Thm_PDE_eqn4},
\begin{eqnarray}\label{Thm_PDE_eqn4b}
\left\| \int_0^t f^i_2\, \dot{U}^{k-1}_i \right\|_{\hat{C}^{2}}  
&\leq & \alpha  \left\|{U}^{k-1}\right\|_{\hat{C}^1}  +  T\, \beta\,  \left\|U^{k-1} \right\|_{\hat{C}^2}.
\end{eqnarray}
Substituting \eqref{Thm_PDE_eqn4a} and \eqref{Thm_PDE_eqn4b} into \eqref{Thm_PDE_eqn4}, we obtain
\begin{eqnarray}\label{Thm_PDE_eqn5a}
\| U^k \|_{\hat{C}^{2}} 
\leq \| U_0\|_{C^{2}} + \alpha\, \left\|{U}^{k-1}\right\|_{\hat{C}^1}  +  T\, \beta\,  \left\|U^{k-1} \right\|_{\hat{C}^2} .  
\end{eqnarray}
Finally, using \eqref{Thm_PDE_eqn3} to substitute for $\left\|{U}^{k-1}\right\|_{\hat{C}^1} $ and bounding all $\hat{C}^1$-norms in the resulting expressions with $\hat{C}^2$-norms, leads to our final $\hat{C}^2$-estimate:
\beq \label{Thm_PDE_eqn5}
\| U^k \|_{\hat{C}^{2}} 
\leq  \alpha\, \| U_0\|_{C^{2}} +  T\,\beta\, \left( \left\| U^{k-1}\right\|_{\hat{C}^{2}} + \left\| U^{k-2} \right\|_{\hat{C}^{2}} + \left\| U^{k-3}\right\|_{\hat{C}^{2}} \right) .  
\eeq
To conclude with the corresponding estimate in the $C^{0,1}_l$-norm, for $l=2$, compute the norm $\|U^k\|_{\hat{C}^{0,1}_l(\mathcal{N}_+')}$, as introduced in \eqref{norm2}, as follows: Substitute on the right hand side of \eqref{norm2} for the $C^{0,1}$-norm the estimate \eqref{Thm_PDE_eqn3} and for the $\hat{C}^{2}$-norm the estimate \eqref{Thm_PDE_eqn5}, we obtain for $k\geq 3$ that
\beq\label{Thm_PDE_eqn6}
\| U^k \|_{\hat{C}^{0,1}_2} 
\leq  \alpha \| U_0\|_{\hat{C}^{2}} +  T\,\beta\, \left( \left\| U^{k-1}\right\|_{\hat{C}^{0,1}_{2}} + \left\| U^{k-2} \right\|_{\hat{C}^{0,1}_{2}}  + \left\| U^{k-3}\right\|_{\hat{C}^{0,1}_{2}}\right) .
\eeq

\noindent \textbf{Step 4:}  We now show that there exists a constant $\mathcal{C}$ independent of $k$ which bounds $\|U^k\|_{C^{0,1}_2} $ for all $k\geq 0$. To begin, we prove by induction that \eqref{Thm_PDE_eqn6} implies
\beq\label{Thm_PDE_eqn7}
\| U^k \|_{C^{0,1}_2}   \leq  \alpha \| U_0\|_{C^{0,1}_{2}}  \sum_{l=0}^{k-1} (3 T \beta)^l , \ \ \ \ \ \forall\, k\geq 1,   
\eeq
for $\alpha$ and $\beta$ from \eqref{Thm_PDE_eqn6}. For simplicity, introduce the notation
$$
y^k= \| U^k \|_{C^{0,1}_2}, \ \ \ \ \ a=\alpha \| U_0\|_{C^{0,1}_{2}} \ \ \  \ \text{and} \ \ x=T\beta.
$$ 
Before we pursue the induction step, we verify \eqref{Thm_PDE_eqn7} for $k=1,2,3$, explicitly. For this, using the formula for the iterates, \eqref{IC_iteration_soln}, and recalling that the first iterate is given by $U^0\equiv 0$, we conclude that $U^1(t,r')=U_0(r')$, which leads to 
$$
y^1 = \| U_0 \|_{C^{0,1}_2} \leq a ,
$$
where we assume, without loss of generality, that $\alpha\geq 1$. To prove \eqref{Thm_PDE_eqn7} for $k=2$, observe that $U^1=U_0$ together with \eqref{Thm_PDE_eqn5a} imply
\beq\nonumber
\|U^2\|_{\hat{C}^2} \leq \alpha \|U_0\|_{C^{0,1}_2} \, (1 + T\beta),
\eeq
and since the $\hat{C}^1$-norm is equivalent to the $C^{0,1}$-norm, we conclude 
\beq\nonumber
y^2 \leq a \, (1 + 3x).
\eeq 
For $k=3$, since $y^0=0$, \eqref{Thm_PDE_eqn6} implies
$$
y^3  \leq\,  a +  x ( y^2 +  y^1 ) ,
$$
and using the above inequalities on $y^2$ and $y^1$, we get
$$
y^3  \leq\,  a \left( 1 +  3x + 9x^2 \right) .
$$
We conclude that \eqref{Thm_PDE_eqn7} holds for $k=1$, $k=2$ and $k=3$. For the induction step, assume that \eqref{Thm_PDE_eqn7} holds for $k-1$, $k-2$ and $k-3$. Then, we find from \eqref{Thm_PDE_eqn6} and our induction assumption that
\begin{eqnarray}\nonumber
y^k  
&\leq & a +  x \left( y^{k-1} + y^{k-2} + y^{k-3} \right) \cr
&\leq & a +  x \left( a \sum_{l=0}^{k-2} (3 x)^l + a \sum_{l=0}^{k-3} (3 x)^l + a \sum_{l=0}^{k-4} (3 x)^l \right) \cr
&\leq & a\sum_{l=0}^{k-1} (3 x)^{l},
\end{eqnarray}
which completes the induction proof and shows \eqref{Thm_PDE_eqn7}. We continue to prove the existence of the $k$ independent constant, by choosing $T>0$ to satisfy the upper bound          
\beq \label{T-bound}
T \,<\, \frac1{3\beta}.
\eeq
(This choice is possible since in the above estimates $\beta$ only depends on $\mathcal{X}$, $\mathcal{Y}$, $\tau$ and the coefficients of the PDE \eqref{Cauchyproblem_PDE}, which are quantities defined in the ambient region $\mathcal{N}_+$ on which \eqref{Thm_PDE_eqn6} holds, so that restricting $T$ to even smaller values leaves $\beta$ unchanged.) We now define
\beq\nonumber
\mathcal{C} = \alpha \| U_0\|_{C^{0,1}_{2}}  \sum_{l=0}^{\infty} (3 T \beta)^l  ,
\eeq 
which is the sought after $k$-independent upper bound for $\| U^k \|_{C^{0,1}_2}$, that is,
\beq \label{Thm_PDE_k-indep_bound}
\| U^k \|_{\hat{C}^{0,1}_2(\mathcal{N}'_+)} \,\leq\, \mathcal{C} \ \ \ \ \forall\, k\geq 0.
\eeq

\noindent \textbf{Step 5:} We now prove convergence of a sub-sequence. Since \eqref{Thm_PDE_k-indep_bound} implies that $\big(U^k\big)_{k\in\mathbb{N}}$ is bounded by $\mathcal{C}$ with respect to the Lipschitz norm, the Arzel\`a-Ascoli Theorem yields that there exists a sub-sequence (which we again denote by $\big(U^k\big)_{k\in\mathbb{N}}$) that converges to some function $U\,\in\, C^{0,1}(\mathcal{N}_+')$ as $k\to\infty$.       Furthermore, since the $U^{k}$ are in $C^{0,1}_2({\mathcal{N}}_+')$ with uniform bound \eqref{Thm_PDE_k-indep_bound}, the $U^{k}$ are in particular in $C^{1,1}({\mathcal{N}}_p')$, so that the Arzel\`a-Ascoli Theorem gives us uniform convergence of some sub-sequence of $\big(U^{k}\big)_{k\in\mathbb{N}}$ to some function $U'$ in $C^{1,1}({\mathcal{N}}_p')$, and the uniqueness of limits implies that $U|_{{\mathcal{N}}_p'} \equiv U'|_{{\mathcal{N}}_p'}$. Moreover, $U^{k}\in C^{0,1}_2({\mathcal{N}}_+')$ together with \eqref{Thm_PDE_k-indep_bound} implies that derivatives of $U^k$ are in $C^{0,1}(\bar{\mathcal{N}}_p')$ and bounded by $\mathcal{C}$ in the Lipschitz norm over the closure $\bar{\mathcal{N}}_p'$. Thus, the Arzel\`a-Ascoli Theorem implies convergence of a sub-sequence of $\big(\partial_\mu U^k\big)_{k\in\mathbb{N}}$ to some Lipschitz continuous function $\Psi_\mu$. By uniqueness of limits we have $\partial_\mu U=\Psi_\mu$ on the interior regions $\mathcal{N}_p'$, while $\lim\limits_{n\rightarrow \infty} \partial_\mu U(q_n)=\Psi_\mu(q)$ for each $q$ in the boundary of $\bar{\mathcal{N}}_p'$ with $q_n\rightarrow q$ follows from the inequality
\begin{align} \nonumber
\big|\Psi_\mu (q) - \partial_\mu U(q_n)\big| \leq  \big|\Psi_\mu(q) - \Psi_\mu(q_n)\big|  + \big|\Psi_\mu (q_n) - \partial_\mu U^k(q_n)\big| \cr + \big|\partial_\mu U^k(q_n) - \partial_\mu U(q_n)\big| \hspace{2.8cm} 
\end{align}
together with a diagonal sequence argument in $k,n\in \mathbb{N}$.                In summary, we proved that a sub-sequence of the iterates $U^k$ converges uniformly to some function $U\in C^{0,1}(\mathcal{N}_+')$ which is in $C^{1,1}(\bar{\mathcal{N}}_p')$, where $\mathcal{N}_p'$ can be any of the regions $\mathcal{N}_L'$, $\mathcal{N}_M^{\prime L}$, $\mathcal{N}_M^{\prime R}$ or $\mathcal{N}_R'$. 

Transforming to $(t,r)$-coordinates by the replacement $r'=\mathcal{Y}(t,r)$ yields that $U$ has the claimed regularity in $(t,r)$-coordinates.               Finally, since uniform convergence implies point-wise convergence, and since we proved the existence of a sub-sequence of the $U^k$ such that its elements and their derivatives both converge uniformly to $U$ and $\partial_\mu U$ respectively, we conclude that $U$ solves the original Cauchy problem \eqref{Cauchyproblem_PDE} -\eqref{Cauchyproblem_initial} point-wise almost everywhere (and point-wise away from the shocks and $\gamma_0$) for initial data $U_0$. \vspace{.2cm}

\noindent \textbf{Step 6:} To complete the proof, it remains to show that $\partial_t U \in C^0(\bar{\mathcal{N}}_M')$ and that the jump of its $r'$-derivative satisfies \eqref{jump_r-deriv_U}. For this, recall that the integral over $F^{k-1}$ in the formula for the $k$-th iterate, \eqref{IC_iteration_soln}, is a sum of terms of the form \eqref{RegularityLemma_eqn1} and \eqref{RegularityLemma_eqn2}. Thus, by Lemma \ref{RegularityLemma}, we obtain $\partial_t U^k \in C^0(\bar{\mathcal{N}}_M)$, and the convergence of $U^k$ in $C^{1,1}(\bar{\mathcal{N}}_M^{\prime L})$ and $C^{1,1}(\bar{\mathcal{N}}_M^{\prime R})$ implies that $\partial_t U \in C^0(\bar{\mathcal{N}}_M)$, as claimed in the theorem. Regarding the jump of $\partial_{r'} U^k$ across $\gamma_0$, we again \eqref{phi_derivative-jump} - \eqref{psi_derivative-jump}. For this, observing that only the coefficients in \eqref{3} multiplied to a Heaviside function contribute to the constant $f(0,r_0)$ in \eqref{psi_derivative-jump}, that is, $M_i$ and $\mathcal{A}_{ji}$, and using that both coefficients are $C^1$ across $\gamma_0$, we conclude that
\beq \label{jump_r-deriv_Uk}
[\partial_{r'} U^k]_0(t)  =  c_0 -  \int_0^t {L}\big|_{\gamma_0}(s) [\partial_{r'} U^{k-1}]_0(s)ds ,
\eeq  
where ${L}$ denotes some linear combination of the $C^{0,1}_l$-coefficients in \eqref{3} and $c_0$ denotes the constant given by
$$
c_0 = 2 \sum_{i=1,2} \left(\partial_{r'} \tau\right)_i\left( M_i(0,r_0) + \sum_{j=1,2} \mathcal{A}_{ji}(0,r_0) \right) U_0(r_0),
$$ 
and $\left(\partial_{r'} \tau\right)_1$ denotes the left-limit of $\partial_{r'} \tau(r')$ and $\left(\partial_{r'} \tau\right)_2$ the right-limit as $r'$ approaches $r_0$. From \eqref{jump_r-deriv_Uk}, we now find that 
\beq \label{jump_r-deriv_Uk_eqn1}
\| [\partial_{r'} U^k]_0\|_{C^0([0,T])}  \leq  |c_0| + T c_1\, \| [\partial_{r'}  U^{k-1}]_0\|_{C^0([0,T])} ,
\eeq  
for $c_1 = \| L \|_{C^0(\mathcal{N}_+')} $ and it follows by induction that
\beq \label{jump_r-deriv_Uk_eqn2}
\| [\partial_{r'} U^k]_0\|_{C^0([0,T))}  \leq  |c_0| \sum_{l=0}^{k-1} \left( T c_1 \right)^l.
\eeq
Choosing, in addition to \eqref{T-bound}, $T>0$ small enough for the above sum to converge in the limit $k\rightarrow \infty$ and defining $\mathcal{C}$ to be the value of this limit multiplied with $2 \max\limits_{i=1,2} \left| \left(\partial_{r'} \tau\right)_i \right| $, then \eqref{jump_r-deriv_Uk_eqn2} implies in the limit $k \rightarrow \infty$ that
\beq \nonumber
\big\| [\partial_{r'} U]_0 \big\|_{C^0([0,T])} \leq \mathcal{C} \sum_{i=1,2} \left| \left( M_i(0,r_0) + \sum_{j=1,2} \mathcal{A}_{ji}(0,r_0) \right) U_0(r_0) \right| .
\eeq   
Changing to $(t,r)$-coordinates on the left hand side of the above equation and absorbing $\big\| \partial_{r} \mathcal{Y}\big\|_{C^0(\mathcal{N}_+)}$ in the constant $\mathcal{C}$, finally yields \eqref{jump_r-deriv_U}. This completes the proof. 
\QED

\subsection{Bootstrapping to $C^1$-Regularity} \label{Sec: Bootstrapping}

We now proceed with the proof of Proposition \ref{soln_IC_Prop}. In more detail, to show the existence of a $C^{1,1}$ regular solution of the integrability condition, \eqref{IC_2shocks}, we consider the $\oone$ solutions proven to exist in Theorem \ref{Thmiteration} and then use a bootstrap argument, which uses the specific structure of \eqref{IC_2shocks} and the RH conditions, to prove that any solution of the form asserted by Theorem \ref{Thmiteration} is in fact $C^{1,1}$ regular. The bootstrapping is accomplished in the following lemma.

\begin{Lemma}\label{C1_lemma}
Assume $U \in \oone(\mathcal{N}_+)$ solves \eqref{IC_2shocks}, and that $U$ has the regularity asserted in Theorem \eqref{Thmiteration}, in particular, that $[\partial_{r} U]_0$  satisfies \eqref{jump_r-deriv_U} across the characteristic line emanating from $(0,r_0)$. Assume the jump conditions, \eqref{jumpone} - \eqref{jumptwo} and \eqref{[Bt]shockspeed=[Ar]}. Then $U$ is in $C^{1,1}(\mathcal{N}_+)$. 
\end{Lemma}

\Proof
Since  $U\: \in \: \oone({\mathcal{N}_+})$ is assumed to be in $C^{1,1}(\bar{\mathcal{N}}_p)$, for $\mathcal{N}_p$ being any of the regions $\mathcal{N}_L$, $\mathcal{N}_M^{L}$, $\mathcal{N}_M^{R}$ or $\mathcal{N}_R$, it suffices to prove that $\partial_t U$ and $\partial_r U$ match continuously across the shock curves and the characteristic curve $\gamma_0$. 

We begin by proving the $C^1$ matching across the shock curves, $\gamma_1$ and $\gamma_2$. Since the above assumptions imply that $U$ is $\oone$ across $\gamma_1$ and $\gamma_2$, Lemma \ref{Lipschitz_across_lemma} yields that 
\beq\label{C1_lemma_eqn1}
[\partial_t U]_i +  \dot{x}_i \, [\partial_r U]_i = 0,
\eeq
c.f. \eqref{Lipschitz across 1; eqn}. On the other hand, taking the jump $[\cdot]_i$ of the integrability condition \eqref{IC_2shocks} yields
\beq\nonumber 
[\partial_t U]_i + c\,[\partial_r U]_i =\,  [F(U)]_i\, ,
\eeq
where we use that $c$ and $U$ are continuous. In the following we prove that the right hand side vanishes, that is, 
\beq\label{C1_lemma_eqn2b}
[\partial_t U]_i + c\,[\partial_r U]_i = 0.
\eeq
\eqref{C1_lemma_eqn1} together with \eqref{C1_lemma_eqn2b} suffice to prove the $C^1$ matching of $U$ across the shock curves, since the shock speed is assumed to be non-characteristic so that \eqref{C1_lemma_eqn1} and \eqref{C1_lemma_eqn2b} impose two linearly independent conditions on $[\partial_t U]_i$ and $[\partial_r U]_i$. 

We now prove $[F(U)]_i =0$. From \eqref{IC_2shocks_def_F} and the continuity of $U$, $(\varphi_j)^t_{0}$ and $(\dot{\varphi}_j)^t_{0}$, we obtain
\beq\nonumber
[F(U)]_i =  [\mathcal{M}]_i\, U +  \sum_{j=1,2} \Big(   [\mathcal{M}]_i |X_j|   + \left(c - \dot{x}_j\right)  [H(X_j)]_i \Big) \left(\begin{array}{c} (\varphi_j)^t_{0} \cr (\varphi_j)^r_{0} \end{array}\right)   , 
\eeq
where $U$, $c$ and $X_j$ in the above equation shall be understood to be restricted to $\gamma_i$. Using  $[H(X_j)]_i = 2 \delta_{ji}$, (where $\delta_{ji}=1$ if $i=j$ and $\delta_{ji}=0$ otherwise), $U= \,^T(\Phi^t_{0}, \Phi^r_{0} )$  and 
\beq\nonumber
\left(\begin{array}{c} J^t_0 \cr J^r_0 \end{array} \right) 
= \sum_{j=1,2}  \left(\begin{array}{c} (\varphi_j)^t_{0} \cr (\varphi_j)^r_{0} \end{array}\right) |X_j| \, 
+ \left(\begin{array}{c} \Phi^t_{0} \cr \Phi^r_{0} \end{array}\right)  \, , 
\eeq
the above equation leads to
\beq\label{C1_lemma_eqn3}
[F(U)]_i =   2\left( c-\dot{x}_i \right)  \, \left(\begin{array}{c} (\varphi_i)^t_{0} \cr (\varphi_i)^r_{0} \end{array}\right) + [\mathcal{M}]_i \,\left(\begin{array}{c} J^t_0\circ\gamma_i \cr J^r_0\circ\gamma_i \end{array} \right)  . 
\eeq
From \eqref{IC_2shocks_def_C}, we now obtain 
\begin{eqnarray}\nonumber
[\mathcal{M}]_i 
&=& \frac{1}{J^t_1} \left(\begin{array}{cc} [J^t_{1,t}]_i & [J^t_{1,r}]_i \cr [J^r_{1,t}]_i & [J^r_{1,r}]_i \end{array}\right) \cr
&=& \frac{1}{J^t_1} \left(\begin{array}{cc} (\mathcal{J}_i)^t_{1 t} & (\mathcal{J}_i)^t_{1 r} \cr (\mathcal{J}_i)^r_{1 t} & (\mathcal{J}_i)^r_{1 r} \end{array}\right) \, ,
\end{eqnarray}
where we used that the jumps in the Jacobian derivatives satisfy the smoothing condition \eqref{smoothingcondt2}. Furthermore, multiplying \eqref{C1_lemma_eqn3} with $J^t_1$ and substituting  \eqref{Jacobiancoeff2_impl}, that is, $(\varphi_i)^\mu_{0}=-\frac12 (\mathcal{J}_i)^\mu_{0 r} $, \eqref{C1_lemma_eqn3} becomes
\beq\nonumber
J^t_1 \, [F(U)]_i
=   -\left( J^r_1 - J^t_1 \, \dot{x}_i  \right)  \, \left(\begin{array}{c} (\mathcal{J}_i)^t_{0 r} \cr (\mathcal{J}_i)^r_{0 r} \end{array}\right) 
+ \left(\begin{array}{cc} (\mathcal{J}_i)^t_{1 t} & (\mathcal{J}_i)^t_{1 r} \cr (\mathcal{J}_i)^r_{1 t} & (\mathcal{J}_i)^r_{1 r} \end{array}\right)\left(\begin{array}{c} J^t_0 \cr J^r_0 \end{array} \right) .
\eeq
Now, as shown in the proof of Lemma \ref{Jacobian2}, the RH conditions imply that 
$$
-\dot{x}_i \, (\mathcal{J}_i)^\mu_{0 r} = (\mathcal{J}_i)^\mu_{0 t}, 
$$ 
and using this identity above gives
\beq\nonumber
J^t_1 \, [F(U)]_i   =   
-  J^r_1 \left(\hspace{-.1cm} \begin{array}{c}  (\mathcal{J}_i)^t_{0r}   \cr   (\mathcal{J}_i)^r_{0r} \end{array} \hspace{-.1cm}\right) 
- J^t_1 \left( \hspace{-.1cm} \begin{array}{c}   (\mathcal{J}_i)^t_{0t}   \cr  (\mathcal{J}_i)^r_{0t} \end{array} \hspace{-.1cm}\right)
+ J^t_0 \left( \hspace{-.1cm} \begin{array}{c}  (\mathcal{J}_i)^t_{1t}  \cr  (\mathcal{J}_i)^r_{1t} \end{array} \hspace{-.1cm}\right)
+ J^r_0 \left(\hspace{-.1cm} \begin{array}{c} (\mathcal{J}_i)^t_{1r} \cr (\mathcal{J}_i)^r_{1r} \end{array} \hspace{-.1cm}\right).
\eeq
Substituting the identity \eqref{smoothingcondt2} for $(\mathcal{J}_i)^\mu_{\alpha\sigma}$, we find that the terms on the right hand side mutually cancel which implies that $[F(U)]_i  = 0$ for $i=1,2$. In more detail, substituting the identity \eqref{smoothingcondt2} for $(\mathcal{J}_i)^\mu_{\alpha\sigma}$, we compute the first component as, (subsequently dropping the index $i$),
\begin{eqnarray}\nonumber
&  & 2 \left(- J^r_1 \, (\mathcal{J}_i)^t_{0r}  - J^t_1 \, (\mathcal{J}_i)^t_{0t} + J^t_0 \, (\mathcal{J}_i)^t_{1t} + J^r_0 \, (\mathcal{J}_i)^t_{1r} \right) \cr
&=&  J^r_1 \left( \frac{[A_r]}{A}J^t_0 + \frac{[B_t]}{A}J^r_0  \right)  + J^t_1 \, \left( \frac{[A_t]}{A}J^t_0 + \frac{[A_r]}{A}J^r_0  \right)  \cr
&& - J^t_0 \, \left( \frac{[A_t]}{A}J^t_1 + \frac{[A_r]}{A}J^r_1  \right) - J^r_0 \, \left( \frac{[A_r]}{A}J^t_1 + \frac{[B_t]}{A}J^r_1  \right)   \cr
&=& \ 0,
\end{eqnarray}
and the second component vanishes by a similar cancellation. This proves the $C^1$ matching across the shock curves.

It remains to prove that $U$ is $C^1$ across the characteristic curve $\gamma_0$. Since $\partial_t U$ is assumed to be continuous across $\gamma_0$, we only need to address the jump in $\partial_r U$, which is assumed to satisfy \eqref{jump_r-deriv_U}, that is,
\beq \label{jump_r-deriv_U_again}
\big\| [\partial_{r} U]_0 \big\|_{C^0([0,T))} \leq \mathcal{C}\left| \sum_{i=1,2}\left( M_i(0,r_0) + \sum_{j=1,2} \mathcal{A}_{ji}(0,r_0) \right) U_0(r_0) \right| .
\eeq   
We now prove that the right hand side of \eqref{jump_r-deriv_U_again} vanishes by the same cancellation leading to \eqref{C1_lemma_eqn2b}. For convenience, we subsequently often omit that all quantities are evaluated at the point $(0,r_0)$ or at $t=0$ respectively. To begin, by comparison of \eqref{IC_2shocks_def_F} with \eqref{3} - \eqref{5}, collecting the \emph{coefficients of Heaviside functions}, we find that 
\begin{eqnarray}\nonumber 
M_i(0,r_0) &=& 
\frac{1}{J^t_1} \left( \begin{array}{cc}   (\varphi_i)^t_{1} \dot{x}_i   & - (\varphi_i)^t_{1} \cr (\varphi_i)^r_{1} \dot{x}_i   &  - (\varphi_i)^r_{1}   \end{array} \right) , \cr
\end{eqnarray}
since, according to \eqref{Jacobian2}, the part in $J^\mu_{1,\sigma}$ multiplied to a Heaviside function is either $\dot{x}_i (\varphi_i)^\mu_{1} $ or  $-(\varphi_i)^\mu_{1}$. Moreover, by the same comparison and since $X_i(0,r_0)=0$, we find that
\begin{eqnarray}\nonumber
\sum_{j=1,2} \mathcal{A}_{ji}(0,r_0)  U_0(r_0) 
&=& \big(c - \dot{x}_i \big)  \left(\begin{array}{c} (\varphi_i)^t_{0} \cr (\varphi_i)^r_{0} \end{array} \right) \cr
 &=& \frac1{J^t_1}\big(J^r_1 - J^t_1\dot{x}_i \big)  \left(\begin{array}{c} (\varphi_i)^t_{0} \cr (\varphi_i)^r_{0} \end{array} \right) .
\end{eqnarray}
Combining the above two equations, multiplying through with $J^t_1$, and using $J^\mu_\alpha(0,r_0) = \Phi^\mu_\alpha(0,r_0)$ and $U = (\Phi^t_0,\Phi^r_0)$, we obtain
\begin{eqnarray}\nonumber
f_0 &:=&   J^t_1(0,r_0) \left( M_i(0,r_0) + \sum_{j=1,2} \mathcal{A}_{ji}(0,r_0) \right) U_0(r_0) \cr 
&=&   \left( \begin{array}{c}   (\varphi_i)^t_{1} \dot{x}_i \Phi^t_0    - (\varphi_i)^t_{1} \Phi^r_0 \cr (\varphi_i)^r_{1} \dot{x}_i \Phi^t_0    - (\varphi_i)^r_{1} \Phi^r_0  \end{array} \right)     + \big(\Phi^r_1 - \dot{x}_i \Phi^t_1 \big)  \left(\begin{array}{c} (\varphi_i)^t_{0} \cr (\varphi_i)^r_{0} \end{array} \right)  \cr
&=&   \left(  \dot{x}_i \Phi^t_0    -  \Phi^r_0  \right) 
\left( \begin{array}{c}   (\varphi_i)^t_{1}  \cr (\varphi_i)^r_{1} \end{array} \right)   
+  \big(\Phi^r_1 - \dot{x}_i \Phi^t_1 \big)  \left(\begin{array}{c} (\varphi_i)^t_{0} \cr (\varphi_i)^r_{0} \end{array} \right)   .
\end{eqnarray}
To proceed, observe that $\mathcal{B}_{ij}(0)=0$, so that \eqref{Jacobiancoeff2_expl_t-0} - \eqref{Jacobiancoeff2_expl_r-1} reduces at the point $(0,r_0)$ to
\begin{eqnarray}\nonumber
(\varphi_i)^t_{0}(0,r_0)   &=&   \frac{[A_r]_i}{4A_i}  \Phi^t_{0}   +   \frac{[B_t]_i}{4A_i} \Phi^r_{0}  ,  \cr
(\varphi_i)^t_{1}(0,r_0)   &=&   \frac{[A_r]_i}{4A_i}  \Phi^t_{1}  +   \frac{[B_t]_i}{4A_i} \Phi^r_{1}    , \cr
(\varphi_i)^r_{0}(0,r_0)    &=&   \frac{[B_t]_i}{4B_i}  \Phi^t_{0}   +   \frac{[B_r]_i}{4B_i} \Phi^r_{0}  ,   \cr
(\varphi_i)^r_{1}(0,r_0)    &=&   \frac{[B_t]_i}{4B_i}  \Phi^t_{1}  +   \frac{[B_r]_i}{4B_i} \Phi^r_{1}  , 
\end{eqnarray}
where all quantities shall be understood to be evaluated at $(0,r_0)$. Substituting the above identities, we obtain
\begin{eqnarray}\nonumber
f_0 &=& \left( \dot{x}_i \Phi^t_0    -  \Phi^r_0  \right) \hspace{-.2cm}
\left(\hspace{-.2cm} \begin{array}{c}      \frac{[A_r]_i}{4A_i}  \Phi^t_{1}  +   \frac{[B_t]_i}{4A_i} \Phi^r_{1}   \cr             \frac{[B_t]_i}{4B_i}  \Phi^t_{1}  +   \frac{[B_r]_i}{4B_i} \Phi^r_{1}  \end{array} \hspace{-.2cm}\right)   
+  \big(\Phi^r_1 - \dot{x}_i \Phi^t_1 \big)  \hspace{-.2cm}
\left(\hspace{-.2cm} \begin{array}{c} \frac{[A_r]_i}{4A_i}  \Phi^t_{0}   +   \frac{[B_t]_i}{4A_i} \Phi^r_{0}  \cr 
\frac{[B_t]_i}{4B_i}  \Phi^t_{0}   +   \frac{[B_r]_i}{4B_i} \Phi^r_{0}   \end{array} \hspace{-.2cm}\right),
\end{eqnarray}
and, by mutual cancellation of terms, a straightforward computation gives  
\begin{eqnarray} \nonumber
f_0 &=&  \left(  \Phi^t_0 \Phi^r_{1}  -  \Phi^t_1 \Phi^r_0  \right)  \left\{  \left( \begin{array}{c}  \frac{[A_r]_i}{4A_i}  \cr  \frac{[B_t]_i}{4B_i}  \end{array} \right)  
+  \dot{x}_i  \left(\begin{array}{c} \frac{[B_t]_i}{4A_i}  \cr  \frac{[B_r]_i}{4B_i} \end{array} \right)  \right\} \cr
&=& 0,
\end{eqnarray}
where we used the RH condition \eqref{jumpone} and \eqref{[Bt]shockspeed=[Ar]} to obtain the the last line. Thus, it follows by \eqref{jump_r-deriv_U_again} that the jump in $\partial_r U$ across $\gamma_0$ vanishes, from which we finally conclude that $U\in C^{1,1}(\mathcal{N}_+)$.
\QED

\subsection{Proof of Proposition \ref{soln_IC_Prop}}

We now complete the proof of Proposition \ref{soln_IC_Prop}. In Theorem \ref{Thmiteration}, we proved existence of a $C^{0,1}$ solution of the non-local PDE \eqref{3}, with $C^{1,1}$ regularity away from the shock curves and $\gamma_0$. In Lemma \ref{C1_lemma}, we then showed that any such solution of the integrability condition \eqref{IC_2shocks} is $C^{1,1}$ regular throughout $\mathcal{N}_+$, which is the regularity claimed in Proposition \ref{soln_IC_Prop}. Moreover, it follows that the $C^{0,1}$ solution is in fact a $C^1$ solution which solves \eqref{3} point-wise (everywhere).  The proof of Proposition \ref{soln_IC_Prop} is complete, once we show that \eqref{IC_2shocks} is indeed a special case of \eqref{Cauchyproblem_PDE} - \eqref{5}, the system of non-local PDE's addressed in Theorem \ref{Thmiteration}, which is achieved in the following lemma.

\begin{Lemma}
The integrability conditions, \eqref{IC_2shocks}, under the regularity assumptions in Proposition \ref{soln_IC_Prop}, are a special case of the non-local system of PDE's \eqref{Cauchyproblem_PDE} - \eqref{5}. 
\end{Lemma}
\Proof
To begin, the characteristic speed $c=J^t_1 / J^r_1$, given in \eqref{IC_2shocks_def_b}, is a function in $C^{0,1}_2(\mathcal{N}_+)$, as becomes clear from the canonical form of the Jacobian, \eqref{Jacobian2}. It thus remains to show that $F(U)$ in the integrability conditions is a special case of the corresponding term in \eqref{Cauchyproblem_PDE}, which we here denote with $\tilde{F}(U)$, to avoid confusion. For this, recall that 
\begin{eqnarray}\label{F_PDE_comprison}
\tilde{F}(U) &=&\tilde{\mathcal{M}}U+\sum_{i=1,2}\left(\mathcal{A}_i\cdot (U\circ\gamma_i)+\mathcal{B}_i\left|X_i\right|\cdot\frac{d}{dt}\left(U\circ\gamma_i\right)\right),
\end{eqnarray}
according to  \eqref{3}-\eqref{5}, with
\begin{eqnarray} 
\tilde{\mathcal{M}} &=&\sum_{i=1,2}M_iH(X_i)+M, \label{M_PDE_comparison} \\
\mathcal{A}_i &=& \sum_{j=1,2}\mathcal{A}_{ij}H(X_j)+A_i , \label{A_PDE_comparison}
\end{eqnarray}
where $\mathcal{B}_i$, $M_i$, $M$, $\mathcal{A}_{ij}$ and $A_i$ are matrix valued functions with components in $C^{0,1}_2(\mathcal{N}_+)$. On the other hand, the term $F(U)$ in the integrability condition is given by
\beq \label{F_IC_comparison}
F(U) = \mathcal{M}\, U    +    \sum_{i=1,2} \left\{ \Big(  |X_i| \mathcal{M}  - H(X_i) \left(\dot{x}_i - c \right)  
\Big) \left( \hspace{-.2cm} \begin{array}{c} (\varphi_i)^t_{0} \cr (\varphi_i)^r_{0} \end{array} \hspace{-.2cm}\right)   
- |X_i| \left(\hspace{-.2cm} \begin{array}{c} (\dot{\varphi}_i)^t_{0} \cr (\dot{\varphi}_i)^r_{0} \end{array} \hspace{-.2cm}\right)    \right\} ,
\eeq
c.f. \eqref{IC_2shocks_def_F}, with
\beq \nonumber 
\mathcal{M} =   \frac{1}{J^t_1} \left( \begin{array}{cc} J^t_{1,t}  &  J^t_{1,r} \cr J^r_{1,t} & J^r_{1,r}   \end{array} \right).
\eeq

Now, from the canonical form of the Jacobian, \eqref{Jacobian2}, it is straightforward to conclude that        $\mathcal{M}$ is a special case of the matrix $\tilde{\mathcal{M}}$, where the $C^{0,1}_2(\mathcal{N}_+)$ regularity holds due to our assumption that the jump in the metric derivatives are $C^3$ functions along the shock curves in Definition \ref{shockinteract} and our choice of $\Phi^t_1$ and $\Phi^r_1$ in $C^3$.          

We next consider the term in \eqref{F_IC_comparison} containing $\,^T((\varphi_i)^t_{0}, (\varphi_i)^r_{0})$, since the Jacobian coefficients $(\varphi_i)^\mu_{0}$ depend on $U\circ\gamma_j$, $j=1,2$, the jumps in the metric derivatives and the metric restricted to $\gamma_i$, we conclude that this term is a special case of $\mathcal{A}_i\cdot (U\circ\gamma_i)$. Note that the $C^{0,1}_2$ regularity again follows from our regularity assumptions for the metric (in Definition \ref{shockinteract}) and for the free functions $\Phi^t_1$ and $\Phi^r_1$.

We finally consider the term $ |X_i| \,^T\left( (\dot{\varphi}_i)^t_{0} , (\dot{\varphi}_i)^r_{0} \right) $ in \eqref{F_IC_comparison}. Since $(\varphi_i)^\mu_{0}$ depend on $U\circ\gamma_j$ itself, the above term is either of the form $\sum_{i=1,2}\mathcal{A}_i\cdot (U\circ\gamma_i)$, namely whenever the derivative on $(\varphi_i)^\mu_{0}$ does not act on $U\circ\gamma_j$, or else, it is of the form $\mathcal{B}_i\left|X_i\right|\cdot\frac{d}{dt}\left(U\circ\gamma_i\right)$. In summary, we proved that the integrability condition \eqref{IC_2shocks} is a special case of the non-local PDE \eqref{Cauchyproblem_PDE} - \eqref{5}, which completes the proof.
\QED

\section{The Shock-Collision-Case: The Matching Conditions} \label{Sec: Matching Conditions}

So far we constructed Jacobians which satisfy the smoothing conditions at the shock curves and are integrable to coordinates on the closed upper half-plane, $\overline{\R^2_+}$. A simultaneous construction gives us such Jacobians on the closed lower half plane, $\overline{\R^2_-}$. To proceed with the proof of Theorem \ref{TheoremMain}, we now ``glue'' these two Jacobians at the $(t=0)$-interface in a way  appropriate to maintain the smoothness of the resulting metric $g_{\alpha\beta}$. 

For the pursue of this gluing, let us introduce some notation. We subsequently denote all objects in \eqref{Jacobian2} - \eqref{Jacobiancoeff2_expl_B} with an additional index ``$+$'' or ``$-$'' to indicate whether they originate from the Jacobian construction on $\overline{\R^2_+}$ or $\overline{\R^2_-}$, respectively. That is,
\begin{eqnarray}\nonumber
J^{\mu\pm}_\alpha(t,r)= \sum_{i=1,2} (\varphi_i^\pm)^\mu_{\alpha}(t) \left| x_i^\pm(t)-r \right| + \Phi^{\mu\pm}_\alpha(t,r),
\end{eqnarray}
is the canonical form of the Jacobian on $\mathcal{N}_\pm=\mathcal{N}\cap \overline{\R^2_\pm}$. Moreover, we denote with $\{\cdot\}$ the jump across the $(t=0)$-interface, that is,  
\beq\label{def_curly_brackets}
\{u \}(r)= \lim_{t\nearrow0} u(t,r) - \lim_{t\searrow0} u(t,r) ,
\eeq
where $u$ is some function for which the above limits are well-defined. Subsequently, we only apply \eqref{def_curly_brackets} for either the metric derivatives, the Jacobian or its derivatives. For those functions the limits in \eqref{def_curly_brackets} are well-defined whenever $r\neq r_0$, however, at $r=r_0$ this is no longer the case. Nevertheless, for our purposes it suffices to define $\{\cdot\}$ at $r=r_0$ as follows: Assume without loss of generality that $x^+_1(t) \leq x^+_2(t)$ and  $x^-_1(t) \leq x^-_2(t)$. We now partition $\mathcal{N} \subset \R^2$ into the four regions $L$, $R$, $M^-$ and $M^+$, where $L \subset \mathcal{N}$ denotes the open region on the left of $\gamma^\pm_1$, $R \subset \mathcal{N}$ denotes the open region to the right of $\gamma^\pm_2$, $M^- \subset \mathcal{N}_-$ denotes the open region in between $\gamma^-_1$ and $\gamma^-_2$, and $M^+ \subset \mathcal{N}_+$ denotes the open region in between $\gamma^+_1$ and $\gamma^+_2$. Then, for some $L^\infty$-function $u$ which is continuous on $L$, $R$, $M^-$ and $M^+$, we define $u_L$, $u_R$, $u_{M^-}$ and $u_{M^+}$ to be the limit of $u(q)$ as $q$ approaches $p$ with $q$ being restricted to the respective region, that is, 
$$
u_L=\lim\limits_{\substack{q\rightarrow p \\ q\in L}} u(q), \ \ \ \ u_R=\lim\limits_{\substack{q\rightarrow p \\ q\in R}} u(q), \ \ \ \ u_{M^-}=\lim\limits_{\substack{q\rightarrow p \\ q\in M^-}} u(q) \ \ \ \text{and} \ \ u_{M^+}=\lim\limits_{\substack{q\rightarrow p \\ q\in M^+}} u(q).
$$
Finally we define $\{u\}(r_0)$ as
\beq\label{def_curly_brackets_r0}
\{u \}(r_0) = u_{M^-} - \, u_{M^+}.
\eeq

With the notation introduced above, we now derive the conditions for the matching of $J^{\mu\pm}_{\alpha}$, conditions which are necessary and sufficient for the metric in the new coordinates, $g_{\alpha\beta}=J^\mu_\alpha J^\nu_{\beta} g_{\mu\nu}$, to be $C^{1,1}$ regular on some neighborhood of the shock interaction. We refer to these conditions as \emph{matching conditions}. To begin with, for the Jacobian to be in the $\oneone$ atlas, it must match continuously across the $(t=0)$-interface, that is, 
\beq\label{matching_C0}
\{ J^\mu_\alpha  \}(r) \equiv J^{\mu-}_{\alpha}(0,r) - J^{\mu+}_{\alpha}(0,r)  \, = 0, 
\eeq
for all $r$. These are the $C^0$-matching conditions.

To determine the matching of the Jacobian derivatives, we follow the reasoning in Section \ref{intro to method}, (which led to the smoothing condition, \eqref{smoothingcondt1_SSC}), but now with respect to the $(t=0)$-interface. In more detail, the condition that $g_{\alpha\beta}$ is continuously differentiable across the $(t=0)$-interface is given by 
\beq\label{matching_metric}
\{ g_{\alpha\beta,\sigma} \} = 0.
\eeq
Substituting $g_{\alpha\beta} = J^\mu_\alpha J^\nu_\beta g_{\mu\nu}$ into \eqref{matching_metric} and using \eqref{matching_C0} as well as the SSC-metric being $C^1$ regular away from the shocks, that is, $\{g_{\mu\nu,\sigma}\}(r)=0$ for all $r\neq r_0$, we conclude that the $C^1$-matching conditions are given by
\beq\label{matching_C1_r=r}
\left( \{J^\mu_{\alpha,\sigma}\} J^\nu_\beta + \{J^\nu_{\beta,\sigma}\} J^\mu_\alpha \right) g_{\mu\nu} = 0, \ \ \ \ \ \forall \, r\neq r_0,
\eeq
and by
\beq\label{matching_C1_r=r0}
\left( \{J^\mu_{\alpha,\sigma}\} J^\nu_\beta + \{J^\nu_{\beta,\sigma}\} J^\mu_\alpha \right) g_{\mu\nu} = -J^\mu_\alpha J^\nu_\beta \{g_{\mu\nu,\sigma}\}, \ \ \ \ \ \text{at} \ r=r_0.
\eeq

It remains to prove that the above matching conditions can be met for the Jacobian in \eqref{Jacobian2}, by appropriately matching the free functions $\Phi^{\mu+}_\alpha$ and $\Phi^{\mu-}_\alpha$ as well as their derivatives. A potential obstacle for the pursue of this matching is that the functions $\Phi^{t\pm}_{0,t}$ and $\Phi^{r\pm}_{0,t}$ are not free to assign, since they are prescribed by the integrability conditions \eqref{IC_2shocks}. Nevertheless, the RH conditions and the integrability conditions imply exactly the matching of $\Phi^{\mu+}_\alpha$ and $\Phi^{\mu-}_\alpha$, which is required for \eqref{matching_C1_r=r} and \eqref{matching_C1_r=r0} to hold.  All this is achieved in the next lemma.

\begin{Lemma}\label{matching_lemma}
Let $J^{\mu\pm}_\alpha$ be two Jacobians of the canonical form \eqref{Jacobian2}, defined on $\mathcal{N}_\pm = \mathcal{N}\cap \overline{\R^2_\pm}$ respectively, with corresponding free functions $\Phi^{\mu\pm}_\alpha$. Assume the integrability condition \eqref{IC} and the jump conditions, \eqref{jumpone} - \eqref{jumptwo} and \eqref{[Bt]shockspeed=[Ar]}, hold, and assume that $J^t_1(0,r)\neq0$ by an appropriate choice of $\Phi^t_1(0,r)$. If the $\Phi^{\mu\pm}_\alpha$ match at $t=0$, such that 
\begin{eqnarray}
\{\Phi^\mu_\alpha\}(r) &=& 0 , \label{matching_C0_a} \label{matching_C1_r} \\
\{\partial_t \Phi^t_{1} \} (r) &=& 
-\Big( (\dot{\varphi}_1^-)^{t}_{1} + (\dot{\varphi}_2^-)^{t}_{1} - (\dot{\varphi}_1^+)^{t}_{1} - (\dot{\varphi}_2^+)^{t}_{1}  \Big) |r-r_0|,  \label{matching_C1_t_first}  \\ 
\{\partial_t\Phi^r_{1} \} (r) &=& 
-\Big( (\dot{\varphi}_1^-)^{r}_{1} + (\dot{\varphi}_2^-)^{r}_{1} - (\dot{\varphi}_1^+)^{r}_{1} - (\dot{\varphi}_2^+)^{r}_{1}  \Big) |r-r_0| \label{matching_C1_t_second}
\end{eqnarray}
hold for all  $(0,r)\in \mathcal{N}$, then $J^{\mu\pm}_\alpha$ satisfies the matching conditions \eqref{matching_C0}, \eqref{matching_C1_r=r} and \eqref{matching_C1_r=r0}.
\end{Lemma}

\Proof
We first address the $C^0$-matching conditions, \eqref{matching_C0}. Taking the jump across the $(t=0)$-interface of the canonical form for the Jacobian, \eqref{Jacobian2}, gives us
\beq \nonumber
\{J^\mu_\alpha \}(r) = \Big( (\varphi_1^-)^{\mu}_{\alpha} + (\varphi_2^-)^{\mu}_{\alpha} - (\varphi_1^+)^{\mu}_{\alpha} - (\varphi_2^+)^{\mu}_{\alpha}  \Big) |r-r_0| \, + \, \{\Phi^\mu_\alpha\}(r)
\eeq
and assuming that \eqref{matching_C0_a} holds, the above equation reduces to 
\beq \label{matching_pf_eqn1}
\{J^\mu_\alpha \}(r) = \Big( (\varphi_1^-)^{\mu}_{\alpha} + (\varphi_2^-)^{\mu}_{\alpha} - (\varphi_1^+)^{\mu}_{\alpha} - (\varphi_2^+)^{\mu}_{\alpha}  \Big) |r-r_0| 
\eeq
In the following we show that 
\beq \label{matching_pf_eqn1b}
 (\varphi_1^-)^{\mu}_{\alpha} + (\varphi_2^-)^{\mu}_{\alpha} - (\varphi_1^+)^{\mu}_{\alpha} - (\varphi_2^+)^{\mu}_{\alpha}   = 0.
\eeq
We first derive simplified expressions for the $(\varphi_i^\pm)^{\mu}_{\alpha}$. Since $x^\pm_1(0)=r_0=~ x^\pm_2(0)$, we conclude from \eqref{Jacobiancoeff2_expl_B} that $\mathcal{B}_{ij}=0$, and thus, we find from \eqref{Jacobiancoeff2_expl_r-0} - \eqref{Jacobiancoeff2_expl_r-1} that
\begin{eqnarray}\label{matching_pf_eqn2}
(\varphi_i^\pm)^{r}_{0}   &=&  
 \frac1{4B} \Big(  [B_t]^\pm_i  \Phi^{t\pm}_0|_i   +   [B_r]^\pm_i \Phi^{r\pm}_0|_i   \Big)  \cr
(\varphi_i^\pm)^{r}_{1} &=& 
\frac1{4B} \Big(  [B_t]^\pm_i  \Phi^{t\pm}_1|_i  +   [B_r]^\pm_i \Phi^{r\pm}_1|_i   \Big)   \, ,
\end{eqnarray}
where we write $B$ instead of $B(p)$ and $\Phi^{\mu\pm}_\alpha|_i$ instead of $\Phi^{\mu\pm}_\alpha\circ\gamma_i(0)$. Using again the hypothesis \eqref{matching_C0_a}, we conclude that the $\Phi^{\mu\pm}_\alpha|_i$ match continuously, that is, there exist a value $\phi^\mu_\alpha$ such that
\beq\nonumber
\Phi^{\mu+}_{\alpha}|_i(0) = \Phi^{\mu-}_{\alpha}|_i(0) = \phi^\mu_\alpha \ \ \ \ \ \ \forall\, i=1,2 .
\eeq
Thus, \eqref{matching_pf_eqn2} simplyfies to 
\begin{eqnarray}\label{matching_pf_eqn2a}
(\varphi_i^\pm)^{r}_{0}   &=&  
 \frac1{4B} \Big(  [B_t]^\pm_i  \phi^t_{0}   +   [B_r]^\pm_i \phi^r_{0}   \Big)  \cr
(\varphi_i^\pm)^{r}_{1} &=& 
\frac1{4B} \Big(  [B_t]^\pm_i  \phi^t_{1}  +   [B_r]^\pm_i \phi^r_{1}   \Big)   \, .
\end{eqnarray}
The expressions for the remaining coefficients follow again from \eqref{Jacobiancoeff2_expl_t-0} - \eqref{Jacobiancoeff2_expl_t-1}, that is,
\begin{eqnarray}\label{matching_pf_eqn2c}
(\varphi_i^\pm)^{t}_{0} &=& -\frac{B}{A} \, \dot{x}^\pm_i \, (\varphi_i^\pm)^{r}_{0}  \ = \  \frac1{4A} \Big(  [A_r]^\pm_i  \phi^t_{0}   +   [B_t]^\pm_i \phi^r_{0}   \Big)  ,  \cr
(\varphi_i^\pm)^{t}_{1} &=& -\frac{B}{A}\, \dot{x}^\pm_i \, (\varphi_i^\pm)^{r}_{1}, \ = \  \frac1{4A} \Big(  [A_r]^\pm_i  \phi^t_{1}  +   [B_t]^\pm_i \phi^r_{1}   \Big)      , 
\end{eqnarray}
where we used the RH conditions \eqref{jumpone} - \eqref{jumptwo} and \eqref{[Bt]shockspeed=[Ar]}, that is,
\beq\label{RH_matching_pf}
\left[B_t\right]=-\dot{x} [B_r], \ \ \ \ \
\left[A_t\right]= -\dot{x}[A_r] \ \ \ \ \
\text{and} \ \ \ \ [A_r]=-\dot{x}[B_t],
\eeq
to eliminate the shock speeds $\dot{x}^\pm_i$ in \eqref{matching_pf_eqn2c}. To prove that the sums in \eqref{matching_pf_eqn1b} vanishes, we now compute them for the different cases $\mu \in\{ t,r\}$ and $\alpha\in\{0,1\}$ separately. From \eqref{matching_pf_eqn2a} we obtain
\begin{eqnarray}\label{matching_pf_eqn2b}
&& 4B \left( (\varphi_1^-)^{r}_{0} + (\varphi_2^-)^{r}_{0} - (\varphi_1^+)^{r}_{0} - (\varphi_2^+)^{r}_{0} \right) \\ &=&   \left( [B_t]^-_1 + [B_t]^-_2 - [B_t]^+_1 - [B_t]^+_2 \right) \phi^t_{0}  \cr &+& \left(  [B_r]^-_1 + [B_r]^-_2 - [B_r]^+_1 - [B_r]^+_2 \right)\phi^r_{0} \nonumber  .
\end{eqnarray}
But now it holds in general for the metric that sums of the above form mutually cancel. In more detail, at $t=0$ we have 
\begin{eqnarray}\nonumber
[g_{\mu\nu,\sigma}]^\pm_1(0) + [g_{\mu\nu,\sigma}]^\pm_2(0) &=& \left(g_{\mu\nu,\sigma}\right)_L - \left(g_{\mu\nu,\sigma}\right)_{M^\pm}  +  \left(g_{\mu\nu,\sigma}\right)_{M^\pm} - \left(g_{\mu\nu,\sigma}\right)_R  \cr
&=& \left(g_{\mu\nu,\sigma}\right)_L - \left(g_{\mu\nu,\sigma}\right)_R,
\end{eqnarray}
from which we conclude that at $t=0$
\begin{eqnarray}\label{cancelation_metric-sum}
[g_{\mu\nu,\sigma}]^-_1 + [g_{\mu\nu,\sigma}]^-_2 - [g_{\mu\nu,\sigma}]^+_1 - [g_{\mu\nu,\sigma}]^+_2 \, = \, 0.
\end{eqnarray}
Now, \eqref{cancelation_metric-sum} together with \eqref{matching_pf_eqn2b} imply that 
\beq\nonumber
(\varphi_1^-)^{r}_{0} + (\varphi_2^-)^{r}_{0} - (\varphi_1^+)^{r}_{0} - (\varphi_2^+)^{r}_{0} = 0, 
\eeq
which together with \eqref{matching_pf_eqn1} proves the $C^0$-matching conditions \eqref{matching_C0}, for $\mu=r$ and $\alpha=0$. Similarly, starting from the expressions \eqref{matching_pf_eqn2a} and \eqref{matching_pf_eqn2c} for the $(\varphi_i^\pm)^\mu_\alpha$ and applying \eqref{cancelation_metric-sum}, we obtain \eqref{matching_pf_eqn1b} for the remaining cases. This proves the $C^0$-matching conditions \eqref{matching_C0}. 

We now address the $C^1$-matching for $r\neq r_0$, that is, \eqref{matching_C1_r=r}. Since the left hand side of \eqref{matching_C1_r=r} is the same linear system as \eqref{smoothingcondt1_SSC} but with a vanishing right hand side, we conclude, in light of Lemma \ref{smoothcondition}, that a solution of \eqref{matching_C1_r=r} is given by
\beq\label{matching_C1_pf_1}
 \{ J^\mu_{\alpha,\sigma} \}(r) = 0  \ \ \ \ \ \forall \, r \neq\, r_0.
\eeq
Furthermore, assuming the $J^{\mu\pm}_\alpha$ satisfying the integrability condition, \eqref{IC}, the above solution is, in fact, the unique solution of \eqref{matching_C1_r=r}, c.f. Lemma \ref{smoothcondition}.

We first prove that \eqref{matching_C1_r} implies \eqref{matching_C1_pf_1} for $\sigma=r$. Observe that \eqref{matching_C0_a} implies that $\{\partial_r\Phi^\mu_\alpha\}(r)=0$ for all $(0,r) \in\mathcal{N}$. Differentiating the canonical form of the Jacobian \eqref{Jacobian2} with respect to $r$ and taking the jump across the $(t=0)$-interface, we get
\begin{eqnarray}\label{matching_C1_pf_2}
\{J^\mu_{\alpha,r} \}(r) 
&=& -\Big( (\varphi_1^-)^{\mu}_{\alpha} + (\varphi_2^-)^{\mu}_{\alpha} - (\varphi_1^+)^{\mu}_{\alpha} - (\varphi_2^+)^{\mu}_{\alpha} \Big) H(r-r_0) \, + \, \{\Phi^\mu_{\alpha,r}\}(r) \cr
&=& -\Big( (\varphi_1^-)^{\mu}_{\alpha} + (\varphi_2^-)^{\mu}_{\alpha} - (\varphi_1^+)^{\mu}_{\alpha} - (\varphi_2^+)^{\mu}_{\alpha} \Big) H(r-r_0),
\end{eqnarray}
where we applied \eqref{matching_C1_r} to compute the last equality. By \eqref{matching_pf_eqn1b}, the above sum in $(\varphi_i^\pm)^{\mu}_{\alpha}$ vanishes and thus \eqref{matching_C1_pf_2} yields
\beq\label{matching_C1_pf_3}
\{J^\mu_{\alpha,r} \}(r) = 0 \ \ \ \ \ \ \ \forall\,r\neq r_0,
\eeq
which proves \eqref{matching_C1_pf_1} for $\sigma=r$. 

To prove \eqref{matching_C1_pf_1} for $\sigma=t$, take the $t$-derivative of the canonical form \eqref{Jacobian2} and take the jump $\{\cdot\}$ of the resulting expression, which leads to
\begin{eqnarray}\nonumber
\{J^\mu_{\alpha,t} \}(r) 
&=& \Big(  (\varphi_1^-)^{\mu}_{\alpha} \dot{x}^-_1 + (\varphi_2^-)^{\mu}_{\alpha} \dot{x}^-_2 - (\varphi_1^+)^{\mu}_{\alpha} \dot{x}^+_1 - (\varphi_2^+)^{\mu}_{\alpha} \dot{x}^+_2  \Big) H(r-r_0) \cr
&& +  \Big( (\dot{\varphi}_1^-)^{\mu}_{\alpha} + (\dot{\varphi}_2^-)^{\mu}_{\alpha} - (\dot{\varphi}_1^+)^{\mu}_{\alpha} - (\dot{\varphi}_2^+)^{\mu}_{\alpha}  \Big) |r-r_0|  +  \{\Phi^\mu_{\alpha,t}\}(r) .
\end{eqnarray}
Using the RH conditions \eqref{RH_matching_pf} to absorb the shock speeds, we can write the term in the previous equation containing the step function as sums of the form \eqref{cancelation_metric-sum} and therefore conclude that it vanishes, leaving us with 
\beq \label{matching_C1_pf_4}
\{J^\mu_{\alpha,t} \}(r) 
= \Big( (\dot{\varphi}_1^-)^{\mu}_{\alpha} + (\dot{\varphi}_2^-)^{\mu}_{\alpha} - (\dot{\varphi}_1^+)^{\mu}_{\alpha} - (\dot{\varphi}_2^+)^{\mu}_{\alpha}  \Big) |r-r_0|  +  \{\Phi^\mu_{\alpha,t}\}(r).
\eeq
Applying now our initial assumption on $\{\Phi^t_{1,t} \}$ and $\{\Phi^r_{1,t} \}$, \eqref{matching_C1_t_first} and \eqref{matching_C1_t_second}, we find that \eqref{matching_C1_pf_4} implies
\beq\label{matching_C1_pf_5}
\{J^t_{1,t} \}(r) = 0 = \{J^r_{1,t} \}(r), \ \ \ \ \ \forall\,r\neq r_0.  
\eeq
To prove the remaining two cases of \eqref{matching_C1_pf_1}, we make use of the integrability condition in SSC, that is,
\beq\nonumber
 J^\mu_{0,\sigma} J^\sigma_1 -  J^\mu_{1,\sigma}  J^\sigma_0 = 0,
\eeq 
c.f. \eqref{IC in SSC}. Taking $\{\cdot\}$, then yields
\begin{align}\label{matching_C1_pf_6}
\{ J^t_{0,\sigma} \} J^\sigma_1 - \{ J^t_{1,\sigma} \} J^\sigma_0 = 0, \cr
\{ J^r_{0,\sigma} \} J^\sigma_1 -\{ J^r_{1,\sigma}\} J^\sigma_0 = 0,
\end{align}
where we used \eqref{matching_C0} to pull the undifferentiated Jacobian out of the curly brackets. Moreover, using \eqref{matching_C1_pf_5} and the continuous matching of $J^{\mu\pm}_{\alpha,r}$, that is \eqref{matching_C1_pf_3}, equations \eqref{matching_C1_pf_6} simplify further to
\begin{align}\label{matching_C1_pf_7}
\{ J^t_{0,t} \} J^t_1  = 0, \cr
\{ J^r_{0,t} \} J^t_1  = 0,
\end{align}
and since by assumption $\Phi^t_1$ is such that $J^t_1(0,r)\neq0$, we conclude that 
\beq\label{matching_C1_pf_8}
\{J^\mu_{\alpha,t} \}(r) = 0 \ \ \ \ \ \ \ \forall\,r\neq r_0.
\eeq
This proves the $C^1$-matching conditions for $r\neq r_0$, that is, \eqref{matching_C1_r=r}.

We now prove the $C^1$-matching conditions at $r=r_0$, that is, \eqref{matching_C1_r=r0}. We first derive the value of $\{g_{\mu\nu,\sigma}\}(r_0)$, where $\{\cdot\}(r_0)$ is defined in \eqref{def_curly_brackets_r0}. For this, compute
\begin{eqnarray}   \label{matching_C1_pf_9a}
\{g_{\mu\nu,\sigma}\}(r_0) &=& \left(g_{\mu\nu,\sigma}\right)_{M^-} - \left(g_{\mu\nu,\sigma}\right)_{M^+}   \cr
&=& \left(g_{\mu\nu,\sigma}\right)_{L}  - \left(g_{\mu\nu,\sigma}\right)_{M^+}  - \left(  \left(g_{\mu\nu,\sigma}\right)_{L} - \left( g_{\mu\nu,\sigma}\right)_{M^-} \right)  \cr 
&=&   [g_{\mu\nu,\sigma}]^+_1(0) - [g_{\mu\nu,\sigma}]^-_1(0),
\end{eqnarray}
or equivalently
\beq   \label{matching_C1_pf_9b}
\{g_{\mu\nu,\sigma}\}(r_0) = - [g_{\mu\nu,\sigma}]^+_2(0) + [g_{\mu\nu,\sigma}]^-_2(0).
\eeq
Now, since \eqref{matching_C1_r=r0} is the same linear system as the smoothing conditions \eqref{smoothingcondt1_SSC} but with the right hand side being a linear combination of the right hand side in \eqref{smoothingcondt1_SSC}, we conclude that the unique solution of \eqref{matching_C1_r=r0} is given by the corresponding linear combination of the solutions to \eqref{smoothingcondt1_SSC}, stated in Lemma \ref{smoothcondition}, namely
\beq\label{matching_condt_2b}
\{J^\mu_{\alpha,\sigma}\} = [J^\mu_{\alpha,\sigma}]^+_1(0) - [J^\mu_{\alpha,\sigma}]^-_1(0) = (\mathcal{J}_1^+)^{\mu}_{\alpha\sigma}(0) - (\mathcal{J}_1^-)^{\mu}_{\alpha\sigma}(0)
\eeq
or equivalently
\beq\label{matching_condt_2b_equiv}
\{J^\mu_{\alpha,\sigma}\} = -[J^\mu_{\alpha,\sigma}]^+_2(0) + [J^\mu_{\alpha,\sigma}]^-_2(0) = -(\mathcal{J}_2^+)^{\mu}_{\alpha\sigma}(0)  +  (\mathcal{J}_2^-)^{\mu}_{\alpha\sigma}(0),
\eeq
which we combine to
\beq\label{matching_condt_2b_equiv}
\{J^\mu_{\alpha,\sigma}\} =\frac12\Big( (\mathcal{J}_1^+)^{\mu}_{\alpha\sigma}(0)  -  (\mathcal{J}_1^-)^{\mu}_{\alpha\sigma}(0) - (\mathcal{J}_2^+)^{\mu}_{\alpha\sigma}(0) +  (\mathcal{J}_2^-)^{\mu}_{\alpha\sigma}(0) \Big),
\eeq
with the right hand side given by \eqref{smoothingcondt2}. \eqref{matching_condt_2b_equiv} is equivalent to the $C^1$ matching condition at $r=r_0$, \eqref{matching_condt_2b}.

We now prove that the Jacobian in the canonical form \eqref{Jacobian2} satisfies \eqref{matching_condt_2b_equiv} and thus meets the $C^1$ matching conditions at $r=r_0$. For this, observe that\footnote{Even though $\Phi^{\mu\pm}_{\alpha,t}$ do not match up continuously, \eqref{matching_C1_t_first} - \eqref{matching_C1_t_second} imply that $\{\Phi^{\mu}_{\alpha,t}\}(r)$ vanishes as $r$ approaches $r_0$, and thus $\left(\Phi^{\mu\pm}_{1,t}\right)_{L}$ and $\left(\Phi^{\mu\pm}_{1,t}\right)_{R}$ are well-defined. Likewise, using \eqref{matching_C1_pf_8}, the RH conditions \eqref{RH_matching_pf} and \eqref{matching_pf_eqn1b}, we find that $\left(\Phi^{\mu\pm}_{0,t}\right)_{L}$ and $\left(\Phi^{\mu\pm}_{1,t}\right)_{R}$ are well-defined.}
\begin{eqnarray}\label{matching_C1_pf_10}
\{ \Phi^\mu_{\alpha,\sigma}\}(r_0) &=& \left(\Phi^\mu_{\alpha,\sigma} \right)_{M^-} - \left(\Phi^\mu_{\alpha,\sigma} \right)_{M^+} \cr
&=& \left(\Phi^\mu_{\alpha,\sigma} \right)_{M^-} - \left(\Phi^\mu_{\alpha,\sigma} \right)_R - \left( \left(\Phi^\mu_{\alpha,\sigma} \right)_{M^+} - \left(\Phi^\mu_{\alpha,\sigma} \right)_R    \right)   \cr
&=& [\Phi^\mu_{\alpha,\sigma}]^-_2 - [\Phi^\mu_{\alpha,\sigma}]^+_2 \cr
&=& 0,
\end{eqnarray}
where we used \eqref{nojumps2} to conclude the last equality. Taking $\{\cdot\}$ of the $r$-derivative of the Jacobian in \eqref{Jacobian2} and applying \eqref{matching_C1_pf_10}, we obtain
\begin{eqnarray}\nonumber
\{J^\mu_{\alpha,r}\} &=&  \left(J^\mu_{\alpha,r} \right)_{M^-} - \left(J^\mu_{\alpha,r} \right)_{M^+} \cr 
&=& - \left( -(\varphi_1^-)^{\mu}_{\alpha}  + (\varphi_2^-)^{\mu}_{\alpha}  + (\varphi_1^+)^{\mu}_{\alpha}  - (\varphi_2^+)^{\mu}_{\alpha} \right) 
\end{eqnarray}
since for $r\in M^{\pm}$ we have $H(x^\pm_1(t)-r)=-1$ and $H(x^\pm_2(t)-r)=1$. Thus, recalling the implicit definition of the Jacobian coefficients, \eqref{Jacobiancoeff2_impl}, the above equations implies \eqref{matching_condt_2b_equiv} at $r=r_0$ for the case that $\sigma=r$.

We complete the proof by showing that the canonical Jacobian satisfies \eqref{matching_condt_2b_equiv}, at $r=r_0$, for the case $\sigma=t$. Taking $\{\cdot\}$ of the $t$-derivative of the Jacobian in \eqref{Jacobian2}, applying \eqref{matching_C1_pf_10}, and using that terms of the form $$
\dot{\varphi}^{\mu\pm}_{\alpha;\, i}\ |r-x^\pm_i(t)|
$$ 
vanish in the limit to $p$, we obtain
\begin{eqnarray}\nonumber
\{J^\mu_{\alpha,t}\} &=&  \left(J^\mu_{\alpha,t} \right)_{M^-} - \left(J^\mu_{\alpha,t} \right)_{M^+} \cr 
&=&  - (\varphi_1^-)^{\mu}_{\alpha}\, \dot{x}^-_1  + (\varphi_2^-)^{\mu}_{\alpha} \,\dot{x}^-_2  + (\varphi_1^+)^{\mu}_{\alpha}\, \dot{x}^+_1  - (\varphi_2^+)^{\mu}_{\alpha}\, \dot{x}^+_2  ,
\end{eqnarray}
since for $r\in M^{\pm}$ we have $H(x^\pm_1(t)-r)=-1$ and $H(x^\pm_2(t)-r)=1$. Using again the implicit definition of the Jacobian coefficient, \eqref{Jacobiancoeff2_impl}, we obtain
\begin{eqnarray}\nonumber
\{J^\mu_{\alpha,t}\} 
&=& -\frac12 \left(- (\mathcal{J}_1^-)^{\mu}_{\alpha r}\, \dot{x}^-_1  + (\mathcal{J}_2^-)^{\mu}_{\alpha r} \,\dot{x}^-_2  + (\mathcal{J}_1^+)^{\mu}_{\alpha r} \, \dot{x}^+_1  - (\mathcal{J}_2^+)^{\mu}_{\alpha r} \, \dot{x}^+_2  \right) \cr
&=& \frac12 \left(- (\mathcal{J}_1^-)^{\mu}_{\alpha t} + (\mathcal{J}_2^-)^{\mu}_{\alpha t}  + (\mathcal{J}_1^+)^{\mu}_{\alpha t} - (\mathcal{J}_2^+)^{\mu}_{\alpha t} \right) ,
\end{eqnarray}
where we used the relation $- \dot{x} \mathcal{J}^{\mu}_{\alpha r} =\mathcal{J}^{\mu}_{\alpha t}$, (which follows from the RH condition, c.f. \eqref{RH_smoothingcondt}), to obtain the last equality. This proves \eqref{matching_condt_2b_equiv} and concludes the proof of Lemma \ref{matching_lemma}.
\QED

\section{The Shock-Collision-Case: The Proof of Theorem \ref{TheoremMain}} \label{Sec: Proof_MainThm_Wrap-up}

We now use the construction of Section \ref{Sec: Canonical Jacobian} - \ref{Sec: Matching Conditions} to prove Theorem \ref{TheoremMain}. To begin with, recall our main theorem: \vspace{.2cm}

\noindent \textbf{Theorem 1.1}
{\it Suppose that $p$ is a  point of regular shock wave interaction in SSC between shocks from different families, in the sense that condition (i) - (iv) of Definition \ref{shockinteract} hold, for the SSC metric $g_{\mu\nu}$.  Then the following are equivalent:
\begin{enumerate}[(i)]
\item There exists a $\oneone$ coordinate transformation $x^\alpha\circ(x^\mu)^{-1}$ in the $(t,r)$-plane, with Jacobian $J^\mu_\alpha$, defined in a neighborhood $\mathcal{N}$ of $p$, such that the metric components $g_{\alpha\beta} = J^\mu_\alpha J^\nu_\beta g_{\mu\nu}$ are $\oneone$ functions of the coordinates $x^{\alpha}$.
\item The Rankine Hugoniot conditions, \eqref{RHwithN1} - \eqref{RHwithN2}, hold across each shock curve in the sense of (v) of Definition \ref{shockinteract}.
\end{enumerate}
Furthermore, the above equivalence also holds for the full atlas of $C^{1,1}$ coordinate transformations, not restricted to the $(t,r)$-plane.} \vspace{.2cm}

\noindent \emph{Proof.} \ By Proposition \ref{canonicalformforJ}, it is immediate that (i) implies (ii). The extension of the equivalence to the full atlas of $C^{0,1}$ coordinate transformations follows from Proposition \ref{canonicalformforJ} as well.

We now prove that (ii) implies (i), which then completes the proof. To construct the coordinate transformations asserted in (i), we use the canonical form of the Jacobian, \eqref{Jacobian2}, such that the free functions $\Phi^{\mu\pm}_\alpha$ satisfy the integrability condition, \eqref{IC_2shocks}, and the matching conditions, \eqref{matching_C0} - \eqref{matching_C1_t_second}. In more detail, choose two functions 
$$
\Phi^{t}_1,\ \Phi^{r}_1\, \in C^{0,1}(\mathcal{N}) \cap C^3\left(\mathcal{N}\cap \overline{\R^2_-}\right) \cap C^3\left(\mathcal{N}\cap \overline{\R^2_+}\right) ,
$$ 
such that they match across the $(t=0)$-interface according to \eqref{matching_C0_a} - \eqref{matching_C1_t_second}, and pick $\Phi^t_1$ such that $J^t_1 \neq 0$ on $\mathcal{N}\cap \{t=0\}$, where $J^t_1$ is related to $\Phi^t_1$ through the canonical form \eqref{Jacobian2}.     


Before we continue with the proof, let us first elaborate on why one can choose such functions $\Phi^{t}_1$ and $\Phi^{r}_1$. Observe that the right hand side of \eqref{matching_C1_t_first} - \eqref{matching_C1_t_second} vanishes at $r=r_0$. Thus, \eqref{matching_C1_t_first} - \eqref{matching_C1_t_second} simply requires $\Phi^{\mu}_{1,t}$ to match continuously at $r=r_0$, (for $\mu=t,r$). Moreover, the coefficient of $|r-r_0|$ on the right hand side of \eqref{matching_C1_t_first} - \eqref{matching_C1_t_second} depends only on the values of $\Phi^\mu_1$ and $\Phi^\mu_{1,\sigma}$ at the point of interaction $p=(0,r_0)$. Thus, choosing for $\Phi^{t}_1$ and $\Phi^{r}_1$ Lipschitz continuous functions for which first derivatives exist continuously at $p=(0,r_0)$, the right hand side of  \eqref{matching_C1_t_first} - \eqref{matching_C1_t_second} is determined \emph{explicitly} and condition \eqref{matching_C1_t_first} - \eqref{matching_C1_t_second} is satisfied at $p=(0,r_0)$. To satisfy  \eqref{matching_C1_t_first} - \eqref{matching_C1_t_second} for $r\neq r_0$, we modify the value of $\Phi^\mu_1$ on $\overline{\R^2_+}\cap \mathcal{N}$, without affecting the (fixed) right hand side of \eqref{matching_C1_t_first} - \eqref{matching_C1_t_second}, by adding two functions $\Psi^\mu_1 \in C^3(\overline{\R^2_+}\cap \mathcal{N})$ with the properties $\Psi^\mu_1(0,r)=0$ and $\Psi^\mu_{1,\sigma}(0,r_0)=0$, while the value of $\partial_t \Psi^\mu_\alpha(0,r)$ is given by the right hand side of  \eqref{matching_C1_t_first} - \eqref{matching_C1_t_second}. Denoting the resulting function again by $\Phi^\mu_1$, the conditions  \eqref{matching_C1_t_first} - \eqref{matching_C1_t_second} are satisfied.    

To continue, choose a non-vanishing function 
$$
U_0 =\,^T(U^1_0,U^2_0)\,\in\, C^2(\mathcal{N}\cap\{t=0\},\R^2),
$$
such that
\beq \label{nonzero_initial_determinant}
U_0^1(r_0) \Phi^r_1(p) - U_0^2(r_0) \Phi^t_1(p)  \neq 0,
\eeq 
which is the condition that the resulting Jacobian has a non-vanishing determinant. Now, in light of the existence theory for the integrability conditions, \eqref{IC_2shocks}, given in Proposition \ref{soln_IC_Prop}, we define the functions $\Phi^{t+}_0$ and $\Phi^{r+}_0$ to be the solutions of \eqref{IC_2shocks} on $\R^2_+$ for initial data $U_0$, and likewise, we define $\Phi^{t-}_0$ and $\Phi^{r-}_0$ to be the solutions of \eqref{IC_2shocks} on $\R^2_-$ for the same initial data $U_0$. In particular, the functions $\Phi^{t\pm}_0$ and $\Phi^{r\pm}_0$ are then in $C^{1,1}\left(\mathcal{N}\cap \overline{\R^2_\pm}\right)$ and satisfy the matching conditions \eqref{matching_C0_a} and \eqref{matching_C1_r}, since $\Phi^{\mu+}_0$ and $\Phi^{\mu+}_0$ have the same initial data.

Now, define $J^{\mu\pm}_\alpha$ to be the Jacobian in the canonical form \eqref{Jacobian2}, on $\mathcal{N}\cap\overline{\R^2_\pm}$, with the free functions $\Phi^{\mu\pm}_\alpha$ introduced above, that is, 
\beq\nonumber
J^{\mu\pm}_\alpha(t,r) = \sum\limits_{i=1,2} (\varphi_i^\pm)^{\mu}_\alpha (t) |x^\pm_i(t)-r| + \Phi^{\mu\pm}_\alpha(t,r). 
\eeq
Then $J^\mu_\alpha$ satisfies the integrability conditions, \eqref{IC_2shocks}, and thus also the original integrability condition, \eqref{IC}, c.f. Appendix \ref{sec IC}, so that $J^\mu_\alpha$ can be integrated to coordinates $x^\alpha$ defined on $\mathcal{N}$. In particular, by the continuity of the $J^\mu_\alpha$ and the initial condition \eqref{nonzero_initial_determinant}, it follows that the Jacobian determinant is non-vanishing in some neighborhood of $p$. Furthermore, by Proposition \ref{canonicalformforJ}, $J^\mu_\alpha$ satisfies the smoothing conditions \eqref{smoothingcondt2} across each of the shock curves and in the limit to $t=0$. From this we conclude that the transformed metric 
$$
g_{\alpha\beta} = J^\mu_\alpha J^\nu_\beta g_{\mu\nu}
$$
is $C^{1}$ across each shock curve with respect to the new coordinates, that is,
$$
[g_{\alpha\beta,\gamma}]^\pm_i = 0 \ \ \ \ \ \text{for} \, i=1,2.
$$
By Lemma \ref{matching_lemma}, $J^\mu_\alpha$ satisfies the $C^0$- and $C^1$-matching conditions, \eqref{matching_C0} - \eqref{matching_C1_r=r0}, from which we conclude that $g_{\alpha\beta}$ is $C^1$ across the $(t=0)$-interface, 
$$
\{g_{\alpha\beta,\gamma}\}(r)=0 \ \ \ \ \ \forall\, r\in(r_0-\epsilon,r_0+\epsilon).
$$
From the above two equations, we conclude that all directional derivatives of $g_{\alpha\beta}$ vanish across each of the shock curves and, in particular, at $p$, and the $C^1$ matching of $g_{\alpha\beta}$ implies that $g_{\alpha\beta}\,\in \oneone(\mathcal{N})$, (c.f. argument in the proof of Lemma \ref{smoothingcondt => J is oone across}).  In summary, we constructed coordinates $x^\alpha$ in which the metric is $C^{1,1}$ regular, which completes the proof of Theorem \ref{TheoremMain}.  \hfill $\Box$

\section{Conclusion}

The result of this paper shows that one can extend Israel's result to shock wave solutions of the Einstein equations containing points of shock wave interactions in spherical symmetry between shocks from different characteristic families, by using a new constructive method to construct Jacobians (integrable to coordinates) which raise the $C^{0,1}$ metric regularity to $C^{1,1}$. Our method differs drastically from Israel's approach, the latter being based on choosing a special coordinate system (Gaussian Normal coordinates) in which the metric is in $C^{1,1}$. In our method it is possible to characterize all Jacobians capable of smoothing the metric and it seems possible to extend our method to more complicated shock wave solutions. 

Our result here shows that no regularity singularities exist at points of shock wave interactions between shocks from different characteristic families, which corrects our wrong conclusion in \cite{ReintjesTemple}. Nevertheless, our result here still does not resolve the question whether such regularity singularities exist in more complicated shock wave interactions, neither in spherical symmetry nor in spacetimes without any symmetries. We believe that resolving this open question is fundamental to General Relativity, since perfect fluid matter models are basic to describe many astrophysical phenomena, and since one cannot avoid the formation of shock waves in the compressible Euler equations which describe such fluid models \cite{Christodoulou,Lax}. Moreover, if regularity singularities exists, it begs the question as to whether there are new general relativistic gravitational effects. In \cite{ReintjesTemple3}, we propose a mechanism by which the hypothetical structure of a regularity singularity could cause scattering effects in gravitational radiation.  In summary, our work here is a first step in resolving the problem we proposed in \cite{ReintjesTemple}, as to whether or not regularity singularities can be created by GR shock wave interactions, but, in general, the existence of regularity singularities remains an interesting open problem which lies at the very basis of GR fluid models.

\begin{appendix}

\section{The Integrability Condition}\label{sec IC}

In this section, we review the equivalence of the integrability condition for $J^\alpha_\mu$, \eqref{IC}, and the existence of an integrating factor $x^\alpha$, such that $J^\alpha_\mu$ is indeed the Jacobian of the coordinate transformation from $x^\mu$ to $x^\alpha$, that is,  
\beq \label{proper_Jacobian} 
J^\alpha_\mu = \frac{\partial x^\alpha}{\partial x^\mu} \, ,
\eeq
with indices $\alpha\in\{0,1\}$ and $\mu\in\{t,r\}$. In particular we also show that this equivalence holds for Lipschitz continuous $J^\alpha_\mu$. For simplicity, we only discuss this issue on $\R^2$ and for a single surface across which $J^\alpha_\mu$ is Lipschitz continuous.

\begin{Lemma}\label{equiv of IC and exi of integrable function} 
Let $\Omega$ be an open set in $\R^2$ with coordinates $x^\nu=(t,r)$. Suppose we are given a set of functions $J^\alpha_\mu(x^\nu)$ in $\oone(\Omega)$, $C^1$ away from some curve $\gamma(t)=(t,x(t))$, satisfying $\det(J^\alpha_\mu)\neq0$. Then the following is equivalent:
\begin{enumerate}[(i)]
 \item There exist locally invertible functions $x^\alpha(t,r) \, \in \,\oneone(\Omega)$, for $\alpha=0,1$, such that \eqref{proper_Jacobian} holds.
\item The set of functions $J^\alpha_\mu \, \in \, \oone(\Omega)$ satisfy the integrability condition
\beq\label{correctIC, appendix}
J^\alpha_{\mu,\nu} = J^\alpha_{\nu,\mu} \, .
\eeq
\end{enumerate}
\end{Lemma}
\Proof The implication from (i) to (ii) is trivial, since (weak) partial derivatives commute. We now prove that (ii) implies (i). Without loss of generality, we assume that $\Omega$ is the square region $(a,b)^2 \, \subset \, \R^2$. For ~$(t,r) \, \in \, \Omega$, introduce
\beq
x^\alpha(t,r)= \int_{a}^r J^\alpha_r(t,x) dx + \int_a^t J^\alpha_{t}(\tau,a)d\tau \, ,
\eeq
then $\partial_r x^\alpha(t,r)= J^\alpha_r(t,r)$ follows immediately. Furthermore, using \ref{correctIC, appendix} we get
\begin{eqnarray}\nonumber
\frac {\partial x^\alpha} {\partial t}(t,r) &=& \int_a^r J^\alpha_{r,t}(t,x)dx + J^\alpha_t(t,a) \cr
&=& \int_a^{x(t)} J^\alpha_{t,r}dx + \int_{x(t)}^r J^\alpha_{t,r}dx + J^\alpha_t(t,a) \, ,
\end{eqnarray}
where $x(t) \in (a,r)$ is the point of discontinuity of $J^\alpha_{t,r}$. We apply the fundamental theorem of calculus to each of the above integrals separately and finally obtain $\partial_t x^\alpha(t,r) =  J^\alpha_t (t,r)$.
Moreover, the Inverse Function Theorem implies that the function $x^\alpha$ is bijective on some open set, since we assumed $\det \left( J^\alpha_\mu \right) \neq 0$.
\QED

Given the existence of coordinates $x^\alpha(t,r)$, the integrability condition, 
\beq \label{IC_appendix}
J^\mu_{\alpha,\beta}=J^\mu_{\beta,\alpha},
\eeq
holds and implies, by the chain rule,
\beq \label{IC in SSC}
J^\mu_{\alpha,\nu}J^\nu_\beta=J^\mu_{\beta,\nu}J^\nu_\alpha \ .
\eeq
We now prove that the reverse implications hold true as well, which is fundamental for our construction of coordinates where the metric is $C^{1,1}$ regular.

\begin{Lemma} \label{Appendix_Lemma2}
Assume we are given $C^{0,1}$ functions $J^\mu_\alpha$ which satisfy $\det J^\mu_\alpha \neq 0$ and \eqref{IC in SSC}. Then, the linear algebraic inverse of $J^\mu_\alpha$, $J^\alpha_\mu$, satisfies \eqref{correctIC, appendix} and there exist a bijective $C^{1,1}$ function $x^\alpha(t,r)$ which satisfies \eqref{proper_Jacobian}. 
Moreover, differentiating with respect to $x^\alpha$, \eqref{IC_appendix} holds. 
\end{Lemma}
\Proof
We begin by proving that \eqref{IC in SSC} implies the integrability condition on the inverse Jacobian $J^\alpha_\mu$, that is,
\beq\label{correctIC}
J^\alpha_{\mu,\nu} = J^\alpha_{\nu,\mu} \, .
\eeq
To begin with, suppose \ref{IC in SSC} holds, which is equivalent to
\begin{eqnarray}
J^r_1 J^t_{0,r} - J^r_0 J^t_{1,r} &=& J^t_0 J^t_{1,t} - J^t_1 J^t_{0,t} \label{ICa, techlemma1} \\
J^r_1 J^r_{0,r} - J^r_0 J^r_{1,r} &=& J^t_0 J^r_{1,t} - J^t_1 J^r_{0,t}  \, . \label{ICb, techlemma1}
\end{eqnarray}
The linear algebraic inverse of $J^\mu_\alpha$,  $J^\alpha_\mu$, is given by
\beq \label{inverse Jacobian}
\left( \begin{array}{cc} J^0_t & J^0_r \cr J^1_t & J^1_r \end{array} \right) = \frac1{\left| J \right|}\left( \begin{array}{cc} J^r_1 & -J^t_1 \cr -J^r_0 & J^t_0 \end{array} \right),
\eeq
where $|J|$ denotes the determinant of $J^\mu_\alpha$. Now, taking the $r$-derivative of $J^0_t$ in \eqref{inverse Jacobian} gives
\begin{eqnarray} \nonumber
J^0_{t,r} &=& \frac{J^t_1}{|J|^2} \left( J^r_1 J^r_{0,r} - J^r_0 J^r_{1,r} \right) - \frac {J^r_1}{|J|^2} \left( J^r_1 J^t_{0,r} - J^r_0 J^t_{1,r} \right).
\end{eqnarray}
Exchanging the first term using \eqref{ICb, techlemma1} and the second term using \eqref{ICa, techlemma1}, we get
\begin{eqnarray}\nonumber
J^0_{t,r} &=&  \frac{J^t_1}{|J|^2} \left( J^t_0 J^r_{1,t} - J^t_1 J^r_{0,t} \right) - \frac {J^r_1}{|J|^2} \left( J^t_0 J^t_{1,t} - J^t_1 J^t_{0,t} \right) \cr
&=& J^0_{r,t} \, ,
\end{eqnarray}
but the right hand side in the equation is exactly what one get when taking the $t$-derivative of $J^0_r$ in  \eqref{inverse Jacobian}. We thus conclude that 
$$
J^0_{t,r} = J^0_{r,t}
$$
holds true. A similar computation verifies the remaining equation in \eqref{correctIC}, that is,
$$
J^1_{t,r}=J^1_{r,t}.
$$ 
This shows that \eqref{IC in SSC} implies \eqref{correctIC}.

Lemma \ref{equiv of IC and exi of integrable function} now implies the existence of coordinates $x^\alpha$ such that $\frac{\partial x^\alpha}{\partial x^\mu} = J^\alpha_\mu$ holds and such that the mapping ~$(x^\mu)\,\mapsto\,(x^\alpha)$ is bijective. Since the coordinate mapping is bijective, we conclude by the Inverse Function theorem that $J^\mu_\alpha$ is indeed the Jacobian of the coordinate transformation, that is,
\beq \nonumber
J^\mu_\alpha =\frac{\partial x^\mu}{\partial x^\alpha}\, ,
\eeq
which also implies that $J^\mu_{\alpha,\beta} = J^\mu_{\beta,\alpha}$ holds.
\QED

\end{appendix}

\section*{Acknowledgments}

I am grateful to Blake Temple for supervising my dissertation and proposing to work on the question of the metric regularity at points of shock wave interaction for my doctoral thesis. In particular, I thank him for many helpful discussions about the existence theory in Section \ref{Sec: Non-local PDE}.

\section*{Funding}

M. R. was supported by the Deutsche Forschungsgemeinschaft (DFG), Grant Number RE 3471/2-1, from January 2013 until December 2014. From January 2015 until December 2016, M. R. was supported by CAPES-Brasil, as a Post-Doc of Excellence at IMPA (Instituto Nacional de Matem{\'a}tica Pura e Aplicada) in Rio de Janeiro, Brazil.

\providecommand{\bysame}{\leavevmode\hbox to3em{\hrulefill}\thinspace}
\providecommand{\MR}{\relax\ifhmode\unskip\space\fi MR }
\providecommand{\MRhref}[2]{%
  \href{http://www.ams.org/mathscinet-getitem?mr=#1}{#2}
}
\providecommand{\href}[2]{#2}

\bibliographystyle{my-h-elsevier}

\end{document}